\documentclass[]{aa} % normal version
\usepackage{amsmath}
\usepackage{graphicx}
\usepackage{caption}
\usepackage{subcaption}
\usepackage{txfonts}
\usepackage{natbib}
\usepackage{xcolor}
\usepackage{multirow}

\bibpunct{(}{)}{;}{a}{}{,}

\def\Oiii{[O~{\sc iii}] $\lambda$5007}
\def\teff{\ifmmode T_{\rm eff} \else $T_{\mathrm{eff}}$\fi}

\def\ltsima{$\buildrel<\over\sim$}
\def\lsim{\lower.5ex\hbox{\ltsima}}

\newcommand{\ha}{\ifmmode {\rm H}\alpha \else H$\alpha$\fi}
\newcommand{\hb}{\ifmmode {\rm H}\beta \else H$\beta$\fi}
\newcommand{\lya}{\ifmmode {\rm Ly}\alpha \else Ly$\alpha$\fi}

\newcommand{\ebv}{\ifmmode E_{\rm B-V} \else $E_{\rm B-V}$ \fi}
\def\micron{$\mu$m}

\def\msun{\ifmmode \rm{M}_{\odot} \else M$_{\odot}$\fi}
\def\msunyr{\ifmmode \rm{M}_{\odot} {\rm yr}^{-1} \else M$_{\odot}$ yr$^{-1}$\fi}
\def\zsun{\ifmmode Z_{\odot} \else Z$_{\odot}$\fi}
\def\lsun{\ifmmode \rm{L}_{\odot} \else L$_{\odot}$\fi}

\def\mup{\ifmmode M_{\rm up} \else M$_{\rm up}$\fi}
\def\mlow{\ifmmode M_{\rm low} \else M$_{\rm low}$\fi}
\newcommand{\oh}{\ifmmode 12 + \log({\rm O/H}) \else$12 + \log({\rm
O/H})$\fi}
\def\Siii{[S~{\sc iii}] $\lambda\lambda$9069,9532}
\def\Oii{[O~{\sc ii}] $\lambda$3727}
\def\Oiii{[O~{\sc iii}] $\lambda\lambda$4959,5007}
\def\hyperz{{\em Hyperz}}
\def\flyf{\ifmmode f_{\rm Lyf} \else $f_{\rm Lyf}$\fi}
\def\pz{\ifmmode P(z) \else $P(z)$\fi}
\def\ki2{\ifmmode \chi^2_\nu \else $\chi^2_\nu$\fi}
\def\zphot{\ifmmode z_{\rm phot} \else $z_{\rm phot}$\fi}

\newcommand{\xphot}{\ifmmode x_\gamma \else $v_\gamma$\fi}
\newcommand{\xobs}{\ifmmode x_{\rm obs} \else $x_{\rm obs}$\fi}
\newcommand{\xcmf}{\ifmmode x_{\rm CMF} \else $x_{\rm CMF}$\fi}
\newcommand{\vexp}{\ifmmode V_{\rm exp} \else $V_{\rm exp}$\fi}
\newcommand{\vmax}{\ifmmode V_{\rm max} \else $V_{\rm max}$\fi}
\newcommand{\nh}{\ifmmode N_{\rm HI} \else $N_{\rm HI}$\fi}
\newcommand{\dv}{\ifmmode \Delta v({\rm em-abs}) \else $\Delta v({\rm em}-{\rm abs})$\fi}

\def\fesc{\ifmmode f_{\rm esc} \else $f_{\rm esc}$\fi}

\def\frellya{\ifmmode f^{\rm rel}_{\rm{Ly}\alpha} \else $f^{\rm rel}_{\rm{Ly}\alpha}$\fi}

\newcommand{\mstar}{\ifmmode M_\star \else $M_\star$\fi}
\newcommand{\muv}{\ifmmode M_{1500} \else $M_{1500}$\fi}
\newcommand{\luv}{\ifmmode L_{\rm UV} \else $L_{\rm UV}$\fi}
\newcommand{\lfir}{\ifmmode L_{\rm IR} \else $L_{\rm IR}$\fi}
\newcommand{\lir}{\ifmmode L_{\rm IR} \else $L_{\rm IR}$\fi}
\newcommand{\lbol}{\ifmmode L_{\rm bol} \else $L_{\rm bol}$\fi}
\newcommand{\liruv}{\ifmmode L_{\rm IR+UV} \else $L_{\rm IR+UV}$\fi}
\newcommand{\liroveruv}{\ifmmode L_{\rm IR}/L_{\rm UV} \else $L_{\rm IR}/L_{\rm UV}$\fi}

\newcommand{\sfrbc}{\ifmmode {\rm SFR}_{\rm BC} \else SFR$_{\rm BC}$\fi}
\newcommand{\sfrir}{\ifmmode {\rm SFR}_{\rm IR} \else SFR$_{\rm IR}$\fi}
\newcommand{\tdust}{\ifmmode T_{\rm dust} \else $T_{\rm dust}$\fi}
\newcommand{\tpeak}{\ifmmode T_{\rm peak} \else $T_{\rm peak}$\fi}
\newcommand{\av}{\ifmmode {\rm A}_{\rm V} \else A$_{\rm V}$\fi}
\newcommand{\afuv}{\ifmmode  {\rm A}_{\rm FUV} \else A$_{\rm FUV}$\fi}
\newcommand{\sfruvir}{\ifmmode {\rm SFR}_{\rm IR} + {\rm SFR}_{\rm UV} \else ${\rm SFR}_{\rm IR} + {\rm SFR}_{\rm UV}$\fi}
\begin{document}
\title{Star formation histories, extinction and dust properties of strongly lensed $z \sim$ 1.5--3 star-forming galaxies from the \textit{Herschel Lensing Survey}}
  \subtitle{}
  \author{P. Sklias\inst{1}, M. Zamojski\inst{1},D. Schaerer\inst{1,2}, M. Dessauges-Zavadsky\inst{1}, E. Egami\inst{3}, M. Rex\inst{3}, T. Rawle\inst{3,4}, 
  J. Richard\inst{5}, F. Boone\inst{2}, J.M. Simpson\inst{6}, I. Smail\inst{6}, P. van der Werf\inst{7}, B. Altieri\inst{4}, J.P. Kneib\inst{8}}
%  \offprints{}						
  \institute{
Observatoire de Gen\`eve, Universit\'e de Gen\`eve, 51 Ch. des Maillettes, 1290 Versoix, Switzerland
         \and
CNRS, IRAP, 14 Avenue E. Belin, 31400 Toulouse, France
 		\and
Steward Observatory, University of Arizona, 933 North Cherry Avenue, Tucson, AZ 85721, USA
         \and
         ESAC, ESA, PO Box 78, Villanueva de la Canada, 28691 Madrid, Spain
         \and
CRAL, Observatoire de Lyon, Universite Lyon 1, 9 Avenue Ch. Andre, 69561 Saint Genis Laval Cedex, France
\and
Institute for Computational Cosmology, Durham University, South Road, Durham DH1 3LE, UK
\and
Leiden Observatory, Leiden University, P.O. Box 9513, NL-2300 RA Leiden, The Netherlands
\and
Laboratoire d'Astrophysique, Ecole Polytechnique Federale de Lausanne (EPFL), Observatoire, 1290 Sauverny, Switzerland
         }

\authorrunning{Sklias et al.}
\titlerunning{Physical properties of strongly lensed $z \sim$ 1.5--3 star-forming galaxies from the \textit{Herschel Lensing Survey}}

\date{Received 1 August 2013; accepted date}

%\abstract{%CONTEXT}
%{%AIMS
%}
%{%METHODS
%}
%{%RESULTS
%}
%{%CONCLUSIONS
%}
% 5 {} token are mandatory
\abstract{Multi-wavelength, optical to IR/sub-mm observations of strongly lensed galaxies identified 
by the \emph{Herschel Lensing Survey} are used to determine the physical properties of high-redshift
star-forming galaxies close to or below the detection limits of blank fields.}
{We aim to constrain their stellar and dust content, determine star formation rates and histories, dust attenuation 
and extinction laws, and other related properties.}
{We study a sample of 7 galaxies with spectroscopic redshifts $z \sim 1.5-3$, which are well detected thanks to gravitational lensing, 
and whose SED is well determined from the rest-frame UV to the IR/mm domain. For comparison, our sample includes
two previously well-studied lensed galaxies, MS1512-cB58 and the Cosmic Eye, for which we also provide updated \textit{Herschel} measurements.
We perform SED-fits of the full photometry of each object as well for the optical and 
infrared parts separately, exploring various star formation histories, using different extinction laws,
and exploring the impact of nebular emission. The IR luminosity, in particular, is predicted consistently
from the stellar population model. The IR observations and emission line measurements, where available, 
are used as \emph{ a posteriori} constraints on the models.
We also explore ``energy conserving models'', that we create by using the observed IR/UV ratio to estimate the extinction.}
{ Among the models we have tested, models with exponentially declining star-forming histories including nebular emission and assuming the Calzetti 
attenuation law best fit most of the observables. SED fits assuming constant or rising star 
formation histories predict in most cases too much IR luminosity.
The SMC extinction law underpredicts the IR luminosity in most cases, except for 2 out of 7 galaxies, where
we cannot distinguish between different extinction laws.
Our sample has a median lensing-corrected IR luminosity $\sim 3\times 10^{11}$ \lsun, stellar masses between
$2 \times 10^9$ and $2 \times 10^{11}$ \msun, and IR/UV luminosity ratios spanning a wide
range.
The dust masses of our galaxies are in the range $[2-17] \times 10^7$ \msun, extending previous studies at the same
redshift down to lower masses. %approximately 10 times
We do not find any particular trend of the dust temperature \tdust\ with \lir, suggesting an overall warmer dust regime at 
our redshift regardless of IR luminosity. 
}
{Gravitational lensing enables us to study the detailed physical properties of individual IR-detected 
$z \sim 1.5-3$ galaxies up to a factor $\sim 10$ fainter than achieved with deep blank field observations.
We have in particular demonstrated that multi-wavelength observations combining stellar and dust emission can 
constrain star formation histories and extinction laws of star-forming galaxies, as proposed in an
earlier paper. 
Fixing the extinction based on the IR/UV observations successfully breaks the age-extinction degeneracy often
encountered in obscured galaxies.}

 \keywords{Galaxies: Starburst -- Galaxies: ISM -- 
Infrared: Galaxies -- ISM: dust, extinction}

  \maketitle

%%%%%%%%%%%%%%%%%%%%%%%%%%%%%%%%%%%%%%%%%%%%%%%%%%%%%%%%%%%%%%%%%%%%%%%%%%%%%%%%%
\section{Introduction}

Strong gravitational lensing offers several interesting opportunities for studies of distant galaxies.
\citep[e.g.\ the review by][]{2011A&ARv..19...47K}. 
The magnification effect allows one to detect galaxies below the detection limits reached in blank fields,
or to significantly improve the S/N of observations of galaxies with the same intrinsic (i.e.\ unlensed) flux.
Lensing provides a gain in spatial resolution in the case of strongly lensed, extended sources.
Furthermore, when targeting massive galaxy clusters known as efficient gravitational lenses,
the confusion limit is reduced in the central region, allowing in particular IR observations
to probe deeper than in blank fields.
Exploiting these advantages for IR observations of distant galaxies is one goals of the
the \emph{Herschel Lensing Survey}, hereafter \emph{HLS} \citep{2010A&A...518L..12E},  targeting 54 galaxy clusters
known for being efficient gravitational lenses.
We examined the dust emission of two IR-bright, highly lensed sources behind the Bullet Cluster \citep{2010A&A...518L..13R}
and A773 \citep[][Rawle et al., \textit{submitted}]{2012A&A...538L...4C}, but even with magnification these sources are 
too faint to be detected in optical bands.

In the present work we study in detail a small sample of bright, strongly lensed galaxies
at redshift $z \sim 1.6-3.2$ detected  between 100 and 500 \micron\ with the PACS and SPIRE instruments 
on board the {\it Herschel} Space Observatory \citep{2010A&A...518L...1P}.
Our sample consists of 5 galaxies drawn from the bright HLS sources described by Rawle et al. (\emph{in prep.}) 
and two well-known star-forming galaxies recently observed with {\it Herschel},
MS1512-cB58 \citep[][hereafter, simply cB58]{1996AJ....111.1783Y}
and  the ``Cosmic Eye"  \citep{2007ApJ...654L..33S}.
The extensive multi-wavelength data available for these galaxies, both in imaging and spectroscopy,
allows us to carry out an empirical study of these strongly lensed galaxies and to
model their spectral energy distribution (SED) in detail, to determine their stellar populations and
dust content. We will discuss their molecular gas content, in a companion paper
(Dessauges-Zavadsky et al., \emph{in prep.}).

Such a study is of interest for a variety of reasons.
For example, direct measurements of the IR and UV luminosity provide the best measurement
of the effective dust attenuation in star-forming galaxies
\citep[cf.][]{burgarella05,buatetal2005,2010MNRAS.409L...1B,kong04,2013ApJ...762..125N,2012ApJ...755..144T}.	 
While previous observations with {\it Spitzer} have often been used to estimate the total IR luminosity
\lir\ from 24 \micron\ imaging, it has become clear with {\it Herschel} that this extrapolation
is inaccurate for redshifts $z \ga 2$ \citep{elbaz2011}. An alternative computation of
the  24 \micron\ to \lir\ conversion has been published by \cite{2013ApJ...767...73R}, 
extending its applicability to z $ \sim 2.8 $.
Ideally, complete IR observations, measuring directly the peak of the IR emission, are therefore
needed to determine accurate IR luminosities.
Whereas such measurements are now becoming available for some individual galaxies
at $z \sim$ 2--4 \citep[e.g.][]{rodighieroetal2011,2012A&A...545A.141B,2012PKAS...27..311B,2012ApJ...759...28P,2012ApJ...754...25R}, 
this is currently restricted to very luminous galaxies, typically to $\lir > 10^{12}$ \lsun\ at $z >2$,
i.e.\ to the regime of ULIRGs (Ultra-luminous IR galaxies).
Alternative, stacking techniques are employed to determine the average properties of fainter
galaxies, as done e.g.\ by \citet{2012ApJ...758L..31L,2013MNRAS.429.1113H,2013arXiv1307.3556I}. % Heinis et al. 2012, Lee et al. 2012
Our, admittedly small, sample of lensed galaxies allows us to push individual galaxy detections
well into the LIRG domain ($10^{11} \le \lir/\lsun \le 10^{12}$).

Direct IR observations of individual dusty galaxies provide also an independent measure
of their total star-formation rate (SFR), and as such important constraints and tests on SFR determinations
e.g.\ from the dust-corrected UV SFR, from SFR(\ha), or from the SFR derived from SED
fits to the commonly available part of the spectrum, i.e.\ the optical to near-IR bands.
For example, it is generally found that dust-corrected SFR(UV) or SFR(\ha) agree
approximately with SFR(IR) for ``not too dusty" galaxies, whereas these UV-optical
features severely underestimate the true SFR for the most dusty galaxies \citep{goldader02,2005ApJ...622..772C,
2011ApJ...738..106W,2013A&A...554L...3O}.
This discrepancy is usually attributed to ``optical depth effects". \cite{Calzetti2001} argues that in the 
case of extremely dust obscured star-forming regions the UV emission can be suppressed to such a level that it would not 
impact the UV spectrum, which would be then dominated by the emission of young stars in less obscured star forming regions, thus
giving the impression of a ``grayer" reddening  that underestimates strongly the global dust obscuration. 
As a consequence, extinction corrected UV-inferred SFRs can miss a large proportion of the star-formation
occurring in such galaxies.

Other examples of the use of SFR comparisons show that the (instantaneous) SFR
determined from the SED fits may show systematic offsets from other SFR indicators.
Such cases are e.g.\ found in the recent studies of \cite{2012ApJ...754...25R,2012ApJ...745...86W}, who
find that SFR(SED) overestimates the ``true SFR'' (derived from the UV+IR luminosity)
by up to a factor ten for young galaxies with $\lbol < 10^{12}$ \lsun, when analyzed with
declining star formation histories.
Similarly results are found by \cite{Wuyts2011} for four lensed galaxies assuming, however, 
constant SFR.
These authors attribute these differences either to an inappropriate extinction law, favoring
e.g.\ the SMC law over the commonly used Calzetti attenuation law for starbursts,
or to assumptions on the star formation histories made in the SED fits. \citet{2012ApJ...754...25R}
suggest also that exponentially rising star formation histories (hereafter SFHs) are
more appropriate to describe galaxies at $z \ga 2$, echoing earlier claims by
several studies and based on different arguments
\citep{renzini2009,marastonetal2010,finlatoretal2007,finlatoretal2010,finlatoretal2011,finkelsteinetal2010,papovichetal2011} .

In a recent analysis of a large sample of Lyman break galaxies (LBGs)
at $z \sim$ 3--6 and using an SED fitting code making consistent predictions for the IR emission,
we have shown that different star formation histories and extinction laws can in principle
be distinguished when \lir\ measurements are available,  and emission line observations
provide further constraints \citep{2013A&A...549A...4S}. So far, however, very few such
data are available for high-$z$ galaxies. Applying, therefore, this method to
somewhat lower redshift galaxies, should be an important proof-of-concept
before larger numbers of galaxies can be observed in the IR with upcoming
facilities such as ALMA. The present sample provides an interesting opportunity
for such tests.

The sample of lensed galaxies studied in this paper allows us also to
carry out other important tests of our recent SED models including the effects
of nebular emission \citep{2013A&A...549A...4S,2012arXiv1207.3663D}. For example,
our SED models predict on average higher specific star formation rates
(sSFR=SFR/\mstar) at $z \ge 3$ than commonly obtained using standard
SED fits neglecting emission lines and assuming constant SFR
and an increase of the sSFR with redshift \citep{debarros2011,2012arXiv1207.3663D}.
How does this trend behave when going to lower redshift? Do
our models yield systematic offsets of the sSFR also at $z \sim 2$, where
a large number sSFR measurements are available, using different
techniques \citep[e.g.][]{2007ApJ...670..156D,elbazetal2007}.
Or do our models naturally ``converge'' towards the available literature
data at $z \sim 2$?
Related to this is the question whether the stellar population ages derived
from our SED models are realistic for $z \sim 2$ galaxies, or whether
models including nebular lines provide too young ages,
as e.g.\ suggested by some authors \citep{2012arXiv1211.1010O}. 
The present sample of strongly lensed, $z \sim$ 1.6--3 galaxies with
a fine multi-wavelength coverage including the optical, near-IR, and IR
domain and (partial) emission line measurements, is ideal
to examine these questions.

The remainder of our paper is structured as follows.
In Sect.\ \ref{s_obs} we present the observational data and our galaxy sample.
Our SED fitting tools is described in Sect.\ \ref{sed_fits}.
The derived IR properties of our galaxies are shown in Sect.\ \ref{s_IR}.
The detailed SED fitting results for each galaxy are discussed in Sect.\ \ref{sec_indiv_sources}.
In Sect \ref{s_discuss} we discuss the global properties we obtain from this sample by topic (e.g. star formation histories,
extinction, the \lir/\luv\ ratio, the dust properties, and so on) and our main results are summarized in Sect.\ \ref{s_conclude}.
We adopt a $\Lambda$-CDM cosmological model with $H_{0}$=70 km s$^{-1}$ Mpc$^{-1}$,
$\Omega_{m}$=0.3 and $\Omega_{\Lambda}$=0.7.

%%%%%%%%%%%%%%%%%%%%%%%%%%%%%%%%%%%%%%%%%%%%%%%%%%%%%%%%%%%%%%%%%%%%%%%%%%%%%%%%%
\section{Observations}
\label{s_obs} 
 
\subsection{Sample}	\label{s_sample}
 
We present a UV-to-FIR SED analysis of seven  star-forming galaxies at redshifts $z \sim 1.5$--3, five of which we selected from the {\it Herschel Lensing Survey}, or HLS. The HLS sources were selected mostly from the {\em Herschel} observations of the galaxy cluster Abell~68, as this cluster is located in the foreground of several high-redshift infrared-bright galaxies two of which are strongly lensed (amplification factors of $\mu = 15$ and $\mu = 30$). 

From this cluster field, we selected all galaxies that have a well-determined spectroscopic redshift in the range $z \sim 1 \textendash 3$ and that are bright in {\em Herschel}.  Formally, we used the PACS $160\mu$m band to select our sources, but our sources are detected in {\em all Herschel} (both PACS and SPIRE) bands.  They also do not suffer from high ``source crowding" which allows an accurate determination of their SED up to 500 $\mu$m.  Although no formal flux limit was imposed, our faintest source has a flux of $S_{\nu} = 25$ mJy at $160\mu$m.  One source with a spectroscopic redshift is not detected in PACS (or SPIRE), and another one falls outside the PACS maps (but is detected in SPIRE).  We did {\em not} include these two sources in the current study.  In total, four galaxies meet our selection criteria in Abell~68.

We augment this sample with another well-known highly-magnified and {\em Herschel}-detected galaxy from the {\em HLS}:  the giant arc in MACSJ0451+0006.  This galaxy has a known spectroscopic redshift of $z = 2.013$ and a magnification factor of $\mu = 49$.  It allows us to extend the span of intrinsic stellar masses of our sample to even lower masses.   Since the purpose of this work is to analyze in detail a small number of objects, we did not attempt, at this point, to extract a larger sample from the {\em HLS}, and limit ourselves to this heterogeneous sample of five galaxies.

Finally, for comparison purposes, we also reanalyze in a homogenous way two well known lensed galaxies,  MS1512-cB58 \citep[][hereafter, simply cB58]{1996AJ....111.1783Y}
and  the ``Cosmic eye" \citep{2007ApJ...654L..33S}, for which we extracted {\it Herschel} data from the archive.  We re-processed the PACS data with the new {\sc unimap} map-maker.  These two objects differ from our main sample in that they were not {\em Herschel}-selected.  They do appear, in fact, fainter in PACS and SPIRE than our other sources, and also suffer from more severe blending.  Nevertheless, we were able to set good constraints on their infrared properties, and so they provide a useful comparison for our sample.

%------------------------------
\begin{table}[htb]

	\centering
	\begin{tabular}{l c c c c }
		\hline
		\hline
		  ID  & RA   & DEC	& z  & $ \mu $\\\hline
       	A68/C0  &	00 37 07.38 	&	+09 09 26.35	& 1.5854 & 30 	\\
       	A68/HLS115	&	00 37 09.50 	&	+09 09 03.97	&  1.5859 & 15	\\
       	A68/h7	 &	00 37 01 47	&	+09 10 22.13		&  2.15& 3 \\
		A68/nn4  &	00 37 10 42	& +09 08 46.05		& 3.19 & 2.3 	\\
 		MACS J0451+0006 &	04 51 57.27 	& +00 06 20.7	&2.013 & 49 \\	
		\hline   
	\end{tabular}
	\caption{ Coordinates (J2000) of the 5 HLS sources. Redshifts of A68/C0 and A68/HLS115 come from CO observations.
	Magnification factors are from the mass model of A68 from \cite{2010MNRAS.404..325R}, and from \cite{2010MNRAS.404.1247J}.}
	              \label{sample_coord}
\end{table}
%------------------------------

\subsubsection{Description of the objects}
\label{s_objects}
We now briefly describe our targets and the available information. The redshifts 
and magnification factors of the HLS sources are given in Table \ref{sample_coord}.
The targets are illustrated in Fig.\ 1.

 \begin{figure*}[t]
 \centering
% \vspace{-0.05\textwidth}
 \begin{subfigure}[c]{0.52\textwidth}
 \centering
 \begin{tabular}{cc}
 \multirow{2}{0.6\textwidth}{
 \includegraphics[width=0.6\textwidth]{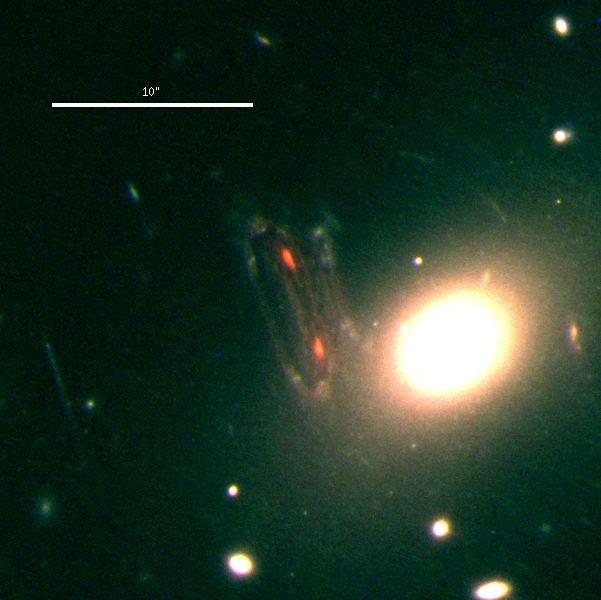}} \\
 & \includegraphics[width=0.3\textwidth]{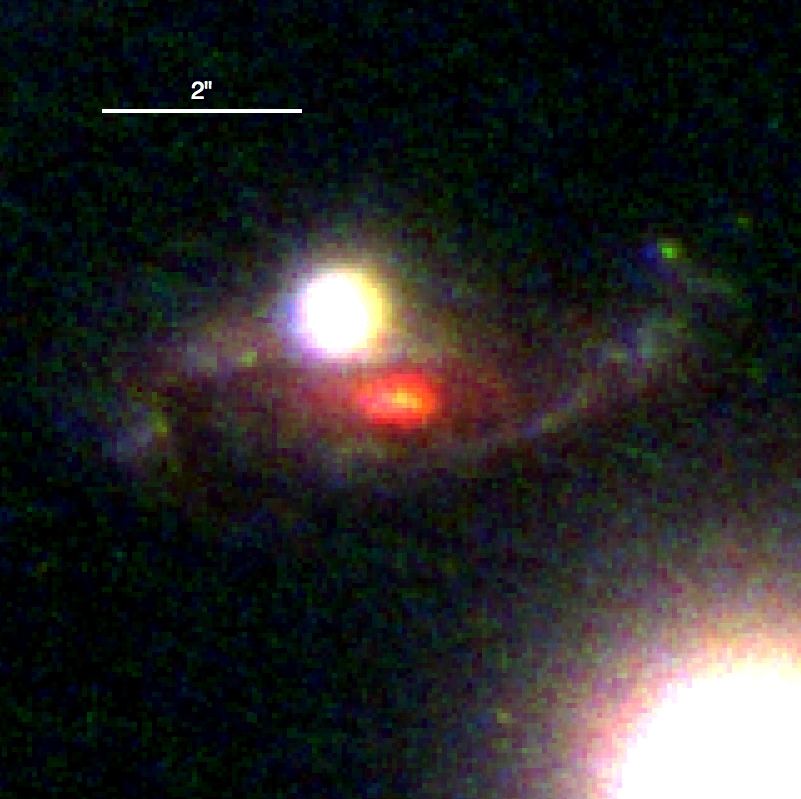} \\
 & \includegraphics[width=0.3\textwidth]{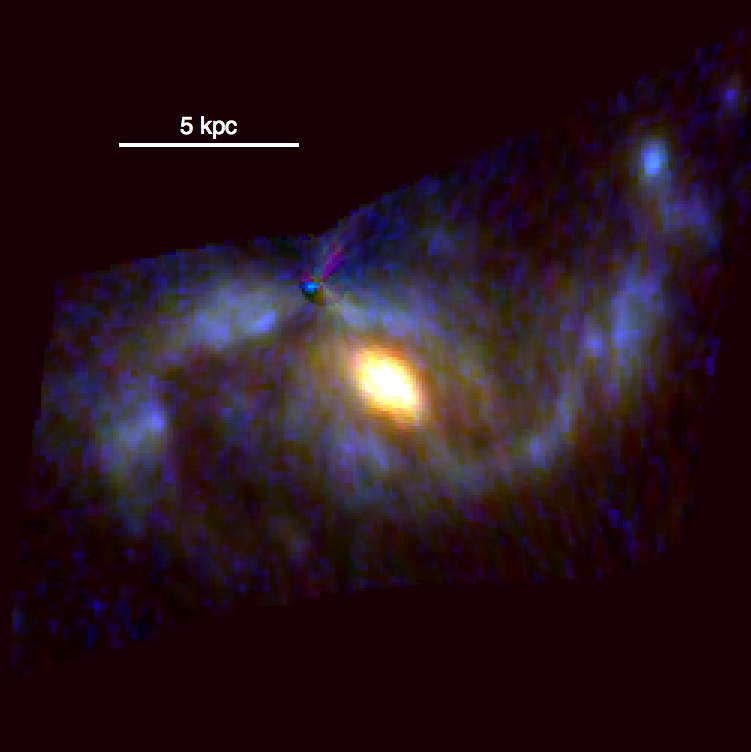} \\
 \end{tabular}
 \caption{A68-C0}
 \label{subfig:A68-C0}
 \end{subfigure}
 \begin{subfigure}[c]{0.46\textwidth}
 %\centering
 \includegraphics[width=0.49\textwidth]{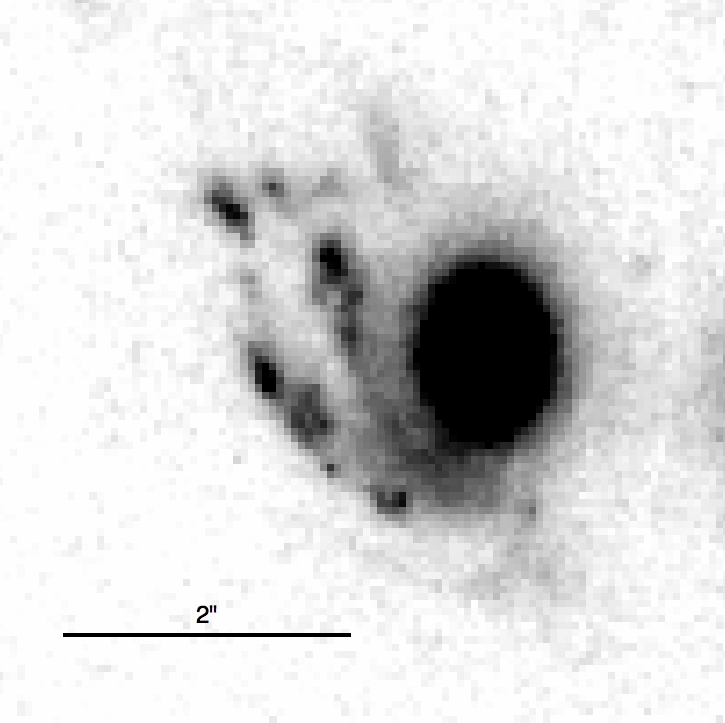}
 \includegraphics[width=0.49\textwidth]{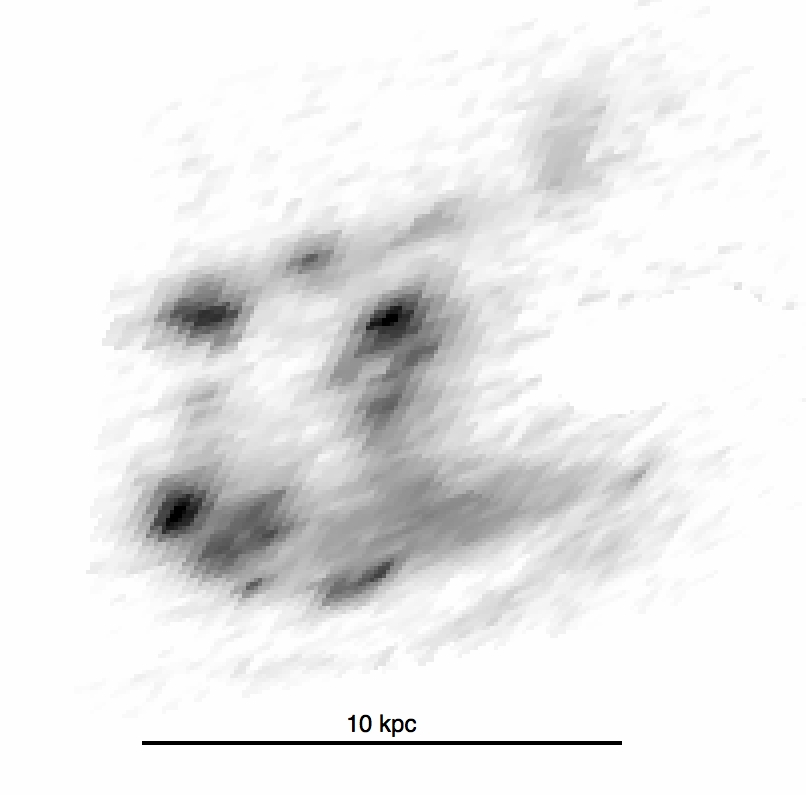}
 \caption{A68-HLS115}
 \label{subfig:A68-HLS115}
 \end{subfigure}
 \begin{subfigure}[b]{0.245\textwidth}
 \includegraphics[width=0.935\textwidth]{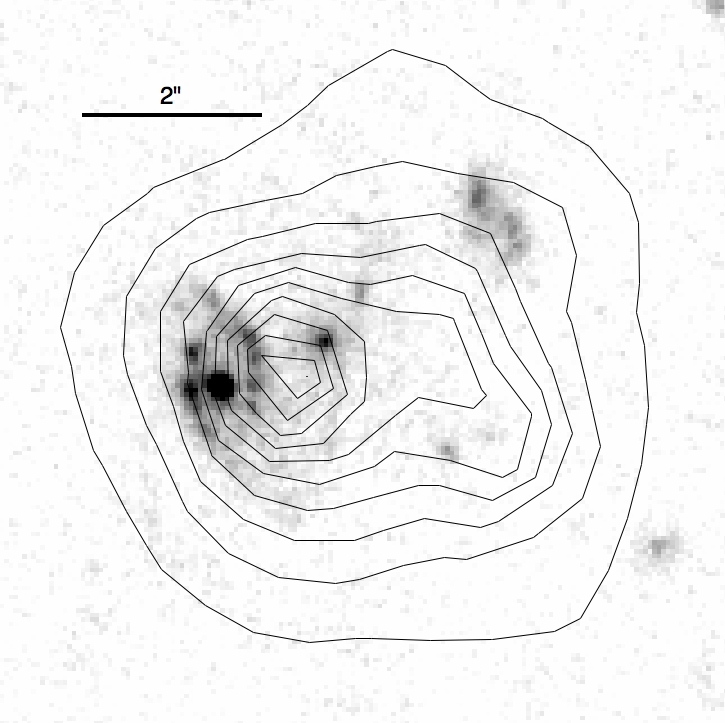}
 \caption{A68-h7}
 \label{subfig:A68-h7}
 \end{subfigure}
 \begin{subfigure}[b]{0.245\textwidth}
 \includegraphics[width=0.935\textwidth]{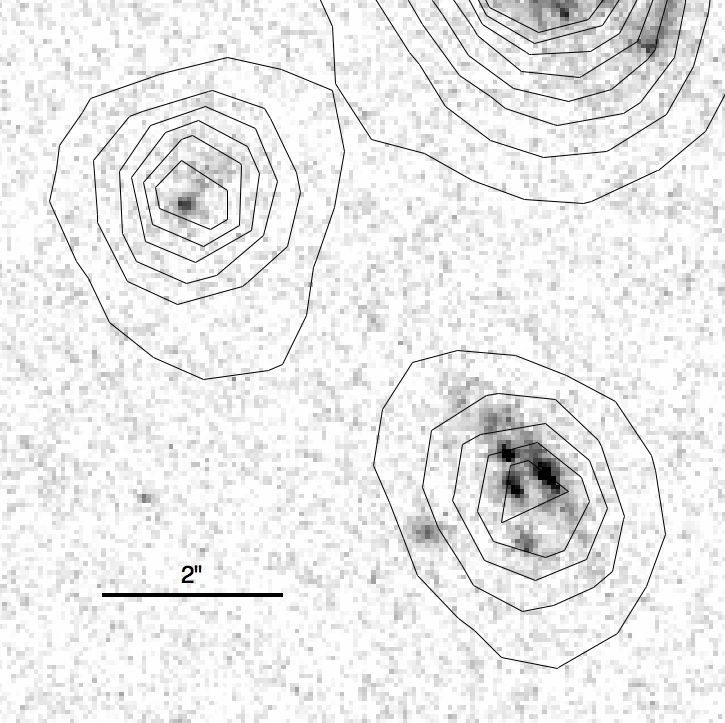}
 \caption{A68-nn4}
 \label{subfig:A68-nn4}
 \end{subfigure}
 \begin{subfigure}[b]{0.49\textwidth}
 \includegraphics[width=0.49\textwidth]{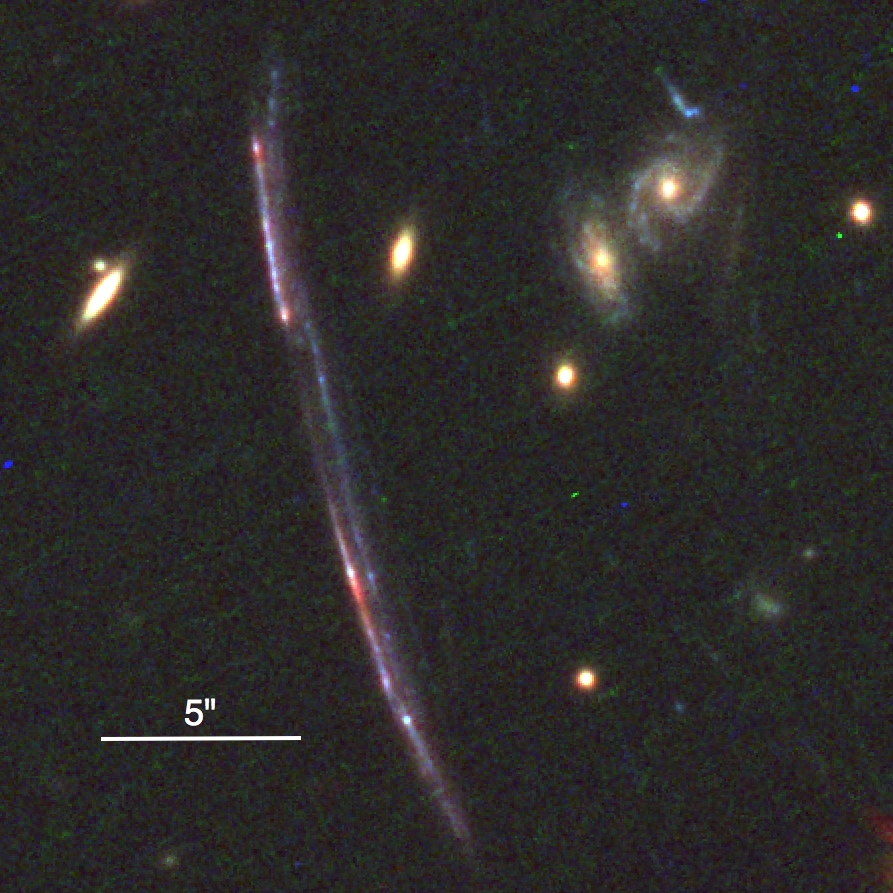}
 \includegraphics[width=0.49\textwidth]{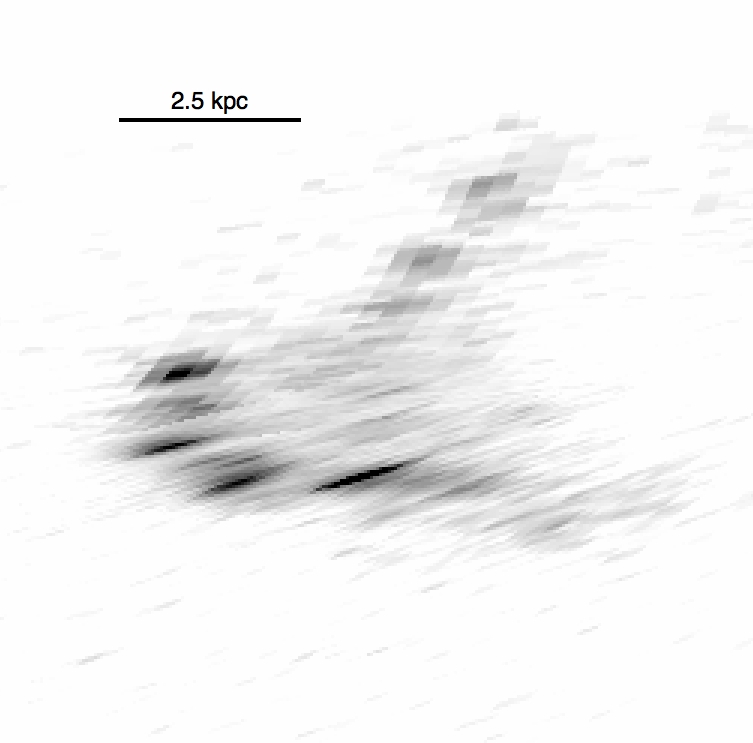}
 \caption{MACS J0451+0006 arc}
 \label{subfig:macs0451}
 \end{subfigure}
 \caption{\textbf{(a)} RGB rendering of A68/C0 using the F160W, F814W and F702W 
 HST bands.
 \textit{Left:}  Images A68-C0a and A68-C0b forming a broad quasi-symmetrical arc 
 near the cluster BCG.
 \textit{Upper right:}  Third, less magnified and less distorted image, A68-C0c.
 \textit{Lower right:}  Source plane reconstruction of A68-C0c after removal of 
 the overlapping elliptical.  Some residuals of the subtraction remain.
 \textbf{(b)} ACS/F814W image and source plane reconstruction of A68-HLS115. 
   Again the neighbouring elliptical has been removed in the reconstruction.
 \textbf{(c)} ACS/F814W image of A68-h7 with IRAC/ch2 contours.  This source 
 consists of an interacting system of four separate components,
 the most extinguished of which is the South-West component.  Although the exact 
 relation of each of the components to one another is unknown, their morphology 
 and photometry is consistent with all of them being related, and forming a 
 coherent system.
 \textbf{(d)} ACS/F814W image of A68-nn4 with IRAC/ch2 contours.  This source 
 consists of a pair of interacting galaxies.  We focus on the North-East 
 component as it is the most extinguished and appears to be the most related to 
 the IR emission.
 \textbf{(e)} \textit{Left:} RGB rendering of the giant arc in MACSJ0451+0006 
 using the F140W, F814W and F606W HST bands.  The arc is 20" long, and can be 
 separated in two main components: the northern part and the southern part.  The 
 northern arc is a double image of the northern part of the source.  The critical 
 line runs through the middle of it.  The southern arc is a single stretched 
 image of the rest of the source.  The two parts can be separated in {\em 
 Herschel} up to $250\mu$m.  The IR emission in the South appears to be dominated 
 by an AGN, so we consider here only the northern, and starburst component, of 
 the IR emission.
 \textit{Right:}  ACS/F606W source plane reconstruction of the arc.  The 
 morphology suggests a merger, despite ambiguious kinematics \citep{2010MNRAS.404.1247J}.}
 \label{C0_image}
 \end{figure*}

{\bf A68/C0: }
C0 is a triply-imaged spiral galaxy lying behind the core of the massive galaxy cluster Abell 68.  The two most magnified images ($a$ \& $b$) form a single continuous broad arc close to the BCG.  They are shown in Figure~\ref{C0_image}.  A third, less magnified and less distorted image ($c$) of this galaxy appears further out on the opposite side of the BCG.  This third image clearly shows the spiral nature of this galaxy.  This galaxy was first reported by \cite{2002MNRAS.330L....1S} from their search for gravitationally lensed EROs and subsequently analyzed in more details in \cite{2002MNRAS.333L..16S}.  These initial papers focused on the bulge component of the galaxy.  This part of the galaxy appears bright red in the composite $JIR$ image shown in Figure~\ref{C0_image}, and is the only readily visible component in the $K$-band.  In this paper, however, we always consider the galaxy in its entirety.  By doing so, the galaxy no longer qualifies as an ERO.  We perform our analysis only on the arc composed of images $a$ \& $b$, as these are the two brightest images, and because image $c$ is blended with a cluster member elliptical galaxy.  The combination of images $a$ \& $b$ represents a linear magnification factor of $\mu = 30$.  The arc has also been referred to as the ``space invader" galaxy, because of its appearance when looked at from the North-West \footnote{http://apod.nasa.gov/apod/ap130308.html}.

In our companion paper (Dessauges-Zavadsky et al., \emph{in prep.}), we present CO observations of this arc, from which we infer a redshift of $z = 1.5854$.  This redshift is consistent with the break detected by \cite{2002MNRAS.333L..16S} in their $z$-band NIRSPEC spectrum as corresponding to the Balmer break.  It also matches our detection of the $H\alpha$ line in the NIR spectrum obtained with LBT/LUCIFER.

{\bf A68/h7:}
This source, also located in the field of the cluster Abell 68,
consists of a system of four galaxies in interaction (Fig. \ref{C0_image}).  We have obtained a VLT/FORS2 spectrum of this object from which we have identified faint C~II and C~IV lines and estimated its redshift to $z = 2.15$.  We then confirmed the redshift with our CO observations (Dessauges-Zavadsky et al., \emph{in prep.}).  Although the FORS2 slit was positioned on the brightest (east most) component, the photometry and SED of all of the individual components is consistent with all of them lying at the same redshift.  The CO spectrum further shows only a single line of FWHM $= 350 $ km/s.  This strongly suggests that the four components are in some form of interaction, but the exact configuration thereof remains uncertain.  It is possible, for example, that the system is made of a weakly interacting pair of two ongoing mergers.  

{\bf A68/HLS115:}
HLS115 is a galaxy lensed by both the cluster itself and a cluster member elliptical galaxy for a total estimated magnification of $\mu = 15$.  We have detected $H\alpha$ from this galaxy with LBT/LUCIFER, and CO with the IRAM/PdB interferometer (Dessauges-Zavadsky et al., \emph{in prep.}), from which we infer a redshift of $z = 1.5859$.  This redshift is nearly identical to that of A68/C0.  The two galaxies, therefore, most likely belong to the same group.  Contrary to A68-C0, however, HLS115 does not show a well-defined spiral structure, but rather consists of a series of clumps.  It has, otherwise, very comparable properties as derived from our SED fitting (cf. Section~5).

{\bf A68/nn4:}
This source consists of a pair of objects in interaction.  It has the highest redshift of the sample,
that is z = 3.19. Here, we study the most obscured of the two components, which is also the one that appears 
to be the most related to the FIR emission. It is undetected from the R-band and bluewards. This source
lies in the outskirts of A68, so it is modestly lensed ($ \mu \approx 2.3 $), and thus intrinsically luminous. 

{\bf MACS0451:}
This is a very elongated arc and highly magnified ($ \mu \approx 49$) source at redshift $z=2.013$ in the field of the cluster MACSJ0451.9+0006  \citep{2010MNRAS.404.1247J}.
The arc measures 20" in length, and so this source is spatially resolved up to 250\micron. 
When examining the FIR SED of this source, we have noticed differences between the northern and the southern parts
of the arc, with the first peaking at 250\micron\ and the latter at 100\micron, indicating very hot dust. 
After careful analysis of this object, we came to the conclusion that the infrared emission of this galaxy includes an AGN component.  However, we are confident that we can separate this AGN component from the star-forming component.  A detailed discussion will be presented in Zamojski et al. (\emph{in prep.}).  Therefore, we chose to retain this object in the current study given the rarity of such highly magnified objects at this redshift, but consider only its star-forming component.
The contribution to the total IR luminosity coming from the two components
is about half and half.
The UV to NIR photometry  has constant colors throughout
the arc, with $ \sim $40\% of the flux coming from the northern part. 
We see no signs of an AGN at these wavelengths. 
A decomposition of the IR emission of southern part of the arc indicates that roughly 90\% of its flux originates 
from the AGN while the remaining 10\% is coming from star formation, the exact number depending on the models used. 
For simplicity, we employ, here, these round numbers as working values, that will be used in particular in 
Sec. \ref{s_nrg_conserv}, and postpone a more detailed analysis for later (Zamojski et al., \emph{in prep.}). 

We note that the detailed photometry of the arc presented here differs non-negligibly from previously published values \citep{2011MNRAS.413..643R}.  This difference stems from the different methods used to make these measurements.  As explained in sections~\ref{hstphot} through~\ref{sec:groundphot}, we model the arc in its entirety starting from the high-resolution {\em HST} images and convolving with the proper PSF, and then solving for the flux simultaneously with all neighbouring objects with a maximum likelihood algorithm.  The flux thus extracted is robust.  Previous values were extracted in a number of apertures along the arc with aperture corrections and color extrapolations applied to these measured fluxes.  Such an approach is prone to larger uncertainties, and we estimate that the inferred factors of $\sim 2\textendash 3$ difference are not incompatible with these uncertainties.  The arc possesses a dense photometric coverage in the optical regime coming from {\em HST}, the Subaru Telescope and the Sloan Digital Sky Survey, with considerable overlap between the bands.  Our method produces a smooth and consistent SED across these bands and across the different instruments, and the physical properties extracted from its SED are consistent with those obtained from other diagnostics (cf. section~\ref{s_arc}). This would not otherwise be the case.  It illustrates the difficulty of working with these highly stretched arcs and the importance of accurate photometry for proper modeling of their SED.

{\bf cB58}  is a well-known very strongly lensed \citep[][$ \mu \sim 30 $]{1998MNRAS.298..945S} galaxy at $
z=2.78$  discovered by \citet{1996AJ....111.1783Y}. We use the CFHT and {\it Spitzer} optical to MIR photometry provided by 
\cite{1996ApJ...466L..71E} and \cite{2008ApJ...689...59S} together with the submm/mm detections 
of \cite{2001dmsi.conf..103V} and \cite{2001A&A...372L..37B}. Spectroscopy of this source is described e.g.\
in \cite{2000ApJ...528...96P} and in \cite{2000ApJ...533L..65T}. 

{\bf Cosmic Eye:} This is an equally strongly lensed Lyman break galaxy (LBG)
 at $z=3.07$ discovered by \citet{2007ApJ...654L..33S}. It is magnified by a factor of 
$\mu =28 \pm 3 $ times by a foreground z = 0.37 cluster and a z = 0.73 massive early-type spiral galaxy \citep{2007MNRAS.379..308D}. 
For our work we use the combined photometry of \citet{2007ApJ...665..936C} and \citet{2009ApJ...698.1273S}.
Spectroscopy of the rest-frame optical emission lines and the UV absorption features is available from \cite{2011MNRAS.413..643R} and
\cite{2010MNRAS.402.1467Q}, respectively.
% % % % % % % % % % % % % % % % % % % % % % % % % % % % % % % % % % % % % % 

\subsubsection{Differential Magnification}
 
One caveat with working with strongly lensed galaxies is that some parts of the galaxy could be magnified more than others.  This so-called differential magnification can modify the balance of the SED if the region being more magnified is particularly bright (or faint) at some wavelengths compared to the rest average of the galaxy, as for example would be the case for a particularly dusty region or cloud.  This could lead to erroneous conclusions when deriving global properties.

The advantage of working with cluster lenses (as opposed to galaxy lenses) is that they have much larger and broader potentials so that the magnification changes little on the scale of a galaxy.  This, however, is true only as long as the source is not located near a caustic.  Sources that cross inside the caustic region are imaged multiple times and could be prone to differential magnification effects.  Within our sample, this happens with A68/C0 and the arc in MACS0451, as well as with our two comparisons objects:  the Cosmic Eye and cB58.

The infrared emission of A68/C0 at 100 \micron\ (highest resolution) is elongated and the ellipse covers well the visible part of the galaxy.  This suggests that it originates from the entire disk rather than being dominated by a bright region near the critical line passing through the center of the object.  Differential magnification does not appear to play an important role in this galaxy.  The northern part of the arc in MACS0451 consists of two mirror images of the same part of the source, and is therefore also crossed by a critical line.  The 100 \micron\ emission, in this case also, does not appear to be any brighter near the critical line region.  The region further appears bluer than the rest of the galaxy in [optical - IRAC] colors, so that, again, the dusty and infrared-bright regions appear to be distributed, just as the optical/NIR light, throughout the whole image.  The FIR emission does not appear to come from a small very magnified region near the critical line.  The case for the Cosmic Eye and cB58 is more difficult, as we do not have the resolution to say anything about the spatial origin of their FIR emission.  Differential magnification effects within these two galaxies can, therefore, not be excluded.

% % % % % % % % % % % % % % % % % % % % % % % % % % % % % % % % % % % % % % 
\subsection{Photometry}
 
The data used in this study comes primarily from the {\it Herschel}, {\it IRAC}, and {\it SCUBA2 Lensing Surveys} 
(\citet{2010A&A...518L..12E}, Smail et al. 2013, \emph{in prep.})
as well as from ongoing efforts to image strong-lensing clusters with the \textit{Hubble Space Telescope} ({\it HST} hereafter).  
In addition, we collected data from various ground-based facilities to complement our wavelength coverage, and better constrain our stellar SEDs. 
The photometry for all sources is given in Table 1 of the appendix.

\subsubsection{{\it HST} photometry \label{hstphot}}
 
Our sources are  strongly lensed, and many of them appear close to large elliptical galaxies, such as the BCG (Brightest Cluster Galaxy), whose light blends with that of the objects we want to study.  To obtain accurate photometric measurements, the light from these neighboring/lensing ellipticals needs to be removed.  We do so by fitting their profile with {\sc GALFIT} \citep{2002AJ....124..266P}.  A large dynamic range in terms of the brightness and extent of sources exists in the center of massive galaxy clusters, in addition to the high density of sources.  It is, therefore, extremely difficult to fit the profile of all cluster galaxies simultaneously.  We thus proceed  in steps  by fitting and removing, first, the light of the brightest galaxies, and then that of the more modest less extended objects\footnote{We note that neighbouring objects within the extent of each large galaxy and to the limit of 5~magnitudes fainter are fit simultaneously with the galaxy we want to subtract, so that to make sure they do not bias the fit of the larger galaxy.  The neighbours are {\em not} themselves subtracted:  their flux is remeasured from the image subtracted of the large galaxy.  They can then be themselves fit out and subtracted if fainter objects exist within the reach of their profile, and so on.}.
 
We use {\sc SExtractor} \citep{1996A&AS..117..393B} to measure the flux of our sources, after subtraction of neighboring cluster galaxies, in elliptical apertures, in a reference {\it HST} image.  We extract our objects in the reddest {\it HST} band available (usually F160W).  In some cases (A68/C0, MACS0451), our sources are stretched so that they take the form of an arc, and ellipses no longer accurately represent their shape.  For those objects, we employ custom apertures.  We then measure the flux of our objects in other {\it HST} bands in those same apertures, after also performing a subtraction of neighbouring cluster galaxies in those bands.  
 
\subsubsection{IRAC photometry}
 
Because of the much coarser resolution of the {\it Spitzer Space Telescope} compared to {\it HST}, we cannot employ the same strategy for IRAC images.  We, instead, perform prior-based photometry.  We adapted the code initially developed by \citet{Guillaume06, Zamojski08, Llebaria08, Vibert09} 
to do prior-based photometry for {\it GALEX}, and applied it to IRAC.  Our code uses the {\em Expectation Minimization} algorithm, a Bayesian algorithm that iteratively adjusts the flux of all objects simultaneously in such a way as to increase, at each iteration, the likelihood that the observed image be drawn from the theoretical image:  the theoretical image, in this case, being the image produced by convolving the prior shape of each object with the IRAC PSF and scaled to the adjusted flux.
 
We use the reddest {\it HST} band to produce ``stamp" images of each object.  These stamp images include only pixels within the {\sc SExtractor} aperture.  They define the prior shape of each object that is then convolved with the IRAC PSF and scaled in flux.  For large elliptical galaxies whose profile include wide wings, we increased the size of the {\sc SExtractor} aperture often by a factor $\sim 2$ or sometimes more, so that to include as much as possible the entire visible flux of the galaxy, up to the surface brightness limit of our images.  This is necessary because of the surface brightness depth of the IRAC observations.  Were we not to do this, we would not subtract these galaxies completely in IRAC, and hence we would not measure their entire flux.  More importantly, however, we could contaminate the flux of neighbouring objects.  The residual maps, in this case, would be dominated by the wings of these large galaxies hollowed out in their centers.  We enlarge the apertures in order to avoid this.
 
There can be overlap between the ellipses of different objects.  We deblend faint and background objects from cluster galaxies by extracting their shape and photometry from the image in which the profile of these cluster galaxies has been subtracted out as explained in section~\ref{hstphot}.  We then use the initial image to extract the shape of the larger galaxies, but only after first subtracting the flux of all the previously extracted fainter objects surrounding them.  For cases where two or more similar size galaxies need to be deblended from the same image, we employ the symmetric part of each galaxy, relative to their center, to deblend the flux in overlap regions as explained in \citet{Zamojski08, Vibert09}.
 
Since the position and shape of our priors are fixed, and only their fluxes are adjusted, our method can naturally recover the flux of objects even when the fluxes of several objects partially overlap (separation $\gtrsim 1$ FWHM $= 1.6\arcsec$) as is the case of most IRAC sources in the crowded field of a massive galaxy cluster.  After subtraction of the theoretical image from the actual image, some residuals can remain.  These residuals are largest for resolved spiral galaxies, most likely due to the intrinsic differences in the shape of the galaxy at $1.6$ and $3.6\mu$m, notably in the size of the bulge and the intensity of the spiral arms and star-forming clumps.  They, nevertheless, remain of the order of $\lesssim 5\%$.
 
\subsubsection{Ground-based photometry}	\label{sec:groundphot}
 
For ground-based images, we use both the procedure we apply to HST images as well as the prior-based method we use for IRAC, 
and retain the one that is most appropriate.  In the case of strong blending with a neighbouring elliptical galaxy (such as for A68-HLS115), prior-based photometry is preferred, whereas for very extended objects (e.g. A68-C0) the combination of {\sc GALFIT} and {\sc SExtractor} or custom aperture is favored.  In all cases, both methods give similar results.
 
\subsubsection{IR-mm photometry: general}

We use aperture photometry (with an appropriate aperture correction) on MIPS and PACS maps, since, at these wavelengths, our sources are well separated from other sources.  We, exceptionally, use a {\sc SExtractor} elliptical aperture for A68-C0 at $100\mu$m since the source is marginally resolved and elongated.  In SPIRE, sources begin to blend, so we use again our prior-based technique, this time with only the positions as priors with each object simply taking the shape of the SPIRE PSF.  We retain as priors only those sources that are detected in at least one of the PACS bands.
 
In the case of the arc in MACS0451, we also use prior-based photometry on the PACS maps, since we want to separate the different components of the arc.  We again trim our list of priors to avoid putting flux in unphysical places.  Here, we simply remove, based on their color and shape, all low-redshift ellipticals, except for the BCG  which may contribute non-negligible flux to the IR \citep{2012ApJ...747...29R}.  Since the resolution of PACS is not as coarse as that of SPIRE, some sources can be marginally resolved (such as is the case of A68-C0), and in particular the giant arc, even after splitting it into two or three components.  Yet, optical/NIR images are hardly representative of the FIR morphology of a galaxy.  We thus opted for the next best approximation and used exponential profiles as shapes for our priors, the effective radius of which is based on that of the optical/NIR light.

Except for A68-C0, our sources are faint, and detected at only a few sigmas, in the SCUBA2 maps.  We therefore, used prior-based photometry, as it performs better than aperture photometry, in terms of both the depth as which it is able to measure fluxes and the reliability of the measurements.  The latter being increased by the fact that the positions of the sources are known and fixed {\em a priori}.  We used pure PSFs as shapes for our priors, and a circular gaussian of FWHM$=14.5"$ to describe the SCUBA2 beam.

\subsubsection{IR photometry:  Cosmic Eye}	\label{s_photo_eye}

\begin{figure}[htb]
%\begin{center}
%\begin{tabular}{|p{0.66\textwidth}|p{0.33\textwidth}|}
%\includegraphics[width=0.33\textwidth]{poll_full_4.pdf}
%\includegraphics[width=0.33\textwidth]{poll_full_10.pdf} &
%\includegraphics[width=0.33\textwidth]{rieke_full_NEW2p5666.pdf} \\ \hline
\includegraphics[width=0.5\textwidth]{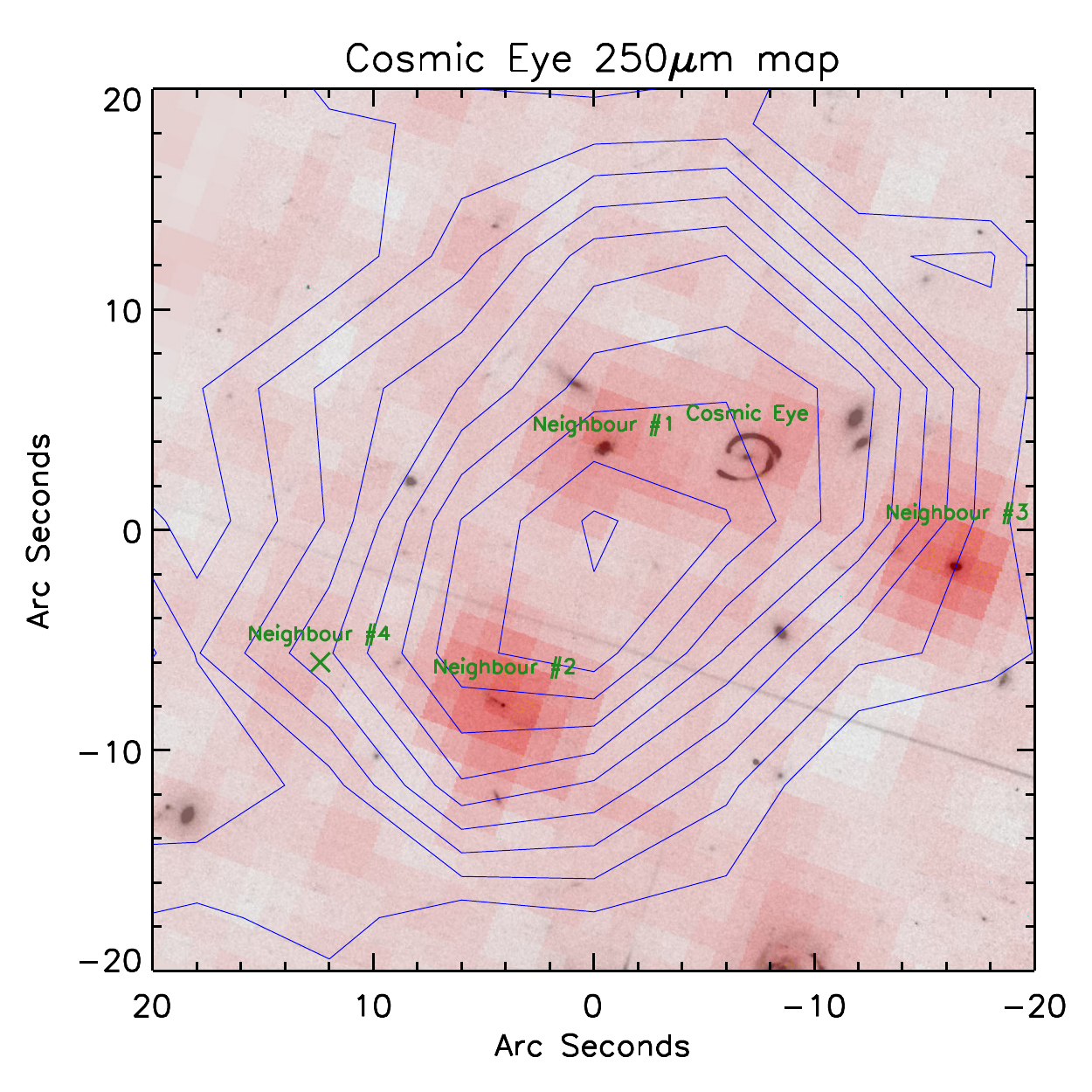}
%\vspace{-5.2in}
%\includegraphics[width=0.33\textwidth]{CE_full_3.pdf}
%\includegraphics[width=0.33\textwidth]{smg2_full_20.pdf}
%\end{tabular}
\caption{HST/ACS image of the region around the Cosmic Eye overlaid with MIPS~$24\mu$m emission in redscale and SPIRE $250\mu$m contours.  Five objects, including the Cosmic Eye, blend the form a single source at $250\mu$m.  Neighbour~\#4 is undetected in any of the optical bands, and faint at $24\mu$m, but appears in IRAC and shows up ever brighter with increasing wavelength in the infrared.  Its SED indicates that it is likely an SMG an $z\sim2.5$.  }
%Also shown on the periphery are SED fits to the MIPS+PACS photometry from which we extrapolate prior fluxes at $250\mu$m, which we then use as inputs to perform prior-based PSF fitting photometry in SPIRE (see text for details).}
\label{fig:eye250}
%\end{center}
\end{figure}

The Cosmic Eye is surrounded by several other equally infrared-bright objects.  Figure~\ref{fig:eye250} shows the SPIRE $250\mu$m contours of the region around the Eye overlaid on top of an HST/ACS optical (F606W) image.  Also overlaid on the same image is a MIPS $24\mu$m redscale, indicating four sources of infrared emission other than the Cosmic Eye within the same SPIRE resolution element.  Neighbour~\#4 is undetected in any of the optical bands, but appears in IRAC.  It is faint but detected in MIPS and PACS, and its SED indicates that it is likely an SMG at high redshift ($z \sim 2.5$).  The most problematic, however, is Neighbour~\#1, as it is only 1~PSW pixel (6") away from the Cosmic Eye.  In such heavily blended situations, even solutions obtained with PSF fitting can be quite degenerate and sensitive to the local noise as well as to initial, prior inputs.  In order to estimate the flux of the Cosmic Eye in the SPIRE $250\mu$m band, we therefore perform our extraction under several added constraints, which we discuss below.

Our strategy has been to first extract the fluxes of all objects in the field up to the PACS $160\mu$m band, fit their SED, and extrapolate it to predict their flux at $250\mu$m.  Because none of the sources are bright in the PACS bands, and because of the crowding in this area, the reliability of the aperture photometry, in terms of centering of the apertures as well as of contamination, is doubtful.  In the case of the Cosmic Eye, we, therefore, chose to extract the PACS fluxes with our prior-based procedure using the MIPS $24\mu$m sources as priors.  MIPS $24\mu$m photometry, we performed in apertures.

We used archival HST and {\em Spitzer}/IRAC data to obtain optical/NIR SEDs of neighbouring objects.  With the photometry in hand, we first fit the redshift of neighbouring objects to the stellar only part of their SED with the exception of neighbour~\#4 which does not have enough photometric data points in this regime, and for which we performed only global SEDs (but all of which returned a redshift of $z \sim 2.5$).  We then fit the thermal SED of these objects, fixing the redshift to that obtained above.  
We used archival HST and {\em Spitzer}/IRAC data to obtain multi-band optical/NIR 
photometry of the objects neighbouring the Cosmic Eye.  We use this photometry to 
first fit for the redshift of the neighbours using only the stellar part of their SED.\footnote{
Neighbour~\#4 does not have enough photometric data points in the stellar regime to constrain its 
redshift from that part alone.  Instead, we performed global (stellar + thermal) SED fits for this object.  
Nevertheless, its redshift appears well-constrained as all libraries returned a best-fit redshift of 
$z \sim 2.5$.  We, thus, use $z = 2.5$ as the nominal redshift of Neighbour~\#4.}
We then fit a preliminary thermal SED, to the MIPS and PACS photometry only, by fixing the redshift to 
that obtained above.  Using the best-fit SED, we obtain an initial guess of the flux at $250\mu$m, and 
run our deblending algorithm with those initial guesses.  Even then, however, the relative contribution 
of the Cosmic Eye and Neighbour~\#1 remains weakly constrained because of the small separation of these two objects.  
The maximum likelihood solution actually assigns more flux to the neighbour than to the Cosmic Eye.  
This is unlikely given their respective SEDs at $\lambda < 250\mu$m.  
We, therefore, re-run the procedure by first fixing the flux of Neighbour~\#1 to that predicted by the best-fit SED, 
subtracting it from the image, and removing it from the catalogue of priors.  
The new solution converges to fluxes for the Cosmic Eye and its three other neighbours close to those 
predicted by their preliminary SEDs.  The residuals are slightly less flat than for the case where all five objects are free to vary, 
but they remain below the noise level.  The difference between the two cases can therefore be said to be of 
little significance.  We thus retain the second, and more physical, solution.   
The errors estimated from the residuals are added in quadrature to the 
dispersion of predicted fluxes for Neighbour~\#1 obtained with different libraries.

At $350\mu$m where the resolution is even worse than at $250\mu$m, the situation becomes even more degenerate, and we were unable to obtain a reliable measurement.  We, therefore, chose to use only photometry up to $250\mu m$, in addition to upper limits at $350\mu$m, $500\mu$m and 3.5mm. We add, however, the 1.2 mm flux from \cite{2013arXiv1309.3281S}.	

%\clearpage

\subsubsection{IR photometry:  cB58}

MS1512-cB58 is very close the cluster cD galaxy, which also shines in the infrared.  Fortunately, its redshift is known spectroscopically to be $z=0.372$.  We can, therefore, extrapolate its flux at $250\mu$m and remove it from the image, before solving for the flux of cB58 itself, in exactly the same way as we proceeded to deblend the Cosmic Eye with its closest neighbour.  cB58 is otherwise not as heavily blended as the Cosmic Eye, and we were able to obtain a reliable flux at $350\mu$m as well.

%%%%%%%%%%%%%%%%%%%%%%%%%%%%%%%%%%%%%%%%%%%%%%%%%%%%%%%%%%%%%%%%%%%%%%%%%%%%%%%%%
\section{SED modelling}

\subsection{SED fits}		\label{sed_fits}

We use an updated version of the \hyperz\ photometric redshift code of \cite{bolzonellaetal2000}, modified to include
the effects of nebular emission in its fitting procedure, as described in \cite{schaerer&debarros2009,schaerer&debarros2010}.  
Designed to derive redshifts from broad-band SED fits of UV--NIR photometry and 
physical parameters of the galaxies, our version was also adapted to use data up to the sub-millimeter range.
The redshift of our sources is fixed to the spectroscopic value and is not considered as a free parameter in the present work.

Using (semi-)empirical and theoretical templates described below, we perform fits of three sets of photometries per object:
\begin{itemize}
 	\item the full photometry (i.e. from the rest-frame UV to the FIR)
 	\item the dust processed FIR emission (from the MIPS 24 \micron\ band longwards)
 	\item the stellar SED photometry up to the IRAC bands. 
\end{itemize}

 These fits of different wavelength intervals are done to provide us with the widest range of parameters that can be deduced from the bulk 
spectral features of our sources, as precisely as possible. They are described in detail in the following paragraphs.

Fits to the full photometry using empirical templates inform us on whether the concerned object resembles a known local object or 
type.  We perform fits to the full photometry only to inform ourselves on whether the object in question resembles a known local galaxy or galaxy type. We do {\em not} use global fits to derive any physical quantity, as they typically reproduce poorly the  observed photometry compared
to the combination of the independent fits to the stellar and thermal components respectively.
The only ``free parameter" that \hyperz\ can explore for empirical templates is eventually adding extinction
on top of the original template used. This affects the template in the wavelength interval [$912\AA - 3\mu$m],
where typically the light emitted by stars gets absorbed by the ISM. This increases the adaptability of the templates used, and comes in 
handy when exploring obscured IR-bright galaxies. Of course, the value of the extinction in this case is of no physical
significance, since it does not consider the intrinsic extinction that comes with every original template.
The total FIR luminosity \lir\ is obtained from integration over the rest-frame interval [$8-1000 $] \micron\ over the fits
to the  {\em FIR only} part of the SED.

We also perform  modified black body fits on the FIR/submm data to derive dust properties such as temperature and mass (see Sect. \ref{s_dust}).

The full and FIR only fits are done using libraries whose templates are defined from the UV to sub-mm wavelengths (typically they are defined from
the Lyman limit to the synchrotron-dominated part of the electromagnetic spectrum). 
The libraries used are:
\begin{itemize}
\item  \citet[][CE01]{2001ApJ...556..562C}: a set of synthetic templates of varying IR luminosity, 
\item  \citet[][hereafter P07]{2007ApJ...663...81P}: a set of templates made out of local observed objects, including spiral 
			galaxies, starbursts, Seyfert and AGN,
			plus templates from synthetic models covering various stages of galaxy evolution,
\item  \citet[][R09]{2009ApJ...692..556R}: a set of templates containing observed SEDs of local 
			purely star forming LIRGs and ULIRGs, and some models 
			obtained as the result of combining the first ones. These templates in particular are defined only down to 3\micron\ or 4000\AA\
			and hence are only considered for the FIR only fits,
\item  \citet[][M10]{2010A&A...514A..67M}: a set of templates made from observations of sub-millimeter galaxies at $z \sim$ 0.08--3.6
			 \citep{2009ApJ...699.1610H,2010AAS...21532601H}.
\end{itemize}
For every set, a free scaling parameter allows matching in terms of intensity.

%------------------------------
\begin{table}[tb]
	\centering
	\begin{tabular}{l c c}
		%\hline
		%\hline
		  SFH  & Extinction law & nebular emission \\\hline
       	  Exp. declining  & Calzetti/SMC  &   with/without	 	\\
       	  constant rate   & 	"	  &     "			\\
		  Exp. rising  &  " &  "	\\
		\hline   
	\end{tabular}
	\caption{Table depicting the various combinations of the basic parameters that we explore in our stellar models. 
		That amounts to ten scenarios, to which we add the commonly used one, that supposes CSFR, 
	Calzetti, no nebular emission and the age prior of $ t_{min} = 100 $ Myr.}         
	     \label{stellar_models}
\end{table}
%------------------------------

The fits to the stellar SED determine the physical parameters such as the SFR, stellar mass, age of the population,
the extinction \av. From the fitted SED we also derive the UV slope $\beta$\footnote{
The commonly used  UV spectral slope $ \beta $, defined as by $f_{\lambda} \propto \lambda^{\beta}$,
is determined between 1800 and 2200 \AA\ (rest-frame) from our best-fit SEDs.},
and UV luminosity \luv\footnote{For \luv\ we use $ \lambda\cdot F_{\lambda} $, averaged over 1400-2200 \AA, 
with $ \lambda_{\rm eff} = 1800$ \AA. }.
They are performed with the \cite{2003MNRAS.344.1000B} library (BC03 hereafter). 
We adopt a Salpeter IMF from 0.1 to 100 \msun.
The extinction laws explored here are the commonly used Calzetti law \citep{2000ApJ...533..682C} and the 
SMC law of \cite{1984A&A...132..389P}, motivated also by recent publications 
\citep{2012ApJ...754...25R,2012arXiv1211.1010O,2012ApJ...745...86W}. 
When having available spectroscopic data for comparison,
we also explore Calzetti's law with stronger line attenuation as prescribed in \cite{Calzetti2001}
(in particular in the case of the Cosmic Eye).

From the fits to the SED, assuming energy conservation,  we also derive the predicted IR luminosity from the difference
between the intrinsic, unobscured SED and the observed one, as described in \cite{2013A&A...549A...4S}. Having access to 
the actual observed IR luminosity allows us to distinguish/constrain different star-formation histories and extinction laws.

For the BC03 library, and following our analysis of a large sample of LBGs from redshift 
3 to 6 \citep{2012arXiv1207.3663D,2013A&A...549A...4S}, we
explore a range of star-formation histories (SFHs), as well as models with or without nebular emission. 
Except otherwise stated, we assume solar metallicity.
The combination of model parameters explored is summarized in Table \ref{stellar_models}.
In practice we have used SFHs with exponentially declining timescales with $ \tau= (0.05, 0.07, 0.1, 0.3, 0.5, 0.7, 1., 3.)$ Gyr,
exponentially rising ones with  $ \tau= (0.01, 0.03, 0.05, 0.07, 0.1, 0.3, 0.5, 0.7, 1., 2., 3.)$ Gyr,
or constant SFR with a minimum age prior of $ t_{\rm min} = 100 $ Myr, as commonly assumed in the literature.
The extinction is allowed to vary from $\av=0$ to 4 in steps of 0.1.
We also apply a foreground galactic reddening correction to our photometry, using the values available on the NED 
\citep{1998ApJ...500..525S}. 

The ratio of \lir\ over \luv\ is known to be an effective tracer of UV attenuation
\citep[e.g.][]{burgarellaetal2005,2010MNRAS.409L...1B,2013MNRAS.429.1113H}. %2012A&A...545A.141B,2005ApJ...619L..51B
From the observed $\lir\ / \luv$ we can therefore determine the extinction needed in \hyperz\ in order to make fits that are 
energy conserving, meaning that the stellar population model produced in this case will reproduce 
the actual observed \lir\ without suffering from the
eventual age-extinction degeneracy often encountered in obscured galaxies. In practice we use the relation between $\lir\ / \luv$ and \av\
from \cite{2013A&A...549A...4S}. These ``energy conserving models''
should thus provide the most accurate physical parameters.

For each object we retain the best-fit SED and physical parameters. 
We also generate 1000 Monte Carlo (MC) realizations of the observed SED,
which are fit and used to determine the median values and the 68\% confidence intervals of the various physical parameters. 
Although for some cases/observed bands the photometry's precision is better, we have imposed a minimal error of 0.1 mag 
(and 0.05 mag for MACS0451 that was overall best constrained) in the SED fitting procedure and the MC catalogs
that is more appropriate when combining the photometry from many different instruments, wavelengths and depths.

%%%%%%%%%%%%%%%%%%%%%%%%%%%%%%%%%%%%%%%%%%%%%%%%%%%%%%%%%%%%%%%%%%%%%%%%%%%%%%%%%
 
%------------------------------
\begin{table*}[!t]

	\centering
	\begin{tabular}{l c c c c c c c c}
		\hline\hline
		\textbf{ID}   & z & $ \mu $&$ \beta $ &  $\luv \times \mu$  &Library of &  $\lfir \times \mu$&  \sfrir\ & \tdust\ \\
		   &&&& [$10^{12}$\lsun] & best FIR fit  &[$10^{12}$\lsun] & [\msunyr] & [K] \\\hline
		
		A68/C0	  	    & 1.5854	& 30  &	$-0.42_{-0.4}^{+0.5} $	& $ 0.19\pm0.02$  & R09 & $ 3.55 \pm 0.2 $ & 20.4 (19.2-21.5) & 34.5  \\	
		A68/h7       	& 2.15	& 3 & 	$-0.01_{-1.0}^{+0.5} $ &  $  0.22\pm0.01 $   & R09 & $ 5.49^{+0.26}_{-0.37} $ &  315 (294-330) & 43.3   \\	
		A68/HLS115   	&1.5859	& 15& 	$-0.31_{-0.18}^{+0.55} $&  $ 0.1\pm0.01 $  & CE01 & $ 5.13^{+0.24}_{-0.23} $ &   59.0 (56.3-61.7) &  37.5 \\	
	
		A68/nn4	& 3.19 & 2.3 &	$2.57_{-1.1}^{+1.3}  $ &  $0.014^{+0.001}_{-0.002} $ & CE01 &$15.8^{+0.4}_{-0.7} $  & 1184 (1132-1214)  & 54.9 \\
		MACS0451 North 	& 2.013	& 49  & $-1.40_{-0.12}^{+0.12} $ & $ 0.55 \pm0.01$ & M10  & $4.26^{+0.3}_{-0.28}$ & 15.0(14.0-16.0) &  47.4 \\
		MACS0451 full arc & " & " & " & $1.2\pm 0.03 $& M10 & $12.0^{+0.0}_{-0.3} $ & 42.2 (41.2-42.2)* & 50-80* \\
		\hline
		cB58 	    & 2.73 & 30 & $-1.15_{-0.1}^{+0.1}  $ &  $ 1.66^{+0.12}_{-0.07} $ & CE01 & $9.12\pm 0.21 $ &  52.4 (51.2-53.6) & 50.1$^a$  \\
		Cosmic Eye   	& 3.07 & 28 &  $-1.41_{-0.08}^{+0.13}  $&  $2.57\pm0.06$   & R09 & $ 9.55^{+0.45}_{0.64} $ & 58.8 (54.9-61.6) & 46.3$^{a}$ \\
		\hline   
	\end{tabular}
	\caption{Main observed and derived properties of our galaxies. $\beta$ stands for the UV slope at 2000 \AA, measured from 
			1800 to 2200 \AA. 
			The values of $\beta$ presented are the average between the best Calzetti based solution and the best SMC one.
			The \lir's are produced by integrating the [8, 1000 \micron] interval on the best fit SED's shown here in the column
			on the left of the \lir values.
			\sfrir\ is obtained then via the \cite{1998ARA&A..36..189K} calibration from the 
			intrinsic (de-lensed) \lir's. Errors and values in parenthesis
			represent the 68\% confidence levels from our MC runs. For the peculiar case of the MACS0451 arc, we show quantities of the northern part
			(that seams to be starburst-dominated), as well as for the whole arc (that may AGN-contaminated, hence the asterisks)
			 for the convenience of the reader and to illustrate the flux ratios between the north segment and the whole arc.
			 See \ref{s_arc} for further discussion.  Uncertainties on the temperatures are of the order of $ \pm 1 $ K, with the 
			 exception of the Cosmic Eye for which it is $\pm 3$ K, due to the uncertainties on the SPIRE photometry and de-blending. \newline
			$ ^a$: Determined from the deblended photometry.
			 }	
							\label{OBSERVED-GLOBAL}		 
\end{table*}

%==============================

%-----------FIG LIR-Z--------------
\begin{figure}[tb]
\hspace{-0.62 in}
\includegraphics[width=12cm]{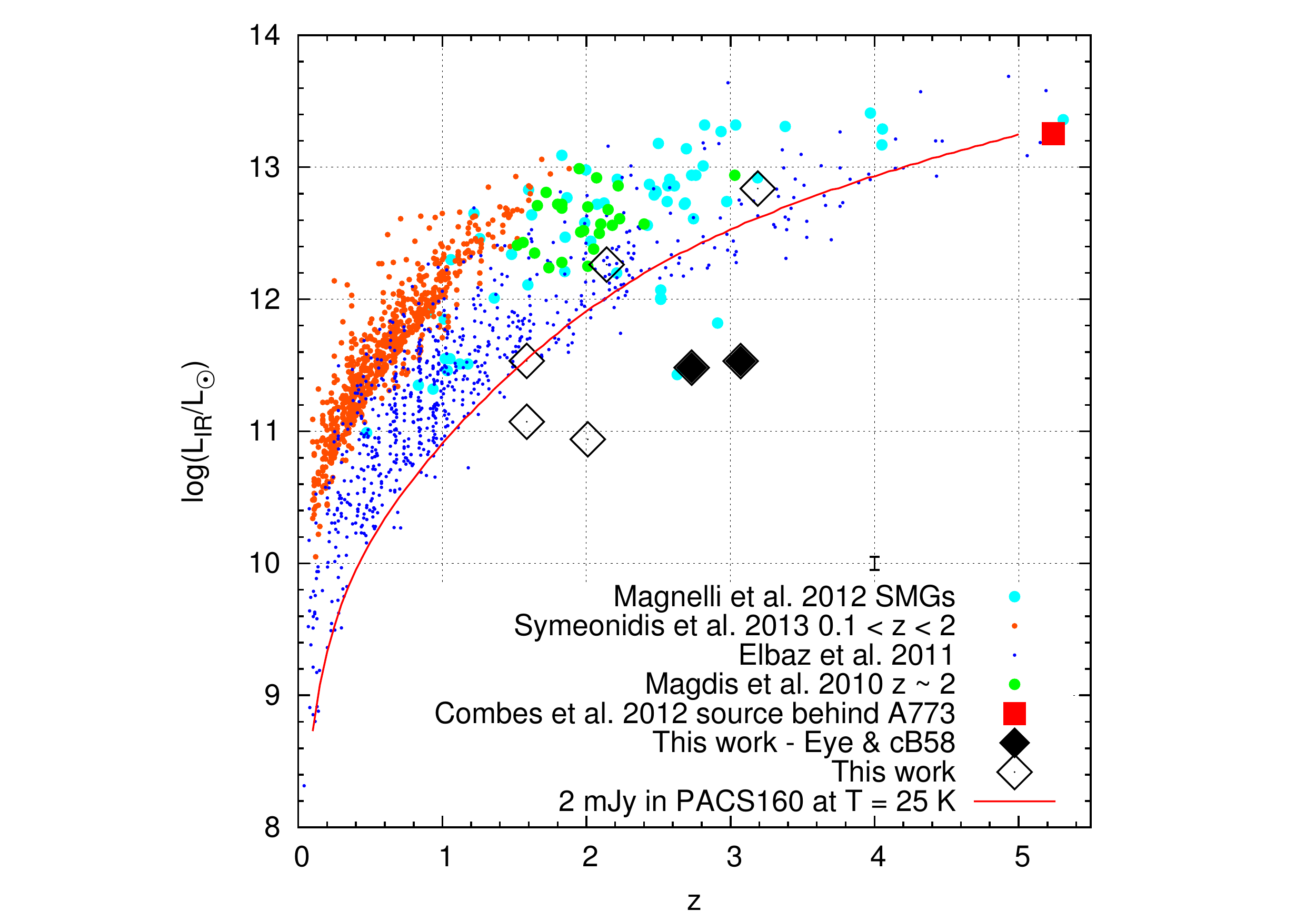}
\caption{IR luminosity of {\it Herschel}-detected galaxies as a function of redshift showing the position of our five lensed galaxies
(open diamonds), two well-studied lensed galaxies from the literature (cB58 and the Cosmic Eye, marked as grey
diamonds), and galaxies from various blank field observations, including data from the GOODS and COSMOS
		 blank fields by \protect\citet{2013arXiv1302.4895S,elbaz2011}. Also plotted is the SMG sample of \cite{2012A&A...539A.155M}, 
		 and the hyLIRG detected by the \textit{HLS} behind Abell 773 \citep{2012A&A...538L...4C}. Clearly, most of the lensed galaxies
		 at $z \sim $ 1.5--3 extend the blank field studies to fainter luminosities, into the LIRG regime.
		 The typical uncertainty of our \lir\ measurements is $\pm 0.1$ dex, or smaller. The red curve shows the minimal \lir\ at each redshift
		 that can produce a flux $ \geq 2 $ mJy in PACS160 (2$ \times $ the confusion limit, $\sim3\sigma$-detection limit in 
		 GOODS-N).}
\label{fig_lir_z}	
\vspace{-0.15 in}
\end{figure}
%-----------FIG LIR-Z--------------
 
%%%%%%%%%%%%%%%%%%%%%%%%%%%%%%%%%%%%%%%%%%%%%%%%%%%%%%%%%%%%%%%%%%%%%%%%%%%%%%%%%
\section{IR properties of the sample}
\label{s_IR}

The main ``observed" quantities of our sample, derived from simple SED fits,  are summarized in Table \ref{OBSERVED-GLOBAL}.
The SPIRE and PACS data provide good constraints on the 
dust emission peak, allowing us to evaluate precisely the FIR luminosity, \lfir,
determined from integration of the best fit SED in the wavelength interval [8,1000] \micron. 
Fits with the different libraries we used agree within typically 0.05 dex
when they accurately reproduce the photometry. 
A comparison with the code {\sc Cmcirsed} of \cite{2012MNRAS.425.3094C} yields a mean 
$ <\log(\lir({\rm CMCIRSED})) -\log(\lir({\rm Hyperz}))>\  = 0.016 \pm 0.079$, showing no systematic offset.
Best-fit SEDs are shown and discussed below. 
Based on these values and correcting for
the lensing magnification factor $\mu$, we then calculate 
the IR-inferred SFRs, \sfrir, adopting the \citet{1998ARA&A..36..189K} calibration.
The temperature \tdust, a measure of the dust temperature, was derived by fitting modified black bodies to the FIR/submm data,
using an emissivity index of $ \beta = $ 1.5. 
Further discussion on the values and parameters used can be found  in Sect.\ \ref{s_dust}.

Overall the observed IR luminosities of our objects range from $(3-16) \times 10^{12}$ \lsun. However, the intrinsic,
lensing-corrected values considerably lower, between $6 \times 10^{10}$ and  $6 \times 10^{12}$ \lsun. 
The intrinsic IR luminosities of our sample are shown in Fig.\ \ref{fig_lir_z}
as a function of redshift, and compared to other galaxy samples observed with Herschel.
Clearly, our sample extends previous blank field studies to lower \lir\ magnitudes, thanks to gravitational lensing.

Concerning our comparison sample, we note that the inferred IR luminosity of the Cosmic Eye, $ \log ($\lfir$ \cdot \mu) = 12.98^{+0.02}_{-0.03}$,
is $ \sim $0.3 dex lower than the estimated value in \citet{2009ApJ...698.1273S}, about $ \sim $0.05 dex below their quoted $1\sigma$ interval, 
which was determined in absence of {\it Herschel} data. Our measure is in exact agreement with the estimation of 
\citet{2007ApJ...665..936C} based on the rest-frame 8\micron\ flux.

For cB58 our IR luminosity, determined from the available FIR measurements (with new PACS and SPIRE data added to the existing
MIPS and 850\micron\ and 1.2mm), is $ \log ($\lfir$ \cdot \mu) = 12.96 \pm 0.01$, which is notably brighter
compared to $ 12.58 \pm 0.08 $ derived  by \citet{2012ApJ...745...86W} and the earlier estimate of (12.48 - 12.78) from  
\citet{2008ApJ...689...59S}. This is mainly due to the detection of a warmer dust temperature made accessible by {\it Herschel}'s
observations (and is discussed further in Sect \ref{s_dust}).

The recent publication of \citet{2013arXiv1309.3281S} has cB58 and the Cosmic Eye
in common in a similar analysis of their IR emission. We find the same \lir\ within our margin of errors for the former. Concerning the Eye
our estimation of \lir\ is $ \sim 0.2 $ dex lower than theirs. This is due to our de-blending work (see Sect. \ref{s_photo_eye}),
that lowered the fluxes attributed to this particular source.

%-----------FIG LIR--------------
\begin{figure}[htb]
%\centering
\hspace*{-0.1 in}
\includegraphics[width=9.2cm]{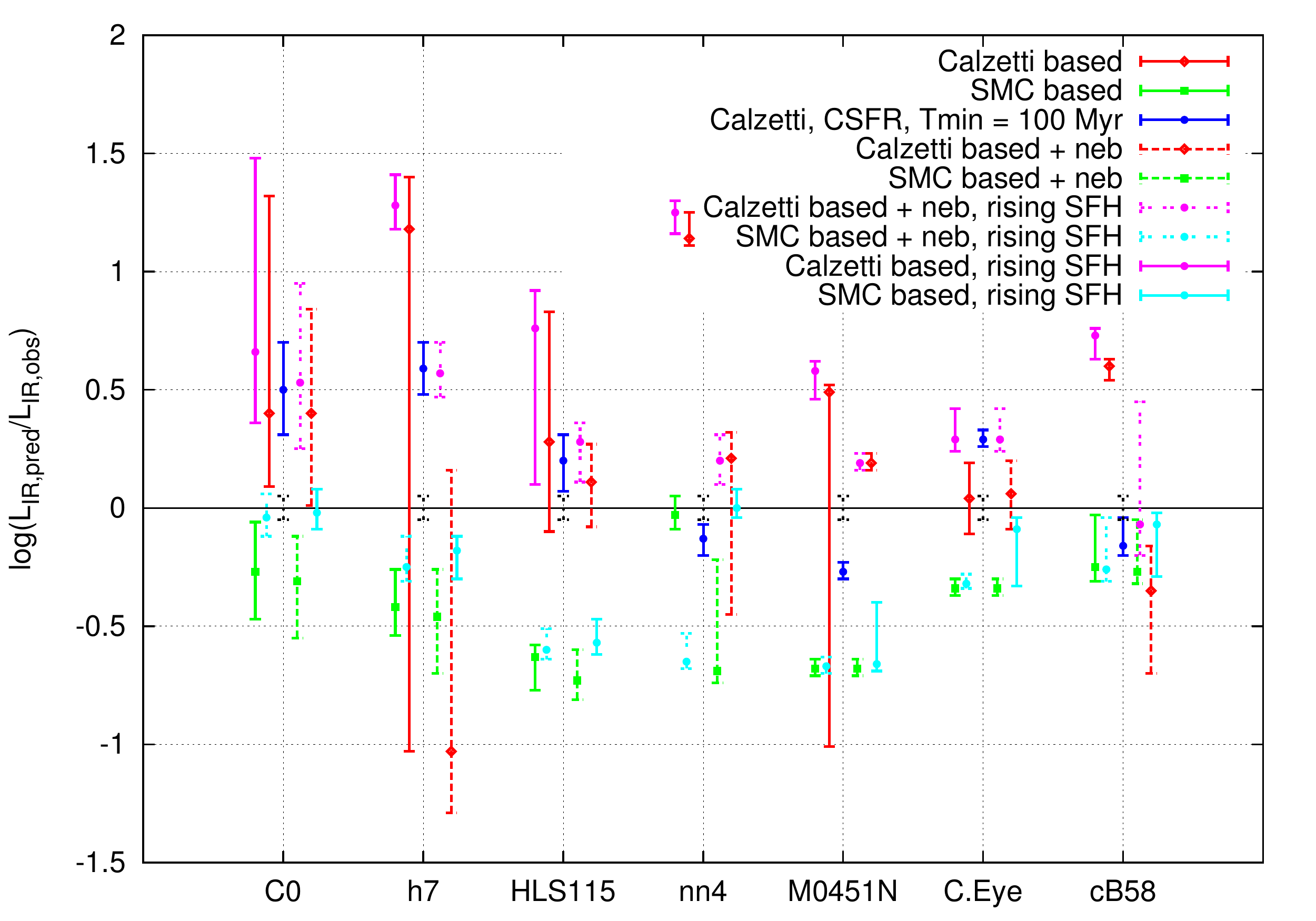}
	\caption{Predicted over observed ratio for \lfir\ for all the galaxies modeled and 
	for the different stellar population scenarios we have explored.
	(dashed symbols stand for declining SFH models including nebular emission, 
	solid lines neglecting this effect, dot-dashed cyan and magenta consider rising
	SFH scenarios including nebular emission). 
	We can see that the SMC based predictions underpredict \lfir\ in almost all cases. 
	The Calzetti based models, although more degenerate, achieve to match most of the objects observed \lfir within the
	68\% confidence range. The rising SFH models predict globally at least as much or more 
	\lfir\ than their corresponding -- in terms of extinction law --  declining SFH ones
	(as shown in \cite{2013A&A...549A...4S}), pushing in particular the Calzetti based 
	models to overpredict the observed quantities. The effect is similar but smaller 
	for the SMC based solutions, and allows a perfect match in the case of C0. \newline
	 Note: for MACS0451N the observed \lir\ of the northern part, representing $\sim 1/3$ of the total
		was adopted here, and the predicted \lir\ compared here are also derived for this same region,
		for coherence. This means, that depending on the model, the predicted \lir\ used are $ \sim 35-40\% $ of the 
		values listed in the Tables \ref{selected_model_table} and \ref{classic_model_table}.
		If one would plot the same for the total arc in the eventuality of a negligible AGN contribution
		the ratios shown would be scalled down by $\sim 0.15$ dex at most.	}
\label{Lir_plot}
\end{figure}
%-----------FIG LIR--------------

%%%%%%%%%%%%%%%%%%%%%%%%%%%%%%%%%%%%%%%%%%%%%%%%%%%%%%%%%%%%%%%%%%%%%%%%%%%%%%%%%
\section{SED fitting results}	\label{sec_indiv_sources}

\subsection{Results for individual HLS galaxies}

We now present and discuss the detailed results from the SED fits for each galaxy and the 
differences obtained for models using different star formation histories and extinction laws,
and with or without nebular emission.
The main derived physical parameters are summarized in Table \ref{selected_model_table}
for variable star formation histories, and in Table \ref{classic_model_table} for ``classical''
models assuming constant SFR and neglecting nebular emission. 
We present in some cases more than one of the different solutions obtained, regardless of the 
reduced \ki2 values 
(almost all of our solutions are in very good agreement with the photometry), where
we deem a discussion interesting when comparing the physical interpretation of our objects with the
different models used.

The energy-conserving models with fixed extinction are discussed separately, after the individual sources discussion,
in Sect. \ref{s_nrg_conserv}.
The physical parameters are discussed and compared to other samples in  Sect.\ \ref{s_stellarpops}.
The IR luminosities predicted from the various SED fits are compared to the observed \lir\
in Fig.\ \ref{Lir_plot}. This comparison provides a consistency check on the dust extinction
and on the age and SFH-dependent luminosity emitted by the stellar population. 

The different sizes of the 
error bars seen are due mainly to different numbers of free parameters. The constant SFR scenario has the smallest number of free parameters 
(on top of allowing only constant SFR, it also forbids ages below 100 Myr, not leaving room for much degeneracy).
The other cases allow varying timescales in star-formation (as stated in Sect. \ref{sed_fits}). This said, only the
declining SFH ones using Calzetti's law are really prone to the age-extinction degeneracy. SMC models tend to mostly favor 
long timescale scenarios, and the rising SFH models converge regardless of timescale as they must be at peak star-formation
at age $ t $, and thus will tend to have the same extinction to reproduce the SED's colors.
After a case-by-case discussion we will come back to this in Sect.\ \ref{ir_lum}.

 In order to better accommodate the reader, the following subsections are organized in a standard pattern, with two main paragraphs
each, a first one discussing the stellar models, and the other the FIR fits. In particular we discuss how the physical parameters
depend on the model assumptions (mostly SFH and extinction law) and we examine how the FIR data allow one to rule out some of the
assumptions.

%------------------------------
\begin{table*}[htb]
	\centering
	\begin{tabular}{l c c c c c c c}
		\hline\hline
			\textbf{ID} &	model	&\ki2  & \av\	& Age [Gyr]	&  $ \mstar\ [10^{10}\msun] $	& $  \lbol [10^{11}\lsun] $ & \sfrbc [\msunyr]\\
				\hline	
		A68/C0$\ast$ 	&Calz+neb+decl&  0.72	&1.6 (1.1-2.0)	& 0.18 (0.05-1.4)	& 2.1 (0.9-3.9) &  2.97 (1.3-7.8) & 35.5 (13.9-117.5)\\
		A68/C0 		  	&SMC+decl	  &  1.02	&0.5 (0.4-0.7)	& 3.4 (2.3-3.5)	& 4.1 (3.3-4.8) &  0.6 (0.4-1.0) & 10.6 (8.1-16.1) \\
		A68/h7$\ast$ 		 &Calz+neb+decl	& 3.67	& 0.3 (0.2-1.4)	& 0.36 (0.18-0.36) & 18.5 (17.2-21.8) & 1.71 (0.98-24.7) & 3.6 (2.8-174.1)\\
		A68/h7 		 &SMC+neb+decl 	& 1.86	&0.5 (0.4-0.6)	& 0.36 (0.25-2.0) & 17.4 (13.6-27.4) & 6.35 (3.8-9.4) & 73.8 (20.1-155.8) \\
		A68/HLS115$\ast$ 	& Calz+neb+decl	& 1.65 & 1.7 (1.5-1.9) &  0.09 (0.05- 0.13)&  1.0(0.7-1.5) &  4.4 (2.9-6.1) &  57.8 (30.3-98.5)\\
		A68/HLS115	   &SMC+neb+decl	& 2.5	& 0.5 (0.5-0.6)	& 3.5 (2.6-3.5)	& 3.0 (2.7-3.6)	& 0.6 (0.5-0.8)	& 10.9 (9.1-13.6)   \\ 
		A68/nn4$\ast$	& SMC+decl & 0.85	&1.9 (1.8-2.0) 	& 0.033 (0.03-0.039) & 5.7(5.3-6.2) & 64.3 (57.3-73.8)	& 1265 (1069-1518) \\
		A68/nn4		& Calz+neb+decl & 4.68 & 2.8 (2.2- 2.9) & 0.017 (0.013-0.18) & 4.9 (4.4-18.9) & 109 (35.3-131)	& 2400 (343.8-3183) \\
		MACS0451$\ast$  	& Calz+neb+decl	& 6.3 & 1.1 (1.0-1.1) & 0.015 (0.013-0.02) & 0.15 (0.15-0.16) & 3.7 (3.1-3.8) & 101.6 (82.3-105.1)\\
		MACS0451  	& SMC+neb+decl	& 8.5	&0.2 (0.2-0.2) & 0.72 (0.72-0.72)  & 0.78 (0.77-0.85) 	& 0.42 (0.42-0.43) & 13.4 (12.9-13.6)  \\
		cB58$\ast$		 & SMC+neb+decl	&  1.57  & 0.3 (0.3-0.4) & 0.09 (0.05-0.09)  & 0.63 (0.4-0.72) & 1.63 (1.5-2.65)& 37.9 (30.9-65.1)\\
		cB58		&Calz+neb+decl	&2.86	&0.5 (0.3-0.6) & 0.128 (0.09-0.18)	& 1.03 (0.8-1.3) &	1.36 (0.62-2.05) & 21.7(9.8-413)   \\  
		Cosmic Eye$\ast$ &Calz+neb\textdagger +decl & 1.35  &0.6 (0.5-0.7)  &  0.18 (0.18-0.25) & 4.03(3.6-4.6) & 3.83(2.84-5.04) & 57.1 (47.8-85.8)  \\
		Cosmic Eye  & Calz+neb+decl	&  1.52  & 0.6 (0.5-0.7) & 0.18 (0.13-0.25)  & 3.86 (3.4-4.4) & 3.92 (2.9-5.0) & 66.1 (47.6-84.7)\\
		Cosmic Eye		 & SMC+neb+decl	&  1.65 & 0.2 (0.2-0.2)	&  1.7 (1.7-2.0)& 6.68 (6.3-7.2) & 1.56 (1.52-1.59)	&  47.3 (46-48.8)  \\
		\hline   
	\end{tabular}
	\caption{Selected variable SFH models, with physical properties derived from fitting with the BC03 library. Values are corrected of lensing.
			\lbol\ stands for the absorbed luminosity in the [912\AA\ $-$ 3\micron] for the given extinction, and is used as a proxy to predict \lir.
			In parenthesis are given the values for the 68\% confidence levels derived from our MC runs. Asterisks show the favored scenarios,
			in general they coincide with the lowest \ki2, except for h7, in which case our considerations discussed in \ref{sec_h7} made us 
			favor the Calzetti based solution. We recall that the results listed for MACS0451 come from the integrated photometry of
			the entire arc, and that a scaling factor 0.4 can be applied to estimate the properties of the northern part separately
			(see Sect. \ref{s_arc}).
			We do not show results of our rising SFH models as they do not provide any more insight to our sample nor better fits. 
			  }              \label{selected_model_table}
\end{table*}
%------------------------------

%------------------------------ 
\begin{table*}[htb]

	\centering
	\begin{tabular}{l c c c c c c c}
		\hline\hline
				\textbf{ID} &	\ki2  & \av\	& Age [Gyr]	&  $ \mstar\ [10^{10}\msun] $	&  \lbol $[10^{11}\lsun] $ & \sfrbc[\msunyr] & \sfrir[\msunyr]\\
			\hline
			
	A68/C0		 & 1.09	& 1.6 (1.4-1.8)	&  0.51 (0.255-1.01)  &  2.76 (2.04-3.57) &	3.74 (2.59-5.53) &	62.8 (42-97.1) & 20.4 (19.2-21.5)  \\
	A68/h7		 & 3.33 & 1.9 (1.8-2.0)	&  0.18 (0.13-0.25) &  20.3 (17.7-24.3) &   71.3 (57.9-85.7) &	1308 (1035-1623) &  315 (294-330) \\
	A68/HLS115 	 & 1.92	& 1.8 (1.7-1.9)	&  0.18 (0.13-0.36)	    &  1.7 (1.4-2.1)&	5.42 (4.2-6.7) &  96.3 (73-128.3) & 59.0 (56.3-61.7)\\
	A68/nn4 	 & 6.16	& 2.4 (2.3-2.4)	 &  0.25 (0.25-0.36)& 20.3 (18.6-24.9)&  51.1 (44.5-56.0) &  878.7 (729.6-962.6) & 1184 (1132-1214) \\ 
	MACS0451 	 & 40.7 & 0.7 (0.6-0.7)	&  0.18 (0.18-0.36) & 0.52 (0.51-0.75) & 1.38 (1.04-1.41) &	33.2 (25.2-33.6) &   15 - 42\\
%	A2218-mult 	 & 	&	&	&  &	&	 \\
	cB58		 & 3.92	& 0.6 (0.6-0.7)	&  0.25 (0.18-0.25)	&  1.07 (0.88-1.14) &   2.1 (1.96-2.7)  &	52.3 (48.8-64) & 52.4 (51.2-53.6)\\
	Cosmic Eye	 & 1.94 & 0.8 (0.8-0.8)	&  0.36 (0.36-0.36)	&  4.14 (4.0-4.25)&   6.65 (6.5-6.8) &	139.4 (136.1-143.1) &  58.8 (54.9-61.6) \\
		\hline   
	\end{tabular}
	\caption{Here we show the fitting results for the ``classic scenario" model, meaning using 
			Calzetti, constant star-formation no nebular emission, and $ t_{min} = 100 $ Myr, obtained like in Table \ref{selected_model_table}.
			Values are corrected of lensing. The \sfrir\ from Table \ref{OBSERVED-GLOBAL} is also listed for comparison. 
			We recall that the results listed for MACS0451 come from the integrated photometry of
			the entire arc, and that a scaling factor 0.4 can be applied to  estimate the properties of the northern part separately
			(see Sect. \ref{s_arc}).
			For MACS0451 \sfrir\ corresponds to the range given by the \lir\ of the northern segment and that of the whole arc.
			 }              \label{classic_model_table}
\end{table*}
%------------------------------

%-----------FIG C0 ste--------------
\begin{figure}[htb]
\centering
\includegraphics[width=8.8cm]{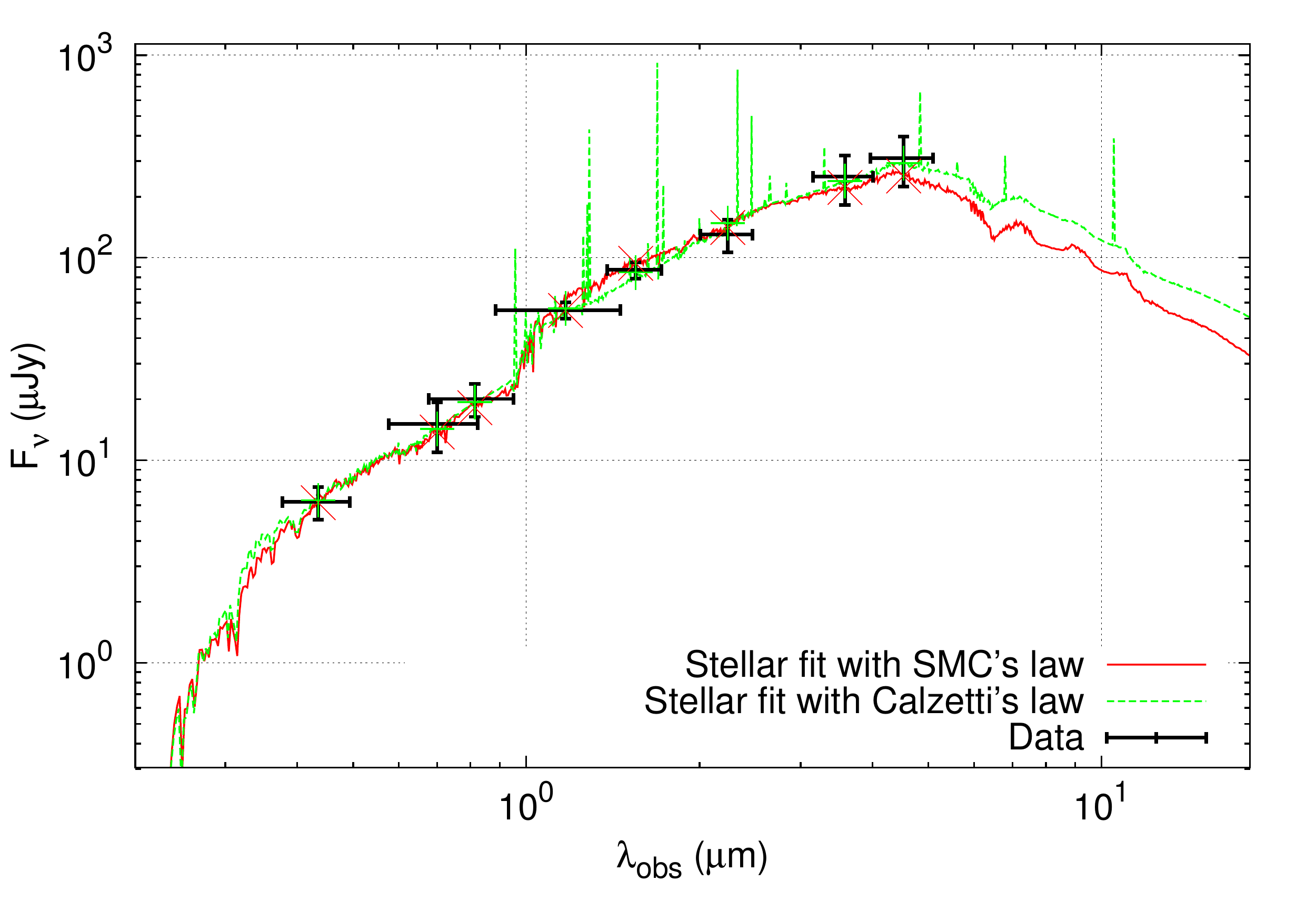}
\includegraphics[width=8.8cm]{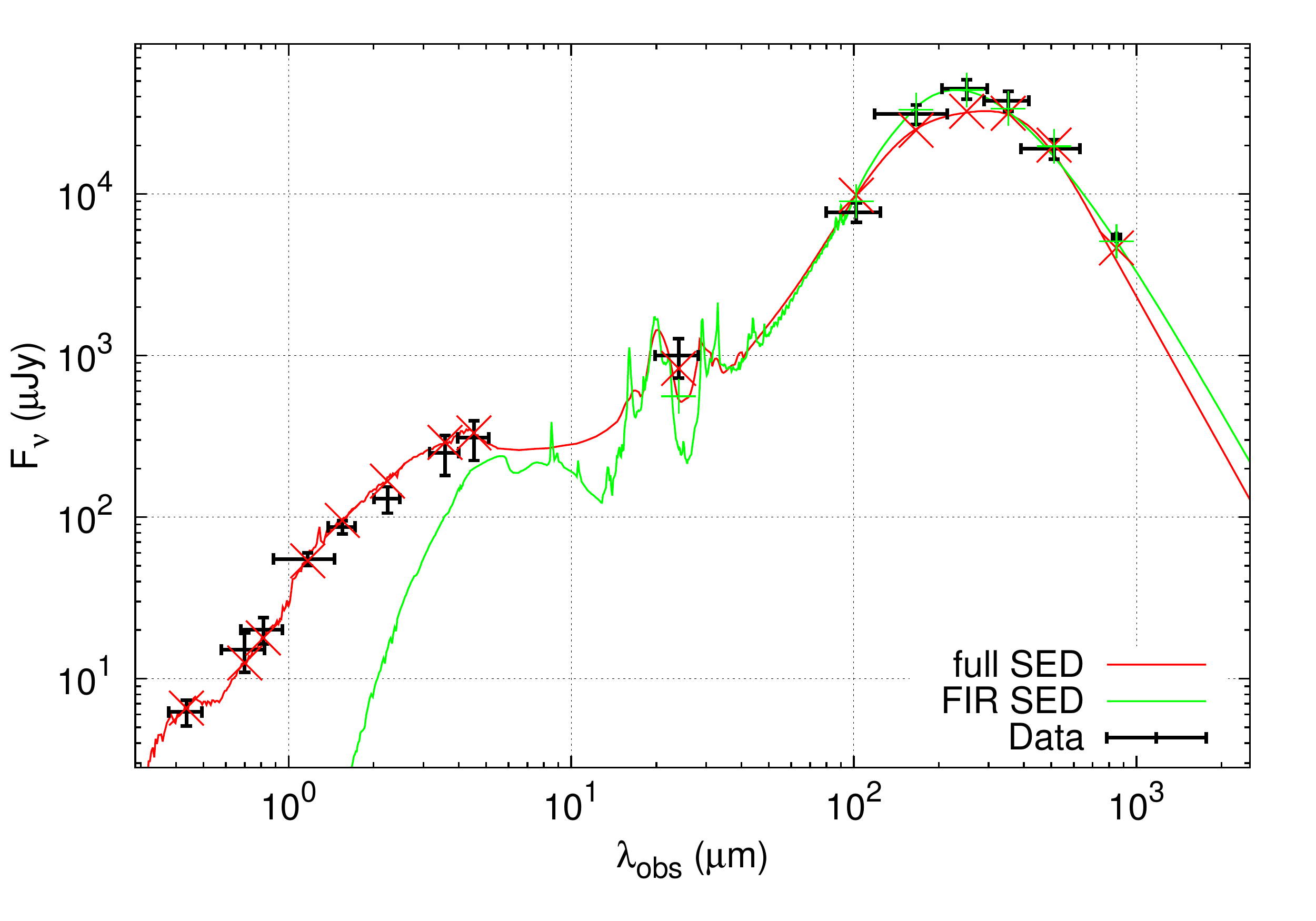}
\caption{{\em Top:} SED plots for A68/C0's best stellar population fits using the 2 extinction laws. In the case
of Calzetti the fit including nebular emission was prefered.
{\em Bottom:} global and FIR-only fits. The best fit for the full photometry uses a 
		Seyfert galaxy template from P07 (shown in red) 
		an extra extinction of   \av\ = 0.6  by the SMC law. The FIR only SED (in green) is by R09.
}
\label{fig_c0_comb}
\end{figure}
%-----------FIG C0 ste--------------

\subsubsection{A68/C0}  \label{s_c0}

The SED fits of A68/C0 performed on the visible to NIR photometry can be seen in Fig. \ref{fig_c0_comb}, and show the best solution for 
Calzetti, and the best one for SMC. 
In terms of \ki2 the overall best fit was obtained with the Calzetti based, declining SFR 
($ \tau = 300$ Myr) model that includes nebular emission. It interprets C0 as an obscured 
($ A_{\rm V} = 1.7 $) population involving a starburst/post-starburst
 in a rather young age,
$ t = 180 $ Myrs, with a well sustained \sfrbc$(t)/\mu$ of $ \sim 36 \msunyr $
(see Table \ref{selected_model_table} for error estimates).
The important extinction overpredicts the observed FIR luminosity by a factor of 4 for the best fit, and 
$2.9^{+5.1}_{-1.7} $ for the MC runs, which puts it on the edge of the derived 
68\% confidence interval, as shown in Fig.\ \ref{Lir_plot}.
A similar discrepancy is also found with ``standard'' SED fits assuming SFR=const and neglecting
nebular emission.

Fits with the SMC law has only a slightly larger $ \chi^2_{\nu} $, 
but give a significantly different interpretation, that is more adequate with the idea that C0 is a quiescently star-forming galaxy, 
it has a very old population of $ t = 3.5 $ Gyr, and an almost constant star formation history ($\tau = 3 $ Gyr) with
 \sfrbc$(t)/\mu \sim 10 \msunyr $. 
In this case the extinction is $\av = 0.5$, and the predicted IR luminosity has its upper 68\% limit slightly
below the observed \lfir\  (cf.\ Fig.\ \ref{Lir_plot}) but matches is at 90\%. 
Perhaps a special mention can be made for the SMC-based rising SFR model, as it reproduces almost perfectly the observed \lir\ (and A68/C0 
is the only case where this happens in our sample). This model actually resembles in most aspects the one just described, with the same age
(oldest allowed at this redshift) and very slowly increasing SFR (same $\tau = 3 $ Gyr, largest among our rising SFHs), and
\sfrbc$(t)/\mu \sim 18 \msunyr $, hence a slightly higher extinction allowing the correct prediction of \lir. 
Based on this model and the
IR observation, the quiescently star-forming galaxy scenario seems indeed well suited, only with more current SFR than in the past, rather
than the opposite.

Despite these differences, the stellar mass of A68/C0, $\mstar \approx (2-4) \times 10^{10}$ \msun, agrees
within a factor $\sim 2$ for all models.

The best full SED fit, shown in Fig.\ \ref{fig_c0_comb}, was obtained with templates from the P07 library,
``modulated" by the SMC extinction law. 
Its steeper slope in the UV allowed for a better match of the B band's photometry (rest-frame UV) without reddening the SED that
much as to degrade the fit in the rest-frame optical.
The fits show a slight underestimation of the dust emission peak, but is in agreement with the $[8-1000 \mu m]$ IR luminosity,
found to be $\log[L_{IR}/L_\odot] = 12.55\pm 0.03$.
The interpretation of the templates's dust emission peak gives a dust temperature of $\sim 35$ K using Wien's displacement law. 
The de-lensed $\sfrir \approx 20$ \msunyr, using the \citet{1998ARA&A..36..189K} calibration. 
The SMC based models that favored long, almost constant SFHs are slightly beneath this value, as is their predicted \lir.
Models with the Calzetti attenuation law overpredict \lir\ by a factor $\sim 2.5$ but marginally reproduce it within their 68\% confidence
level (cf.\ Fig.\ \ref{Lir_plot}).

%-----------FIG h7 ste neb--------------
\begin{figure}[htb]
\centering
\includegraphics[width=8.8cm]{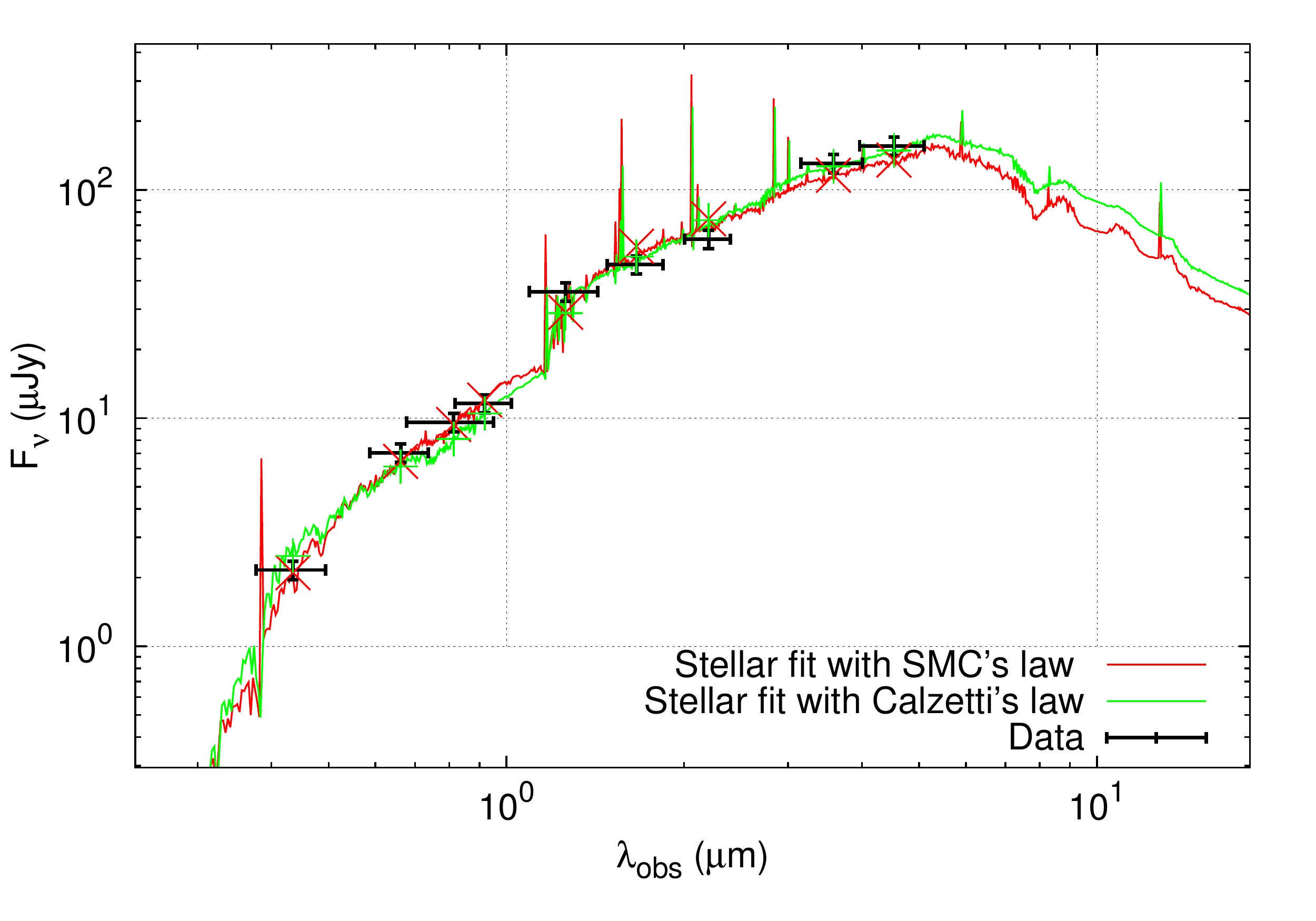}
\includegraphics[width=8.8cm]{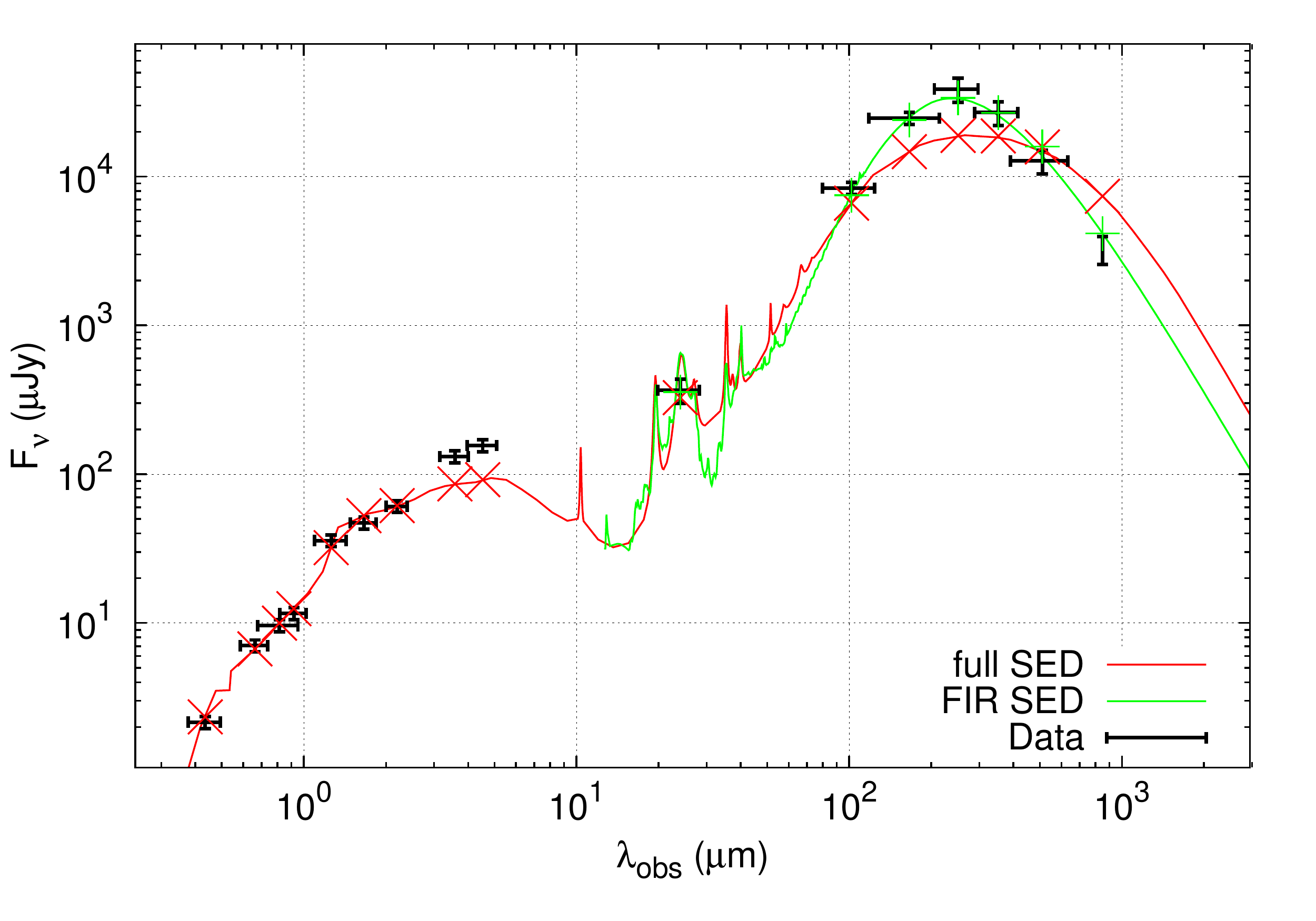}
\caption{{\em Top:} A68/h7 stellar population fits including nebular emission. Here the older Calzetti based population renders
well the photometry longwards of 4000\AA but fits less the UV, and ultimately the SMC has the smallest \ki2. 
{\em Bottom:} full and FIR SED (red and green respectively). Only M10's templates could produce $\ki2<10$ and in particular
\ki2 = 4.2 for the present SED, attenuated with the SMC law. The peak and width of the dust emission distribution are here well fitted
by a R09 template.}
\label{fig_h7_comb}
\end{figure}
%-----------FIG h7 ste neb--------------

\subsubsection{A68/h7} \label{sec_h7}

A68/h7 shows a very red slope in its photometry and is peculiarly prone to a large degeneracy in terms of age and extinction.
Naturally, this translates to large uncertainties in the expected IR emission, as seen in Fig.\ \ref{Lir_plot}.
The best fits are obtained without
nebular emission regardless of the extinction law, and the SMC based model produces ultimately the smallest \ki2. The SMC based solutions 
favor slowly decaying almost constant star formation histories when excluding line emission, and do not seem to favor any particular SFH 
when including it. On the other hand the Calzetti based models favor very rapidly declining SFHs ($\tau =50 $ Myr, the smallest rate in our parameter space) 
for the runs that include nebular emission, whereas the SF timescale is less well constrained without lines.
In all cases, the inclusion of nebular emission degrades the fit quality by a factor of 1.5-1.6 in terms of \ki2, 
which is not very important, but as we can see in the following, affects very strongly the physical interpretation for the Calzetti-based
models.
Best-fit SEDs to the stellar part of the SED are shown in Fig. \ref{fig_h7_comb}. Although differing by
a factor $\sim 2$ in \ki2, the two fits showing different extinction laws are clearly fairly similar, and satisfactory.
Also it can be noted that for the case of this obscured/old population the emission lines are not very strong, which is logical.

How does the inclusion of nebular emission affect the assessment of physical parameters?
For the SMC law the changes are small. However, with the Calzetti attenuation law, there is a 
strong divergence in the solutions, with the one including nebular emission shifting the median age from 10 Myr to 360 
Myr\footnote{This effect of rendering a solution older when including nebular emission seems opposite
to the trend established for z $ \sim 3-6 $ LBGs in \citet{2012arXiv1207.3663D}.}.
The solution without nebular emission
seems highly unrealistic, as it has a median solution for $\sfrbc/\mu$ that is $ \sim 8000 $ \msunyr, and spans at the $ 1\sigma $ level
from $ \sim 3$ \msunyr  up to $ \sim 16000 $ \msunyr. So actually the model including lines lies within a sub-region of the whole 
degenerated parameter space of the former. Also in terms of extinction the model without lines has its $ 1\sigma $ interval for \av\ 
between 0.3 and 2.8, whereas for the model with nebular emission the derived \av\ range (between 0.2 to 1.4)  is less extreme.

Despite the differences between the models just discussed, the stellar mass agrees quite well (within $\sim$ 20\%) 
for the models listed in Tables \ref{selected_model_table} and \ref{classic_model_table}.

The full and FIR SEDs of A68/h7 can be seen in Fig. \ref{fig_h7_comb}. The full SED fit was obtained with an SMG template from M10, with additional
extinction on the rest-frame UV/optical slope. An \av\ of 0.5 with the SMC's law yields a \ki2\ $ \sim $3 times smaller than using 
Calzetti, which is driven by the steep UV slope given by the photometry.
The best fit of the FIR data was obtained with the Rieke templates and gives a lensing corrected \lfir\ of 
$ 1.7_{-0.1}^{+0.08}\cdot10^{12} \lsun$. 
The highly degenerated solution for the Calzetti models makes it hard to produce a robust statement as far as how their \lfir\ predictions 
can help us consider these SFHs to be accurate. Especially the solution without nebular emission which spans across $ \sim $2.5 dex
at the $ 1\sigma $ level actually contains every variant between an extreme young/obscure starburst to a quiescent/old population 
(cf Fig. \ref{Lir_plot}). As already discussed, the addition of line emission reduces in this case the degeneration towards the older solution which
still predicts the observed \lfir\ within $ 1\sigma $. 
The SMC based predictions fall short of the observed \lfir\ at $ 3\sigma $ regardless of line emission, which indicates solutions that are 
probably slightly too old.
Clearly, the standard model SFR, \sfrbc, derived for SED fits assuming constant SFR, is inconsistent (too large) with the \sfrir\ 
derived the standard calibration.  This, and the large degeneracies found for the fits of this object, shows that the
instantaneous SFR of this galaxy cannot accurately be determined 
with the current approach. As we will see with the energy conservation approach discussed in Sect. \ref{s_nrg_conserv}, the
use of the observed \lir\ as a constrain in our population modeling, proves to be a very useful tool in breaking such
degeneracies.

%-----------FIG hls115 ste--------------
\begin{figure}[tb]
\centering
\includegraphics[width=8.8cm]{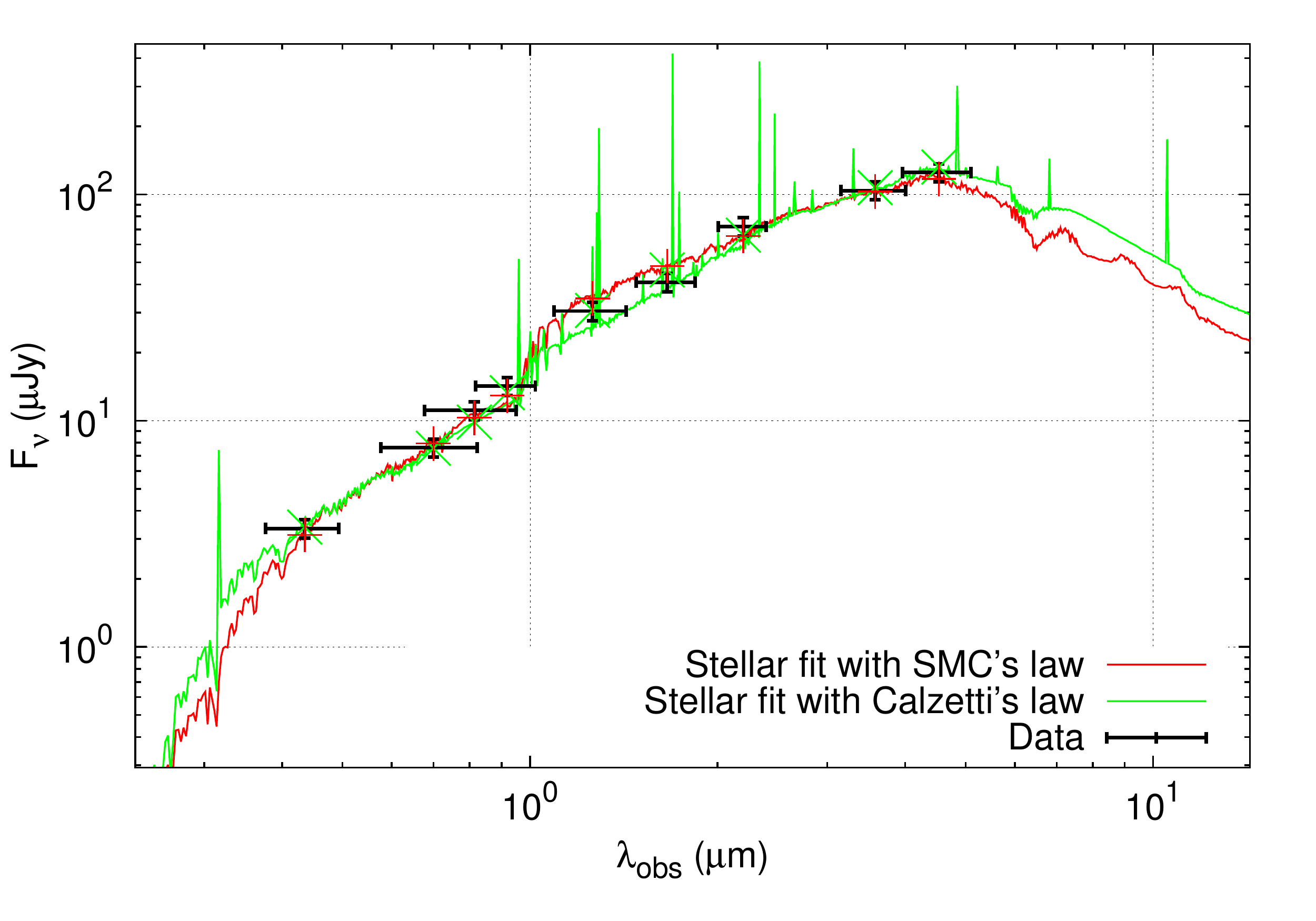}
\includegraphics[width=8.8cm]{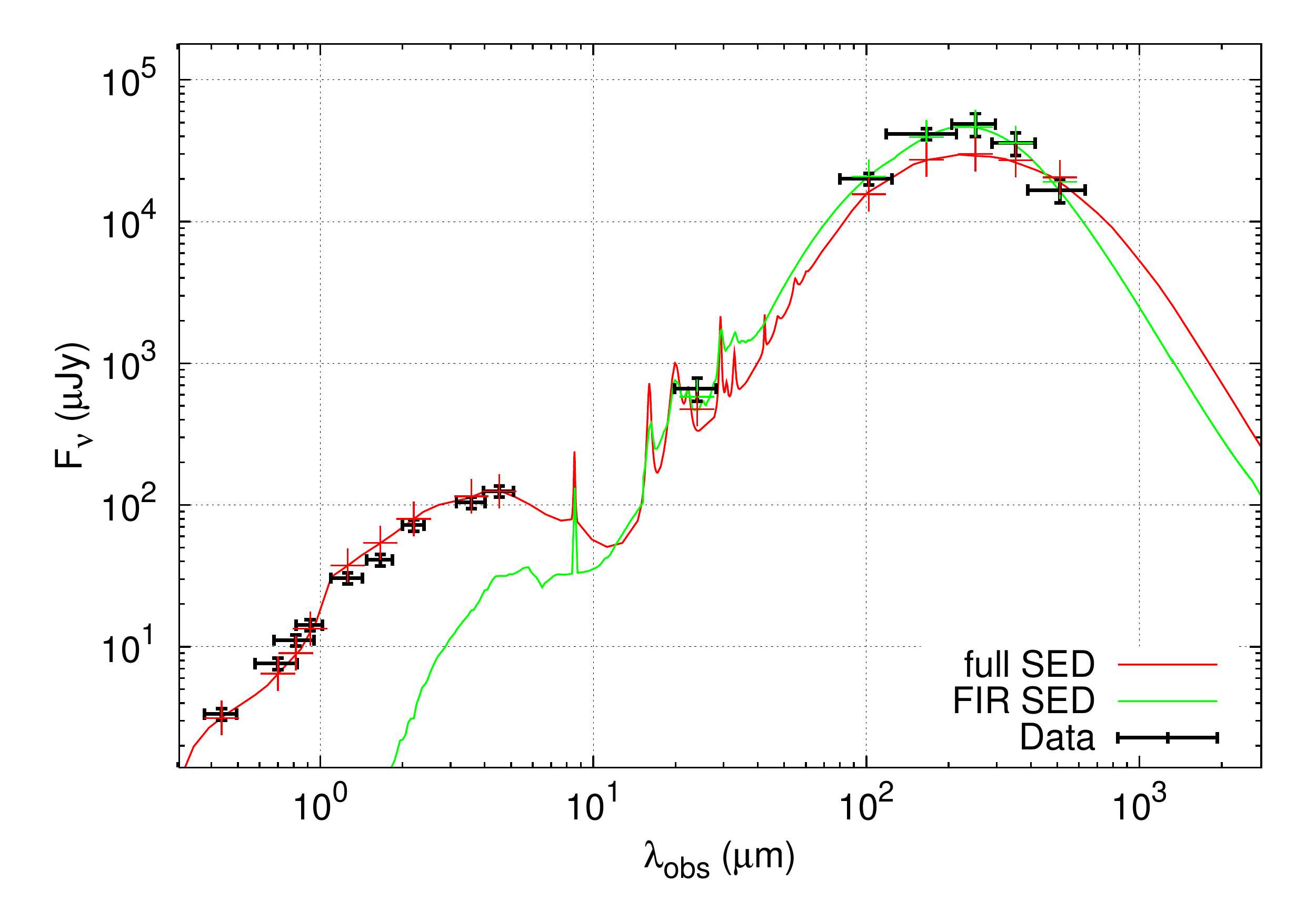}
\caption{{\em Top:} A68/HLS115 stellar population SEDs. The Calzetti based model with nebular emission (green) is plotted together with the 
SMC based one (red) for comparison. Although formally the best \ki2 was obtained with the Calzetti model with no emission, the effects of 
line emission helped lower the continuum, thus downsizing a little the physical properties of the population. The curve of the SMC model
that produces a much older population illustrates well how higher the continuum can be without nebular lines. 
{\em Bottom:} A68/HLS115 full and FIR SED fits, in red and green respectively. The full fit was obtained with an SMG template from M10 with an
additional Calzetti-based extinction of \av = 1.2. The FIR fit is obtained with the CE01 library.}
\label{fig_hls115}
\end{figure}
%-----------FIGhls115  ste--------------

\subsubsection{A68/HLS115}

The SED of HLS115, shown in Fig. \ref{fig_hls115},  is very similar to that of A68/h7,
with a slope almost as red. 
The model based on Calzetti's law, exponentially 
declining SFR ($ \tau = 50$ Myr) and nebular emission is still
a very good fit, and manages also to reproduce \lfir\ within a rather narrow 
uncertainty interval. The SMC based solutions not only produce
fits of lower quality in this case, but they also predict a too weak IR luminosity.

The physical properties of HLS115 with the aforementioned model (see Table \ref{selected_model_table}) describe
it as a young galaxy having passed through a recent starburst. With   $\mstar \approx (0.7 - 1.5) \times 10^{10}$ \msun, 
$\av\ \approx 1.5-1.9$, and $ t \approx 50-130$ Myr it still actively forms stars at $ \sfrbc(t)\approx 30-100 $ \msunyr (at 68\%
confidence level). 
Considering the two models listed in Table \ref{selected_model_table}, the median stellar mass differs by a factor of three,
mostly due to age differences.

The best fit for the UV-to-FIR photometry was again obtained with a M10 template (see Fig. \ref{fig_hls115}). It 
reproduces well the stellar emission with an additional extinction of \av$ = 1.2 $ with Calzetti's law, but misses the IR peak
by a factor $ \sim 1.5 $. The FIR photometry is best fitted by a template from CE01, and gives a lensing-corrected $\lfir\approx
3 \cdot 10^{11}\lsun $. This corresponds to a \sfrir\ of $ \sim 51 $ \msunyr with \citet{1998ARA&A..36..189K}'s calibration, well 
within the 68\% confidence intervals produced by the stellar population fit. 
As already mentioned above, the IR luminosity is well predicted by the Calzetti based 
models, and, as already seen for the other objects, including nebular emission reduces the degeneracy/uncertainty (as 
shown in Fig.\ \ref{Lir_plot}). The restricted scenario based on Calzetti, CSFR and $ t_{min}= 100$ Myr, yields similar results. 
As can be noted 
A68/HLS115 is a clear case, where the predicted IR emission allows one to exclude the SMC extinction law.

%-----------FIG nn4 stel--------------
\begin{figure}[htb]
	\centering
		\includegraphics[width=8.8cm]{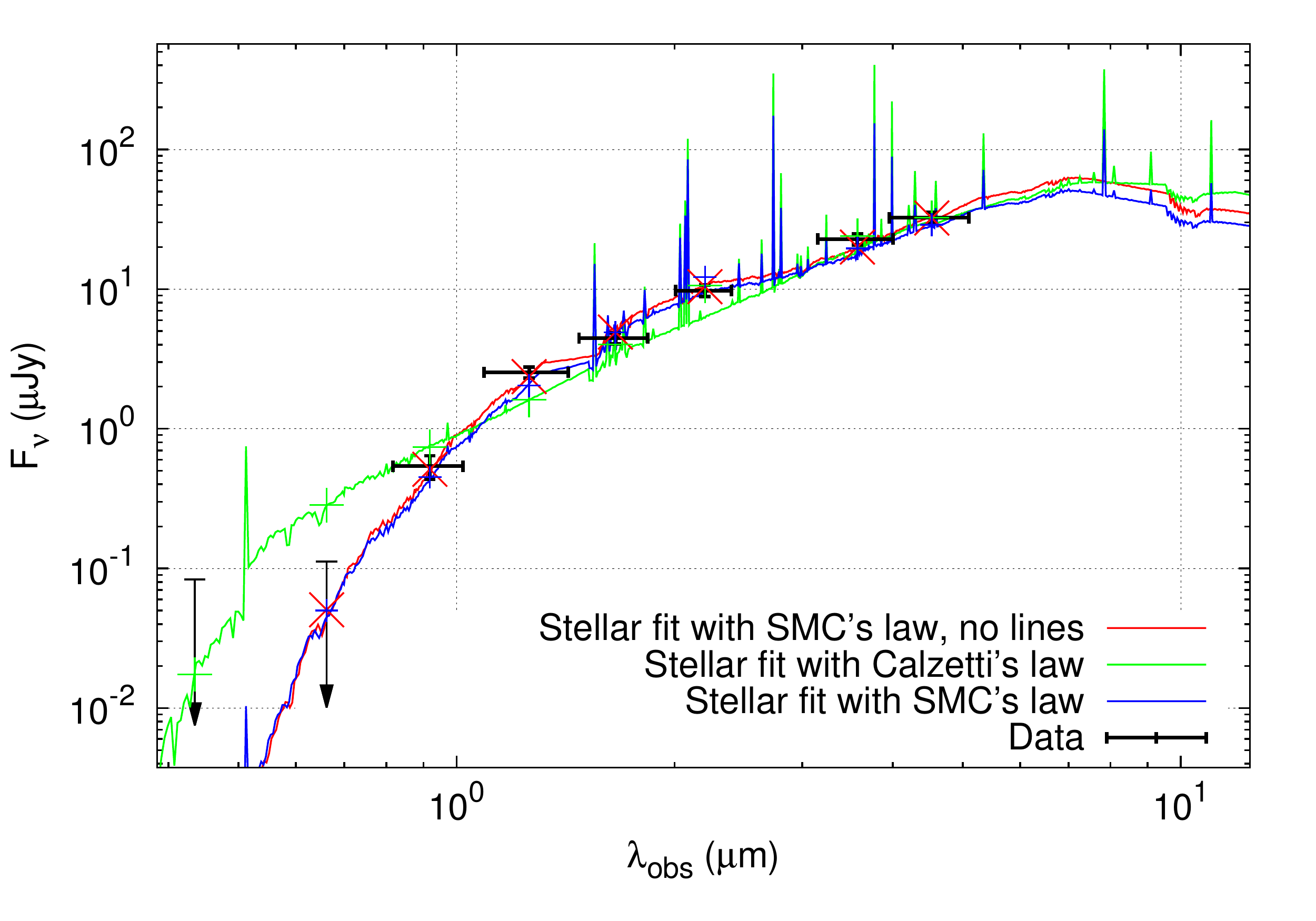}
\includegraphics[width=8.8cm]{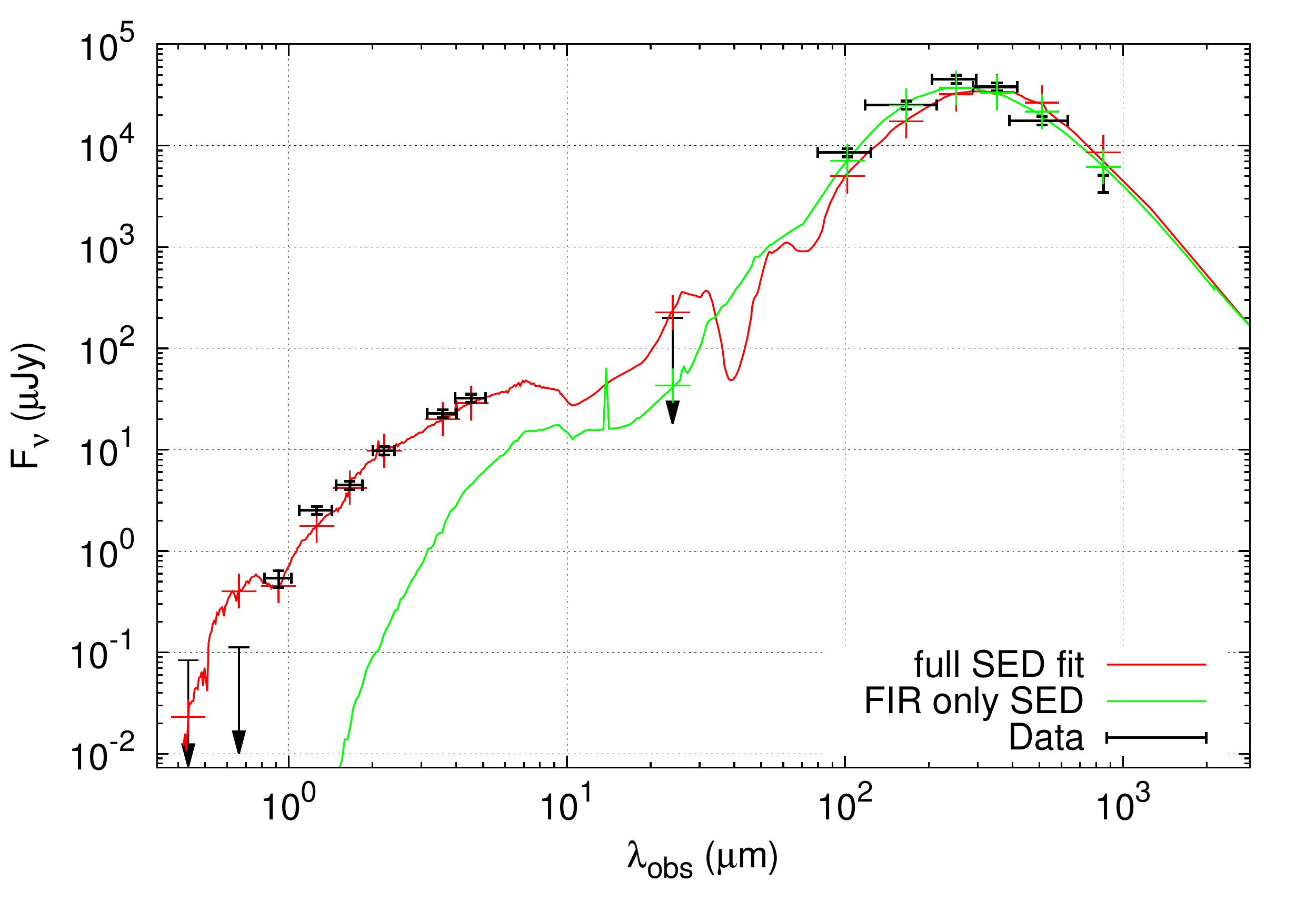}
			\caption{{\em Top:} SED plot for A68/nn4's stellar population with non-detections shown by 1 $\sigma$ upper limits. 
			 For such a strongly obscured galaxy, the two extinction laws produce quite different slopes, with the 
			 Calzetti law  being unable to reproduce the steep photometry at the 1 $\sigma$ level. Allowing for more extinction 
			 (\av$\geq 4$) could not offer a better fit since this law is relatively flat, and it would have worsen the minimization 
			 in the rest-frame optical domain. The SMC base solution on the other hand is steep enough. The nebular lines are relatively
			 weak, since despite the very young age they are also absorbed, and bare little effect on the solution here.
			 {\em Bottom:} Full SED fit (red), obtained with
			the template of IRAS 20551-4250 from the P07 library, with an additional SMC-based extinction of \av = 0.4.  
			Given the intrinsic attenuation of the template and its characteristic 2175 \AA\ bump, adding more extinction in order 
			to pass beneath the 1$ \sigma $-detection limits was not possible. 
			The best fit of the FIR data (green) was obtained with the CE01
			library.}
						\label{fig_nn4}
\end{figure}
%-----------FIG nn4 stel--------------

\subsubsection{A68/nn4}

As mentioned in Sect.\ \ref{s_objects}, we study here the faintest and reddest component of what seems to be a pair in strong
interaction.
Its UV slope is indeed so steep, that no fits were successful when using Calzetti's law, at least at the 
$ 1\sigma $ level. In Fig. \ref{fig_nn4} we show
a plot comparing the two solutions with and without nebular emission for the SMC based models, plus the Calzetti based
solution including nebular emission for comparison. The photometry is very well fitted without 
emission lines (\ki2 = 0.85), and the fit is somewhat degraded when adding them (\ki2 = 2.7), although it's
 mostly the flux in the K band that 
gets overestimated due to the \Oiii\ lines. 
In both cases the models produced describe a powerful (de-magnified \sfrbc $ \sim 1200$
\msunyr) and very obscured (\av $ \sim 1.9 $ for the SMC law!) starburst at young age; 
$ t\approx 30 - 40  $ Myr is indeed the youngest age we've seen in the sample for a SMC based model. The addition of lines downscales
slightly the continuum and the physical quantities, and pushes the age towards $ \sim 60 $ Myr. 
The Calzetti based model with no emission produces an extreme solution with a median age of 2.5 Myr, $\av \approx 3.9-4$, and a
de-magnified \sfrbc\ above $ 10^5$ \msunyr! This solution is very unlikely, but it's interesting to see here again that the 
effects of adding nebular emission to the model reduces slightly its ``extreme" character, and produces a solution with $ t\approx
13-18 $ Myr, $\av \approx 2.2-2.9$ and   \sfrbc $\approx 300-3000$ \msunyr. The solution here is more degenerated, but falls within
reasonable orders of magnitudes, also for the predicted IR luminosity, as can be seen from Fig.\ \ref{Lir_plot}.
This said, we can see in Fig. \ref{fig_nn4} that the spectral slopes produced for this solution are rather 
different than for the SMC based ones, and fall short of the $ 1\sigma $ error bars. 
The Calzetti based solution with higher line attenuation  (E(B-V)$_{\star } =0.44  \times $E(B-V)$_{\rm neb}$) gives a solution that lies 
between the one for E(B-V)$_{\star } =$ E(B-V)$_{\rm neb}$ and the one excluding line emission, both in terms of \ki2 and in terms of 
derived properties.
The predicted median masses differ by a factor $\sim$ 4 between the different fits, with more plausible values 
probably being on the high side, $\mstar \approx 2 \times 10^{11}$ \msun, corresponding also to a more realistic 
``typical" age for such an IR bright galaxy.

The best fit for the whole photometry was obtained with the template of IRAS 20551-4250, a local merger and ULIRG from the P07 library,
with an additional SMC-based extinction of \av\ = 0.4 (Fig. \ref{fig_nn4}). The extreme attenuation in the rest-frame UV range
could not be reproduced without damaging the fit redwards, but most of the photometry is correctly matched at a $\sim 2\sigma $ level,
with the dust peak only slightly colder than what we can see in the FIR only fit.
Given the low gravitational magnification ($ \mu = 2.3 $), its intrinsic luminosity is very high, 
making it the brightest of our sample, a ULIRG
with \lfir $ \approx 6\cdot 10^{12} \lsun $. This luminosity is very well 
reproduced by the SMC based model with no lines, with very little
spread at the $ 1\sigma $ level (cf. Fig. \ref{Lir_plot}). When adding line emission here, the general downsizing of the solution 
makes it miss the observed  \lfir\ by a little, and is a bit more degenerate. 
For the Calzetti attenuation law, we can see that line emission has the same drastic effect of reducing also the \lfir, bringing it from 
a highly overestimated value ($> 1$ dex) within acceptable agreement with the observed value.
Standard fits with constant SFR and neglecting nebular emission also predict correctly \lir, although they fit less well the
stellar part of the SED.
As expected, this fit also yields an \sfrbc $\approx 880$ \msunyr\ in close agreement with \sfrir. 

A68/nn4 has also the warmest dust peak in our sample with \tdust$ \approx 55$ K.

%-----------FIG north line--------------
\begin{figure}[tb]
\centering
\includegraphics[width=8.8cm]{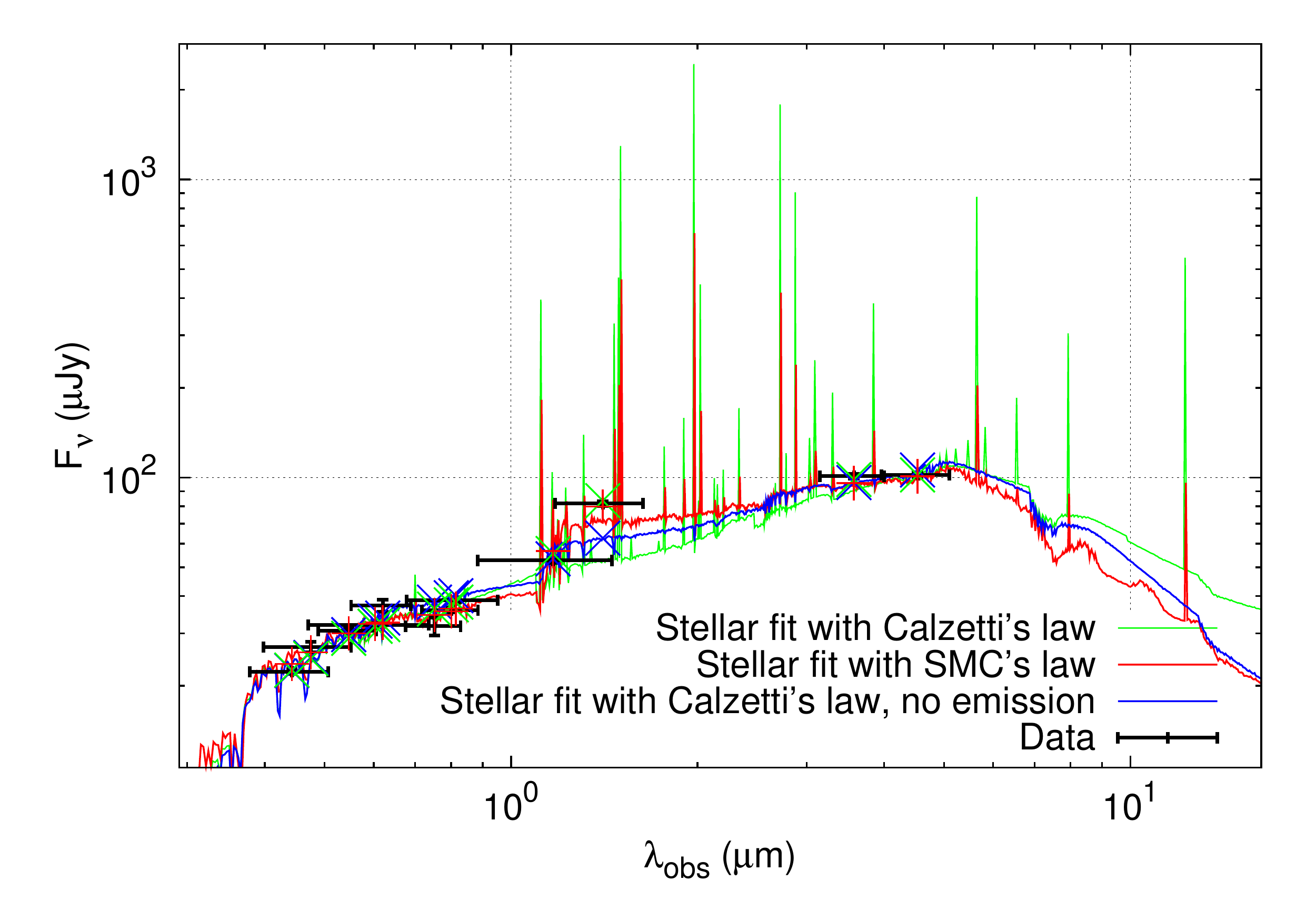}
\caption{Stellar SED plot for the MACS0451 arc. Note that the F140W photometic point can only get well fitted
with the help of nebular emission in the case of Calzetti's law (green). When using the SMC law (red), this flux is matched by
the combination of a Balmer break and the line emission, but the $ \chi^2 $ is slightly larger than the Calzetti solution.
The blue curve shows that fits without nebular emission can't produce enough flux to match the F140W point.
}
																			
\label{fig_north_lines}
\end{figure}
%-----------FIG north line--------------

\subsubsection{MACS0451 Arc}	
\label{s_arc}

This galaxy is a fairly peculiar case and will be discussed in detail in Zamojski et al.\ (2013, \emph{in prep.}).
Here we will only describe its stellar population modeling, and some of the derived properties.
We recall that the UV to NIR SEDs were obtained by using the integrated photometry of the whole arc, that presented the same 
colors throughout its length. If one would be interested in the quantities of the northern part only, the correction 
factor is about $ \sim 0.4 $.

The stellar SED of this arc can be seen in Fig. \ref{fig_north_lines}. It is a clear case of how the inclusion of nebular emission
can successfully fit some photometric points (here F140W) that otherwise could have led to photometric redshift misinterpretations 
in absence of spectroscopic data. The F140W band presents an excess for any stellar continuum emission modeled to fit the whole set, but
falls on the \Oiii\ and \hb\ region, and we see that this strong emission lines can account for the $ \sim 14\% $ of missing flux, thus improving
the \ki2 by a factor $ \sim5 $. 
The physical properties all tend towards a very young age $ \approx 15 $ Myr and very low stellar 
mass \mstar$ \approx 1.5 \cdot 10^9 $ \msun. The very 
blue slope (bluest of the sample) is probably dominated by very young stars. 
These aspects give an instantaneous \sfrbc\ of $ \sim 100 $ \msunyr, which combined to the very young age
shows that our model interprets the photometry as a starburst. 
When using the ``classic" model with $ t_{\rm min} = 100 $ Myr we obtain a constant \sfrbc\ of  $ \sim 34 $ \msunyr\
and a mass approximately 3 times higher ($\mstar \approx 5 \times 10^9$ \msun).
Since the predicted IR luminosity is quite close to the observed one, this SFR value lies 
also within the limits of the $\sfrir \sim 15-42$ \msunyr, where the lower (higher) value comes from the
\lir\ of the northern part (total) of the arc. 
The model that produces the smallest instantaneous SFR is the SMC based one, which models the colors around the F140W
band with a larger Balmer break and nebular emission, and achieves an age of 720 Myr and \sfrbc$ \sim 13 \msunyr$. The downside of this
scenario is that it requires too little dust extinction to fit the data, thus strongly under-predicting the observed \lir.
The stellar mass in this scenario is $ \sim 4 $ times larger than for the Calzetti based one, but still about a factor 2 less than the
one estimated in \cite{2011MNRAS.413..643R}, probably due to a strong overestimation of the IRAC photometry in their work.
Given the uncertainty on \lir\ due to the possible presence of an AGN in this galaxy,
this object is not well suited to obtain good constraints on the star formation history.

%%%%%%%%%%%%%%%%%%%%%%%%%%%%%%%%%%%%%%%%%%%%%%%%%%%%%
\subsection{Other lensed galaxies}
To extend our sample of lensed galaxies, for comparison with other studies in the literature, and
to test in particular the strength of the emission lines predicted by our models against spectroscopic observations
available for these objects, we have also modeled in detail cB58 and the ``Cosmic Eye", two
well-know galaxies at $z \sim$ 2.7--3.1. 
Indeed, once the available (broad-band) photometry of the stellar
part of the SED is fitted, our models still predict the IR luminosity and the strength of numerous emission
lines, which are not included as a constraint in the fitting procedure. They thus represent
very useful {\em a posteriori} tests/constraints of the models, as discussed in \citet{2013A&A...549A...4S}. 
%

%-----------FIG cb58 full--------------
\begin{figure}[htb]
\centering
\includegraphics[width=8.8cm]{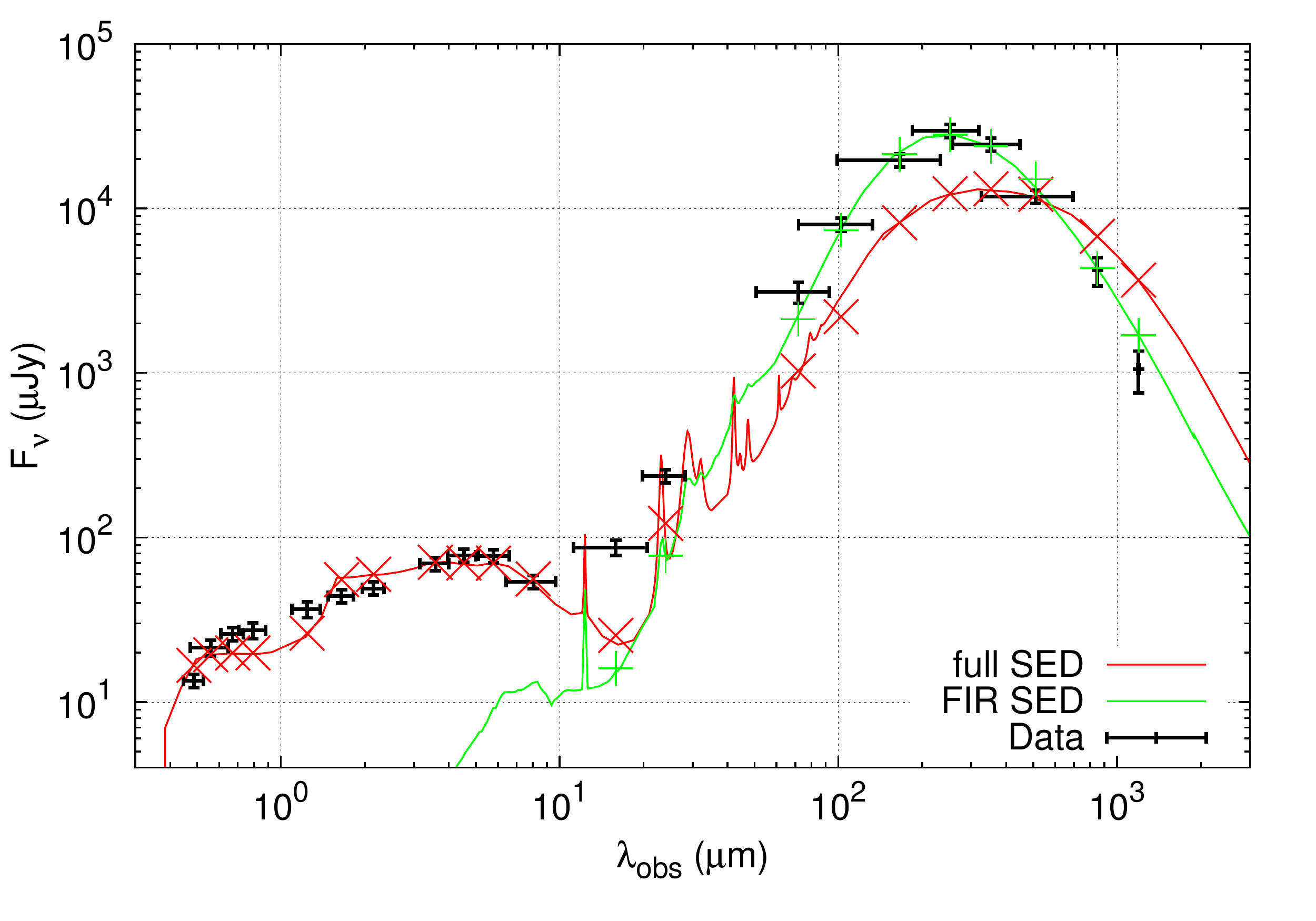}
\includegraphics[width=8.8cm]{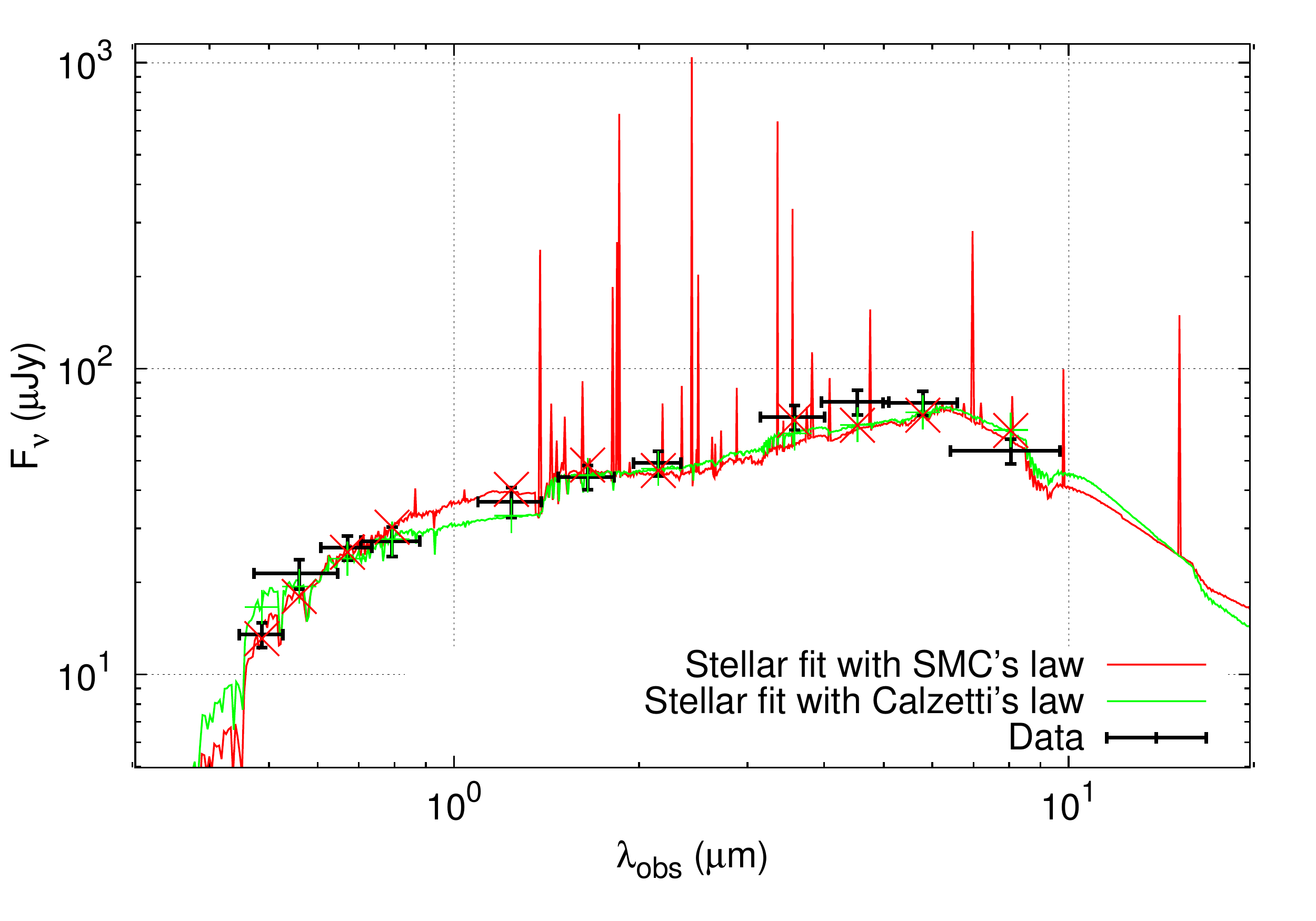}
\caption{{\em Top:} Observed and model SED for cB58. Although not a good fit, the best match to the full, global SED
is found with templates from M10. We can see that the template does not match the photometry in the NUV-blue as it 
does not describe a population as young as is cB58's and misses the IR peak by a factor of $ \sim 2.5 $. The FIR fit is obtained with a
 template among the brightest and warmest from CE01, that matches the IR peak very well.
{\em Bottom:} SED fits to the stellar part of cB58. The red line shows the best-fit for declining SFHs, including nebular emission,
and assuming the SMC extinction law. Green shows the fit neglecting nebular emission, and assuming the Calzetti law (best fit among the 
Calzetti based solutions).
We can see that for the latter the fitted SED falls out of the $ 1\sigma $ uncertainty of the $g$ band photometry as noticed in \protect\citet[][]{2012ApJ...745...86W,2008ApJ...689...59S}. In such a case, UV slopes derived from SED fits can vary
substantially depending on the extinction law used, especially around the 1300-1800 \AA\ interval. 
}
\label{fig_cb58}
\end{figure}
%%-----------FIG cb58 full--------------

\subsubsection{MS1512-cB58} 
\label{s_cb58} 
The global SED of cB58 from the optical to the sub-mm domain is shown in Fig.\ \ref{fig_cb58}.
The best fit was obtained using an SMG template of M10, which still shows some significant 
deviations from the observations: the Balmer break is much stronger in the template, and it 
fails to reproduce the intensity and steepness of the FIR emission, as usually seen with IR-bright galaxies.
The FIR-only best fit is also achieved with one of the brightest templates from CE01, 
 which successfully reproduces the fluxes from 70 \micron\ to
 850 \micron, with only a slight overestimate of the 1.2 mm band. 
 The rather poor fit of the MIR photometry can be explained by the weak PAH features
 of the bright end of the CE01 library.
 This fit, including the new {\it Herschel} observations of cB58,  provides the slightly revised lensing-corrected IR luminosity
 $\lir =3.04 \times 10^{11}$ \lsun\ listed in Table \ref{OBSERVED-GLOBAL}.
 
 Fits to the stellar part of the SED (i.e.\ up to 8 \micron) are generally good, quite independently of the detailed model assumptions
 (SFH and extinction law), and with or without nebular emission.
Two examples, here assuming different extinction laws, are shown in Fig. \ref{fig_cb58}.
The strongest emission lines, \Oii, \Oiii\ and \ha\ lie outside of the observed bands, 
so that they cannot contribute significantly to the photometry. However, the \Siii\ lines lie in the IRAC 3.6\micron\ band, 
where they can slightly contribute to the observed flux.

As seen with \citet{2012ApJ...745...86W} and \cite{2008ApJ...689...59S}, the rest-frame UV photometry is best fitted with the SMC law, 
mainly due to its steepness in the FUV regime, that better accounts for the $ g- V$ and $ V-R $ colors. In this case, the usual power law 
approximation for the UV continuum becomes very ill-defined, unless one would distinguish a (redder) FUV regime and a (bluer) NUV one.
Nonetheless, the mean slope we measure from our SEDs at 2000\AA\, $ \beta =-1.15 \pm 0.1$  is in very good agreement with the photometry
in the $ VRI $ bands and the spectroscopic value of $ -1.28\pm0.14 $ from \cite{2001A&A...372L..37B}.

%-----------------------------
\begin{figure*}[htb]
\centering\includegraphics[width=8.8cm]{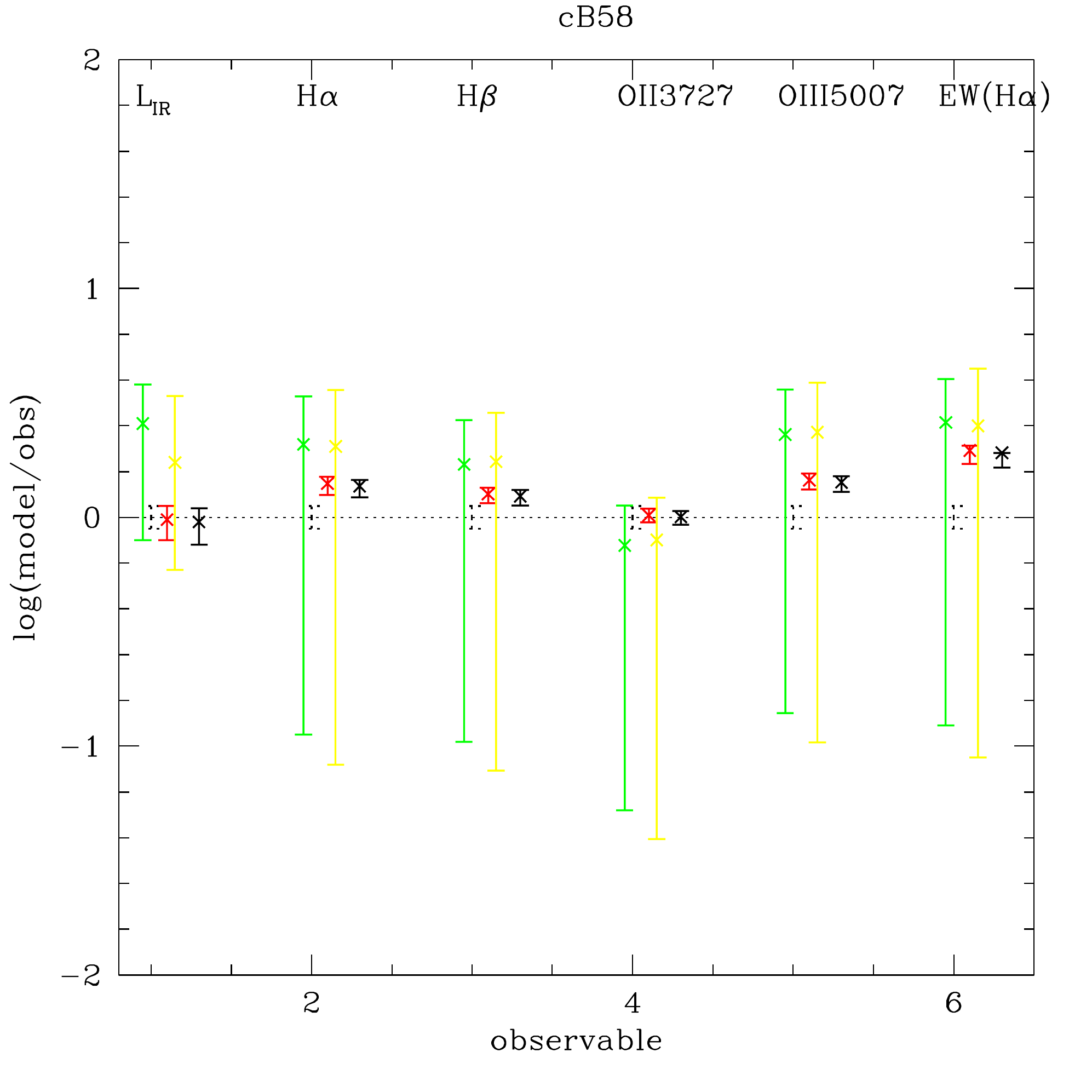}
\centering\includegraphics[width=8.8cm]{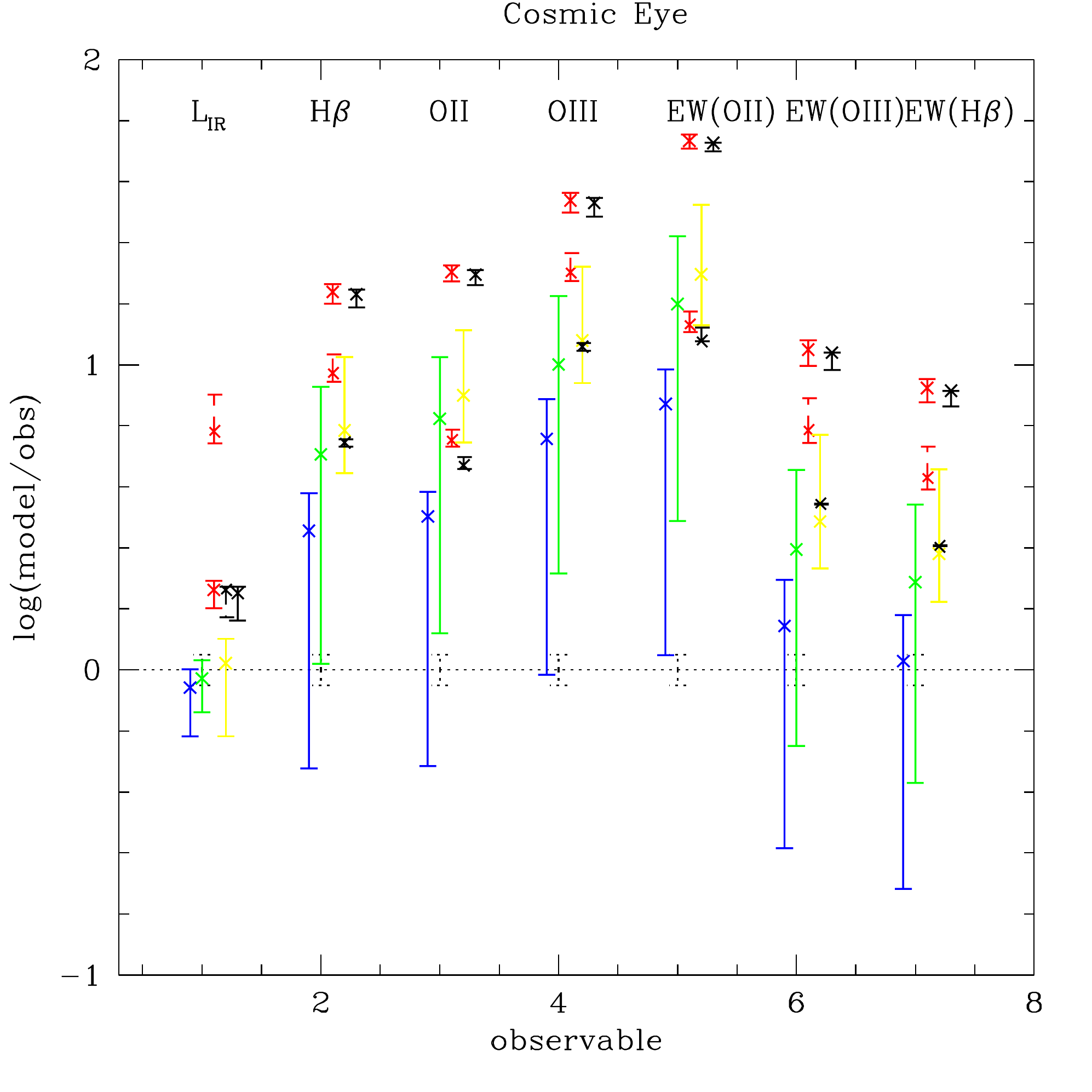}
\caption{{\em Left:} Comparison of additional observables with model predictions for cB58 adopting the Calzetti attenuation law. 
Shown is the logarithmic ratio of the
model prediction with respect to the observable quantity for \lir, the flux in the \ha, \hb, \Oii, and \Oiii\ lines, as well as 
the \ha\ equivalent width. The observed values are  from this paper (\lir), and from
\citet{2000ApJ...533L..65T} respectively. 
Although other emission lines have been detected in this galaxy, we do not include them here because the
information they provide is basically redundant.
Errorbars indicate the 68\% confidence interval predicted by the models;
the typical observational errors, not included here, are shown with the black dotted error bars.
Colors correspond to models with different SFHs (green: exponentially declining, and yellow: delayed SFHs;
both for subsolar metallicity, 1/5 \zsun. Black: SFR=const; red: exponentially rising; both for solar metallicity).
Results for the SMC law are similar and equally reproduce the observables.
{\em Right:} Same as  left panel for the Cosmic Eye adopting the Calzetti attenuation law. 
Sources of the observational data are described in the text.
Blue points show the predictions for models with exponentially declining SFHs and applying 
a higher extinction to nebular emission (i.e.\ E(B-V)$_{\star } =0.44  \times $E(B-V)$_{\rm neb}$).
Red and black dashed symbols show the same for the rising and constant SFR models respectively.
}
\label{fig_res_cb58}
\end{figure*}
%-----------------------------

Since other observables (lines and IR luminosity) have also been measured  for this galaxy, it is interesting to compare 
them with the predictions from our models. Such a comparison is shown in Fig.\ \ref{fig_res_cb58} for the additional key 
observables, not included in the SED fits.
Again, we note that all models are able to reproduce the observed IR luminosity and the main
emission lines within a factor of few (say two, typically). Constant SFR models and rising 
star-formation histories predict slightly too strong emission lines, but applying a different
(stronger) attenuation to the nebular spectrum compared to the continuum 
\citep[as observed in nearby starburst galaxies][]{2000ApJ...533..682C} would reduce this difference.
Note that we have here shown the results allowing for different metallicities, between solar and 1/5 \zsun\
bracketing the range of the observed metallicity of cB58 that is 1/3 \zsun\ \citep{2000ApJ...533L..65T}, 
and adjusted to show the best results for each SFH.
As expected, for star-formation histories allowing for strong variations (declining or delayed SFHs)
the range of predicted line strengths is larger. 
Overall we conclude that the observations of cB58 do not provide a strong test/constraint
for the star-formation histories and attenuation law applicable to this galaxy.

The physical properties derived from a subset of models are summarized in Table \ref{selected_model_table}.
Typically, the median mass is $\mstar \approx 6.5 \times 10^9$ \msun, 
the instantaneous SFR $\approx$ 13--66 \msunyr,
the extinction $A_V \approx$ 0.3--0.9,
with uncertainties of a factor $\sim 2$ approximately.
Ages (defined as time since the onset of star formation) are 30--60 Myr for declining SFHs,
50--90 Myr for delayed SFH,  and older (100--250 Myr) for constant SFR or rising SFHs.
No systematic age shift is found between the two attenuation laws considered.

The physical parameters derived here are very similar to those derived by \citet{2012ApJ...745...86W},
when accounting for the factor 1.7 normalization due to different IMFs used between their and our work
\footnote{The IMF normalization affects stellar mass and SFR; our values need be corrected downwards
by a factor 1.7 to compare with the more realistic Chabrier IMF used by   \citet{2012ApJ...745...86W}.}.
Note that the stellar mass derived by \cite{2008ApJ...689...59S}, a factor $\sim 5$ lower than our value,
is affected by an error in their SED scaling (Siana, 2012, private communication).
Our SFR determination is also in good agreement with the $\sfruvir=52.5 \pm 10$ \msunyr\ 
obtained from the observed UV and IR luminosities (Table \ref{OBSERVED-GLOBAL})\footnote{We produce the quantity \sfruvir\ by summing
the SFRs derived from \luv\ and \lir\ via straightforward applications of the \cite{1998ARA&A..36..189K} respective
calibrations, presenting this way
a value that accounts for both the obscured and unobscured star formation.}. Our values are
in good agreement with \citet{2012ApJ...745...86W}'s SMC based SFR(SED) 
and with \cite{2008ApJ...689...59S}'s LMC based SFR(SED) and their \sfruvir.
From their SED fits using the Calzetti law \citet{2012ApJ...745...86W} find SFR(SED) systematically larger 
than other SFR indicators, 
including SFR(\ha), which they and earlier studies \citep[cf.][]{2012ApJ...754...25R,2009ApJ...698.1273S} consider as an
incompatibility, suggesting a preference for the SMC extinction law.
The fair agreement of most/all of our models with the observed IR luminosity and the \ha\ flux
demonstrates that there is no such inconsistency. 
Apparent differences between the various SFR indicators can naturally be explained by 
simplifying assumptions mostly on age and SF timescale made for these calibrations,
as discussed in depth by \citet{2013A&A...549A...4S}.
Although formally the best fit is found for the SMC attenuation law, the observations of cB58 do not
allow one to draw firm conclusions on the favored extinction law for this galaxy.
The main signal leading to a slight improvement for an extinction law steeper than Calzetti
is the photometry in the $g$ band filter, as already shown in \cite{2008ApJ...689...59S}.

%-----------FIG Eye full--------------
\begin{figure}[tb]
\centering
\includegraphics[width=8.8cm]{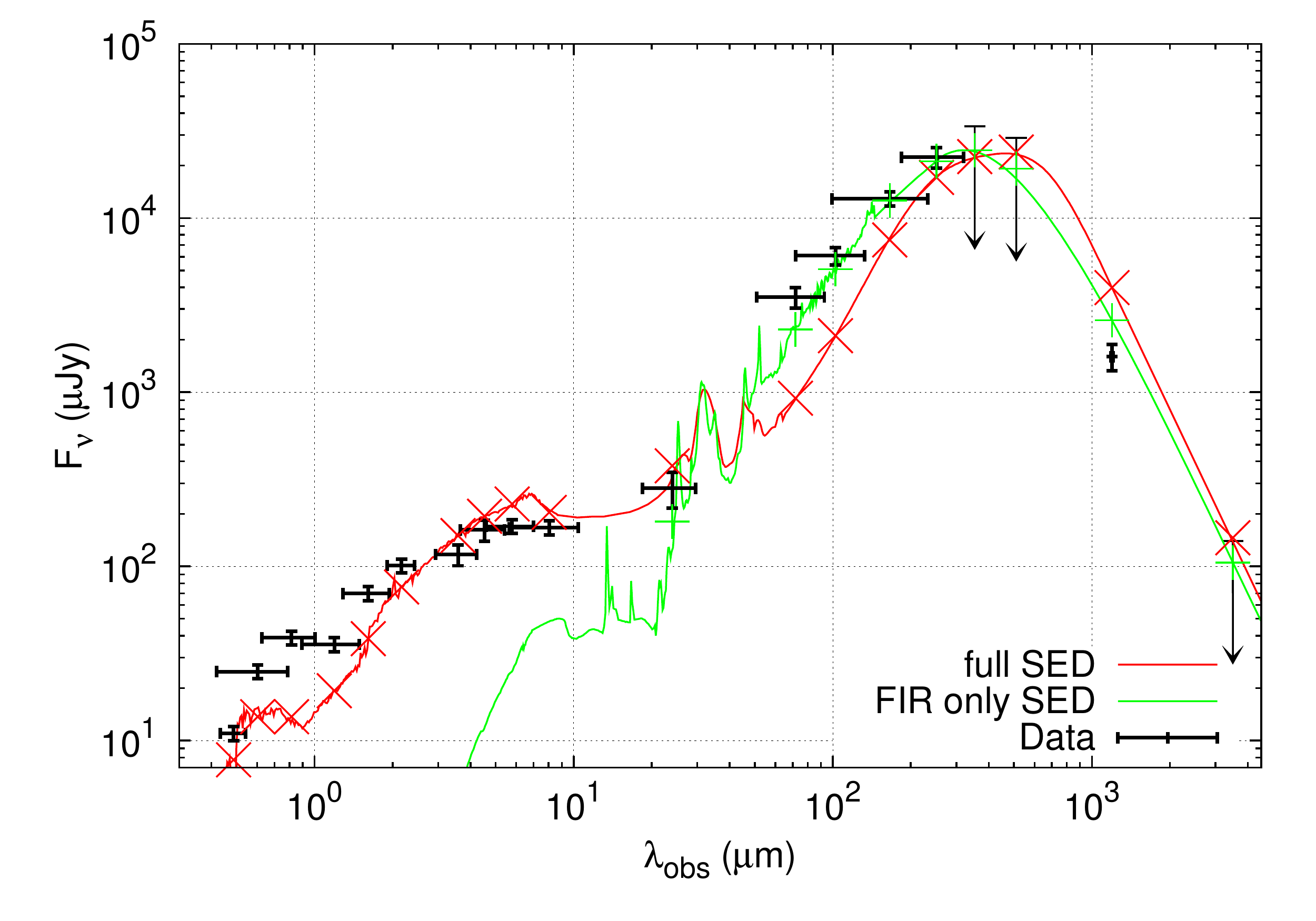}
\includegraphics[width=8.8cm]{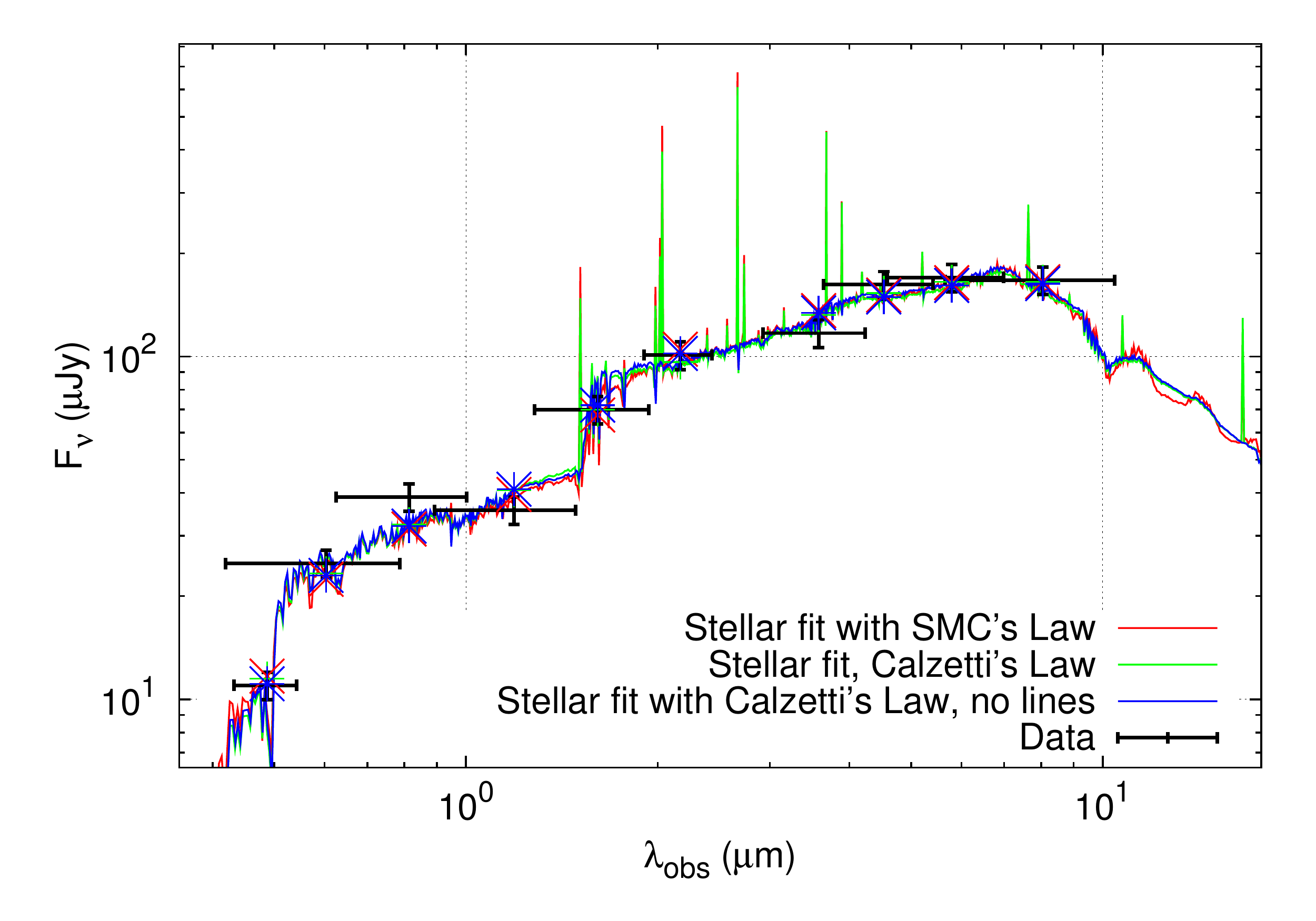}
\caption{{\em Top:} Observed and model SEDs of the Cosmic Eye. 
The global SED is obtained with a Seyfert galaxy template from P07 but is really a poor fit, although it is the best fit
among our libraries (an SMG template from M10 manages a better fit of the visible-NIR photometry, but misses the FIR data by more than one
order of magnitude). The FIR data are best fitted by a R09 template, that accurately matches the photometry and $ 1\sigma $ limits, only slightly
over-estimating the 1.2mm  detection. 
{\em Bottom:} Comparison of the Eye's best Calzetti fits with and without emission lines to the stellar SED.
Also plotted is the very old ($ \sim 1.5 $ Gyr) SMC based model, which despite its large age, has favored an constant SFR and
produces stronger lines than the Calzetti based one. The difference between the three is almost imperceptible. }
\label{fig_eye}
\end{figure}
%-----------FIG Eye full--------------

\subsubsection{The Cosmic Eye}
\label{s_eye}
 The global SED of the Cosmic Eye, from the optical to the sub-mm domain, is shown in Fig.\ \ref{fig_eye}. 
Its full photometry has proven very hard to fit, the best fit was obtained with a Seyfert galaxy template
from P07, but failed to reproduce any fine details. Some fits with SMG templates from M10 matched the stellar photometry well
but could not reproduce at all the dust emission.
Contrary to cB58, its rest-frame visible photometry shows what can be 
interpreted as a Balmer break, which is found in all the stellar model SEDs we explored.
Its steep IR peak is best fitted with templates from R09's library, from which we derive an intrinsic
$\lir = 3.41 \times 10^{11}$ \lsun.

The case of the Cosmic Eye turns out more interesting than cB58 to test different
models. Indeed, in this case models for different SFHs and extinction laws predict
a larger range of emission line strengths and IR luminosities, as shown in Fig.\ \ref{fig_res_cb58}.
In this plot the observed \lir\ is from Table \ref{OBSERVED-GLOBAL} and the emission line measurements
from \cite{2007ApJ...654L..33S,2011MNRAS.413..643R} and from Johan Richard (private communication)\footnote{The emission
line fluxes are listed in \cite{2011MNRAS.413..643R}. The observed equivalent widths are 7.5, 60, and 20 \AA\
for \Oii, \Oiii, and \hb\ respectively, according to \cite{2007ApJ...654L..33S} and as measured from the NIRSPEC
spectrum by Johan Richard.}.
As clearly shown in this figure, the models with constant or rising SFHs predict lines much stronger
than observed. Indeed, the relatively weak observed emission lines indicate that the current SFR must be lower than
the past (or past-averaged) one, which is not the case for SFR=const or rising SFHs.
Although models with a declining star-formation history predict the weakest emission lines (and a relatively
large range due to inherent age uncertainties), our standard models applying the same attenuation
to both stellar and nebular emission (with the Calzetti law) still predict emission lines in excess
of the observations. If we adopt the empirically motivated  relation between stellar and nebular
extinction of E(B-V)$_{\star } =0.44  \times $E(B-V)$_{\rm neb}$ \citep{Calzetti2001}, the 
predicted emission lines are indeed weaker and in reasonable agreement with the observations,
as shown in Fig.\ \ref{fig_res_cb58}. This effect is larger here than for cB58, since the Cosmic Eye
has a higher extinction (as found from the SED fits and also confirmed by the higher IR/UV ratio, cf.\ below).
We note that the models with declining SFHs also perfectly reproduce the observed IR luminosity
when the Calzetti law is adopted.

We have also computed the SED fits for different SFHs adopting the SMC law.
For all cases the model over-predicts the emission lines shown in Fig.\ \ref{fig_res_cb58}
by a factor 10--20,
and the IR luminosity is under-predicted by a factor 2--3 with
a 68\% confidence interval of  $\approx \pm 0.1$ dex. This is also the case for 
declining SF histories, which -- for the SMC law -- choose solutions with long star-formation
timescales and old ages, giving results close to models of constant SFR.
In short, the {\em a posteriori} comparison of the IR and emission lines of the Cosmic Eye
show that declining SFHs and the Calzetti law reproduce well these observables, whereas
constant or rising SFR and the SMC law fail to do so.
The physical parameters obtained for the declining SFH model including nebular lines
are listed in Table \ref{selected_model_table}.

The current SFR(SED) $\approx 60$ \msunyr (from the preferred solution, involving the Calzetti law, a declining SFH and nebular emission),
is close to the \sfrir = 70 \msunyr\ 
obtained from the observed IR luminosity (Table \ref{OBSERVED-GLOBAL}) using the standard \citet{1998ARA&A..36..189K} relation.
Our \sfrir\ estimate is a factor $\sim 2$ lower than the SFR(IR+UV)=$140 \pm 80$ \msunyr\
derived by \citet{2009ApJ...698.1273S} using the Kennicutt relation. This is mostly due to a slight downward
revision of the IR luminosity from the {\it Herschel} data, although both measurements agree within
the errors. \cite{2007ApJ...665..936C} had predicted SFR $\sim$ 60 \msunyr\ from extrapolation of the 24 \micron\ flux, and 
were accurate within their margins of error.  

We find \sfruvir\ to be $ \approx 102 \pm 10 $ \msunyr\, which is in perfect agreement with the 
SFR$_{\rm UV-corrected}\sim100$ \msunyr\ infered by \cite{2007ApJ...654L..33S}.
Our physical parameters (extinction, stellar mass, instantaneous SFR) are also in reasonable agreement with those derived 
by \citet{2011MNRAS.413..643R} from SED fits.

We note that the stellar mass derived by \citet{2007ApJ...665..936C}, lower than ours by factor $\sim 6$,
appears clearly too low. Their mass was derived using the K-band mass/light ratio predicted by the
{\sc Starburst99} models \citep{leitherer99} for a young population of 10--30 Myr. This assumption of
a very young age may explain part of the difference; another factor may be related to a different
IMF, although \citet{2007ApJ...665..936C} do not specify their assumptions. Indeed,  the default
IMF used by {\sc Starburst99} covers only the range of 1 to 100 \msun, leading thus to an
underestimate of the stellar mass by a factor 2.56. In any case, the physical properties of our
entire sample was determined in a consistent manner. Of course, one must remember that
due to our choice of a Salpeter IMF from 0.1 to 100 \msun, our masses and SFR values are too high
by a factor 1.7 compared to the probably more realistic Chabrier IMF. 

Later we discuss the position of the Cosmic Eye and all our objects in 
the well-known IRX--$\beta$ plot.

\subsection{Energy conserving models}
\label{s_nrg_conserv}

Now we discuss the results we obtain with our energy conserving models. As presented in Sect. \ref{sed_fits}, we narrowed down the 
free parameters of our SED fitting by fixing the extinction using the knowledge of the the total IR luminosity, and assuming it is only 
due to the obscuration of the SED between 0.912 and 3 \micron. We convert the observed $\lir/\luv$ ratio into \av\ using the calibration presented
in \cite{2013A&A...549A...4S}. 
These energy conserving models
should thus provide the most accurate physical parameters. 

The results are shown in Table \ref{energy_consist_table}. We can see that the strong age-extinction degeneracies that appeared before
in many cases are greatly reduced, as are uncertainties on the physical parameters we derive. 
We have checked the energy conservation for these models which is in general 
verified within 10\% of the observed \lir.
 Another property that is better constrained here is the e-folding timescale of the decreasing
models, $ \tau $.  Our sample shows a strong tendency to prefer the smallest $ \tau $ among the ones we tested (from 50 to 100 Myr), with the
exception of cB58 and A68/C0 which prefer long timescales/constant SFR (the latter being less constrained than the others hence the persisting 
degeneracy).
In particular, in the case of the previously highly degenerated solution for A68/h7, we find a suiting solution 
indicating a post-starburst regime, with rapidly decaying SFR. This is in agreement with   the strong IR emission and the hypothesis
that it's the result of a recent merger. The Cosmic Eye also is seen to be in a post-starburst phase (defined by $ t/\tau>1 $), 
whereas cB58 seems to be starbursting ($ t/\tau<1 $). 

A short discussion on the case of MACS0451 may be useful here to clarify what $\lir/\luv$ ratio was used. As stated in Sect. \ref{s_sample}, 
the IR emission of the southern part of the arc is dominated by an AGN, whereas the northern part seems to be a clean starburst (Zamojski et al.,
 \textit{in prep.}). In order to consider the global $\lir/\luv$ in a manner coherent to the approach of studying the integrated properties of
our sources and take into account the presence of the AGN, we have set as total \lir\ of the arc the sum of the northern \lir\ and 10\%
of the southern one. We then compare this total \lir\ to the total \luv\ to derive the \av\ used for our model. Approximating such an 
elongated arc with a global value for the extinction may seem coarse, but the stellar photometry we have is indeed very constant
all along the arc. Making the exercise of estimating two values for the extinction of its two parts (while still correcting the south for the AGN 
contribution) and summing the physical parameters derived in the end, yields sensibly the same mass and SFR as the ones shown
in Table \ref{energy_consist_table}, and are in agreement with the observationally derived values of SFR discussed in Sect. \ref{s_indics}.

%------------------------------	 
\begin{table*}[htb]
	\centering
	\begin{tabular}{l c c c c c c}
		\hline\hline
			\textbf{ID} &	z & \av\	& Age [Gyr]	&  t/$ \tau $	& $ \mstar\ [10^{10}\msun] $ & \sfrbc[\msunyr]  \\ 
			\hline
	A68/C0		 & 1.5864 & 1.1	& 1.01(0.25-1.7)  &  2.03 (0.67-3.64)  &   3.4 (2.3-4.5)     &	15.5 (8.8-21.2)\\ 
	A68/h7		 &  2.15 & 1.26 	& 0.25 (0.18-0.25)  & 3.63 (3.61-3.63)	 & 26.1 (19.6-27.7) & 129.2 (123.5-134.9)\\   
	A68/HLS115 	 & 1.5869	& 1.58 	& 0.13 (0.09-0.13)  &  1.83 (1.81-2.5)    &  1.37 (1.03-1.65)   &	42.45 (32.3-46.9)\\
	A68/nn4$\ast$  & 3.19 & 2.17	& 0.033 (0.033-0.036)  &  0.66 (0.66-0.72)  &	5.6 (5.2-6.0)   & 1243 (1176-1314)\\  
	MACS0451 	 &  2.013 & 0.63 &  0.13 (0.13-0.18)  & 1.28 (0.6-1.28)  & 0.49 (0.49-0.52)  & 22.3 (22.1-24.9)	\\ 
	cB58 	 	 &	2.73	& 0.7 	& 0.13 (0.13-0.13)  & 0.13 (0-0.43)  &  0.75 (0.71-0.81)&	63.2 (58.3-67.3)\\ 
 	Cosmic Eye	 & 3.07 & 0.58 	& 0.18 (0.18-0.18)  &   2.58 (2.58-2.58)    & 4.0 (3.9-4.1)&	56.1 (55.2-57.2)\\  

\hline   
	\end{tabular}
	\caption{  Physical parameters derived from the energy conserving models,
				where \av\ is fixed, obtained from the observed ratio of  \lir\ over \luv\ as discussed in  
				\citet{2013A&A...549A...4S}. All models shown here use Calzetti’s extinction law and 
				nebular emission, except for A68/nn4 whose extreme attenuation makes it lie outside the  \lir/\luv\  range where the  
				\citet{2013A&A...549A...4S} relation was calibrated. Applying SMC’s law instead works very well for this source.
				t/$ \tau $ is the ratio of the age of the population over the characteristic e-folding timescale. Zeros indicate an 
				``infinite” timescale, meaning that the preferred solution in such case is the constant star-forming rate. 
		     	 }              \label{energy_consist_table}
	\end{table*}
%------------------------------

%%%%%%%%%%%%%%%%%%%%%%%%%%%%%%%%%%%%%%%%%%%%%%%%%%%%%%%%%%%%%%%%%%%%%%%%%%%%%%%%%
\section{Discussion}
\label{s_discuss}

\subsection{Constraints on the SFH and extinction law}	
\label{ir_lum}

An important aspect of the present work is the comparison of the predicted IR emission from the stellar 
population synthesis (see Sect.\ \ref{sed_fits} ) with the observed IR luminosity (this section doesn't consider
the previously described energy conserving models as they are conceived to reproduce the observed \lir). 
In Fig. \ref{Lir_plot} we can see the overall results in comparison with the actual observed values.

The general result concerning the extinction laws tested is that the models based on the SMC law predict much less IR luminosity that the
ones using Calzetti.
Globally, SMC-based models predict too low IR luminosities \lir\ as compared to the observations (up to $ \sim 0.5 $ dex). The Calzetti based
models make accurate predictions at the 1$ \sigma $ confidence level. The SMC models, in 4 out of 7 galaxies, fail to predict the  observed
luminosities at the 99\% confidence level.

The main difference between the these two interpretations is explained by the fact they seem to ``prefer'' different timescales for the stellar
models they produce (see Table \ref{selected_model_table} and individual object sections), with the SMC-based models yielding systematically
older populations that the Calzetti-based ones, and this directly results in smaller instantaneous SFRs, less extinction to fit the
rest frame UV-optical slope, and hence a smaller output in the IR. Only in the case of cB58 and for one model for A68/nn4 do
the predicted \lfir\ match well the observed luminosities with the SMC extinction law. They are also the only objects for which the SMC models
produce a young population ($< $ 100 Myr). For these cases, Calzetti-based models without nebular emission produce extremely young ages  
($ \sim 10  $ Myr) and along with this a very high SFR, reminiscent of cases discussed by 
\citet{2012ApJ...754...25R,reddy2012_herschel}.

In most cases the impact of nebular emission  slightly
lowers the predicted \lir, or leaves it unchanged, and tends to reduce the strong age-extinction degeneracy. This has 
a positive effect for the Calzetti based models as it produces better predictions for all cases (except for the very peculiar MACS0451 N,
for which the observed FIR properties are discussed in detail in Zamojski et al. (2013, \emph{in prep.})). The inclusion of nebular lines
also creates a stellar population model for nn4 and cB58 that successfully lowers the predicted \lir\ to match the observations (by 
preferring a slightly less extreme age). 

As discussed in \cite{2013A&A...549A...4S}, when using rising SFHs, our models usually produce at least as much or more predicted \lir\ in
comparison with a constant or declining SFR; by definition, rising SFHs are always at their peak of star formation, meaning there's a 
maximum of young blue stars at any given age $ t $. This usually implies more extinction than in the other cases to produce a correct 
fit of the photometry. We observe the same phenomenon with our present sample. For the Calzetti based models we can safely exclude the 
rising SFH scenario for our objects, as it overpredicts the observed \lir. This effect is not as strong with the SMC-based models, who
are still underpredicting the luminosities, with the exception of A68/C0 that finds just the right amount (Sect. \ref{s_c0}).

To sum up, the exercise of using the observed \lir\ as an \textit{a posteriori} consistency check for our models
on our present sample, shows us that the Calzetti based,
exponentially declining SFH models are in best agreement with the observations. 
The SMC-based solution reproduces the observed \lir\ only when the fits yield young ages. For the two galaxies where the SMC law matches
the observed \lir\ and models with the Calzetti attenuation law would overpredict it, we find that including the effect of nebular lines reduces
the age-extinction degeneracy, leaving thus both the SMC and Calzetti laws as similarly good explanations.
%

%-----------FIG IRX-beta--------------
\begin{figure}[tb] 
	%\centering
	%\includegraphics[width=14cm]{irxbeta_light2.pdf} 
	\hspace{-0.8 in}
	\includegraphics[width=13cm]{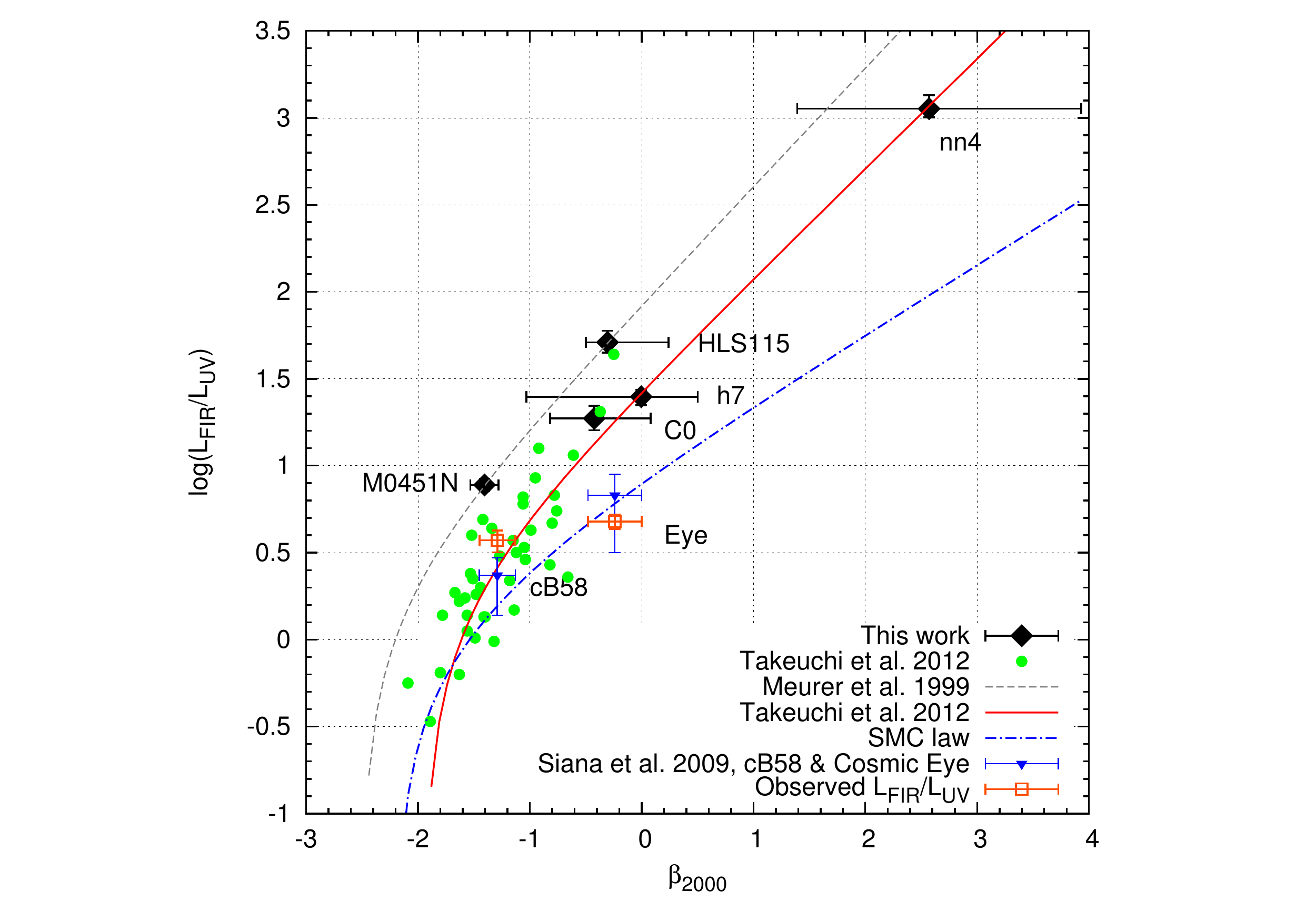}
	\caption{Observed \lir/\luv\ ratio as a function of the UV slope  $\beta$ (at 2000 \AA\ rest-frame)
         for our objects (black symbols), and for the local starbursts of  \protect\cite{1999ApJ...521...64M} (M99), revised
         by \citet{2012ApJ...755..144T}
         in green (the corrected IRX-$\beta$ curve is shown with the solid red line, and the M99 relation
         is shown for reference in the gray dashed line).
        All our values for $\beta$ are the mean from the Calzetti and the SMC based models.
        For the Cosmic Eye and cB58 we plot the spectroscopically measured slopes available in the literature. 
       The expected relation  for constant SFR and age $\protect\ga 100$ Myr for the SMC (blue dashed-dotted) attenuation/extinction law is 
       also plotted. High-z ULIRGs \citep[e.g.][]{2013A&A...554L...3O} tend to populate the region above the M99 relation.
        Discussion in text.}
                \label{UVslope}
\end{figure} 
%-----------FIG IRX-beta-------------- 

\subsection{IRX--$\beta$ plot}
In Fig.\ \ref{UVslope} we show the observed IR/UV luminosity ratio as a function of the UV slope, the so-called
IRX--$\beta$ plot, for the objects of our study. For comparison we also plot the sample of nearby starbursts
from \cite{1999ApJ...521...64M}, respectively the updated version of this paper by
\cite{2012ApJ...755..144T}, and the relations expected for stellar populations with constant SFR and age $\ga 100$ Myr
both for the Calzetti (approximated here by Meurer's curve, with which it closely coincides) and the SMC attenuation/extinction law.
The majority of our objects lie close to the  ``Meurer'' relation, defined by the local starbursts. 
The reddest galaxy, A68/nn4, is at a somewhat intermediate location between the constant SFR sequences
corresponding to the two extinction laws.

The case of the Cosmic Eye, found below the Meurer relation, is worth discussing it separately.
Indeed, previous studies have argued that this deviation indicates that the 
SMC extinction law should be more appropriate for the Cosmic Eye \citep[cf.][]{2009ApJ...698.1273S,2012ApJ...745...86W}.
However, we have just shown above that the observed IR luminosity and emission lines cannot be understood with
the SMC law. How can this be reconciled? The basic argument invoked to 
argue for an SMC law based on the IRX-$\beta$ plot rests on the assumption of the constant SFR
over typically 100 Myr, which determines an intrinsic UV slope and the UV output per unit SFR.
Assuming then a specific attenuation law leads to a simple relation between IRX and $\beta$ 
\citep[cf.][]{1999ApJ...521...64M,2010MNRAS.409L...1B,2012A&A...545A.141B}.
However, other parameters such as the star formation history
and age can affect the relation between these quantities, as e.g. shown by \citet{2004MNRAS.349..769K}.
For example, IR/UV ratios below the observed starburst sequence can be obtained for galaxies
with a low present to past-averaged SFR.
Our above results, showing that the observed emission line strengths of the Cosmic Eye 
can only be understood with such a star-formation history, agree perfectly with this conclusion,
hence demonstrating that the IRX-$\beta$ relation does not imply the SMC extinction is favored
for this galaxy.

Concerning the case of cB58, we can see that although its rest-frame UV is better fitted with a steeper SMC-like
law, it lies in a region where the SMC curve and the updated M99 of \cite{2012ApJ...755..144T} are still too close 
together to allow a proper distinction.

In passing we also note that the determination of the UV slope, especially from photometry, can be
quite uncertain, both due to systematics (e.g.\ the precise sampling of the UV spectrum by the filters,
deviations of the spectrum from a pure power-law etc.) and to random errors, making the 
uncertainty on individual $\beta$ slopes fairly large. For illustration see Fig.\ \ref{fig_eye},
showing that good SED fits can yield UV slopes (here $\beta_{2000}=-1.41 \pm 0.2$) differing 
quite strongly from the various published values of the UV slope of the Cosmic Eye that are usually
found between -0.45 to $ \sim 0 $ \footnote{\citet{2009ApJ...698.1273S} determine $\beta =-0.45$ and $\sim 0$ from photometry and 
spectroscopy respectively, and they use the mean of these two values in their analysis. We adopt this value in Fig.\ \ref{UVslope}.}. 
Indeed, an extensive look into the UV-spectrum
publications \citep{2007ApJ...654L..33S,2010MNRAS.402.1467Q} indicates that the slope in the range [1300\AA,1800\AA] should 
be -0.45$ \pm 0.05 $, in agreement with the photometric slope between the V and I bands. \cite{2007ApJ...654L..33S} infer a
value of $ -1.6\pm0.1 $ which is incompatible with their spectrum, and \cite{2010MNRAS.402.1467Q} do not publish
a value. Depending on where precisely the spectrum was sampled the slope can be vary from $ \beta\approx -0.8 $ 
up to $ \beta\approx 0 $. Our SED inferred slope centered at 1500\AA\ is in agreement with this bluest value, and can be accounted
for as the fit's slight deviation from the photometry at these wavelengths.

Given these uncertainties
and the necessary underlying assumption on star formation history, we suggest that the 
IRX--$\beta$ plot should not be overinterpreted, e.g. to distinguish different extinction/attenuation laws.

\subsection{SFR indicators}
\label{s_indics}

%-----------FIG SFR_comparisons--------------
\begin{figure}[*tb]
	\vspace*{-0.3 in}
	\hspace*{-.2 in}
	\includegraphics[width=11cm]{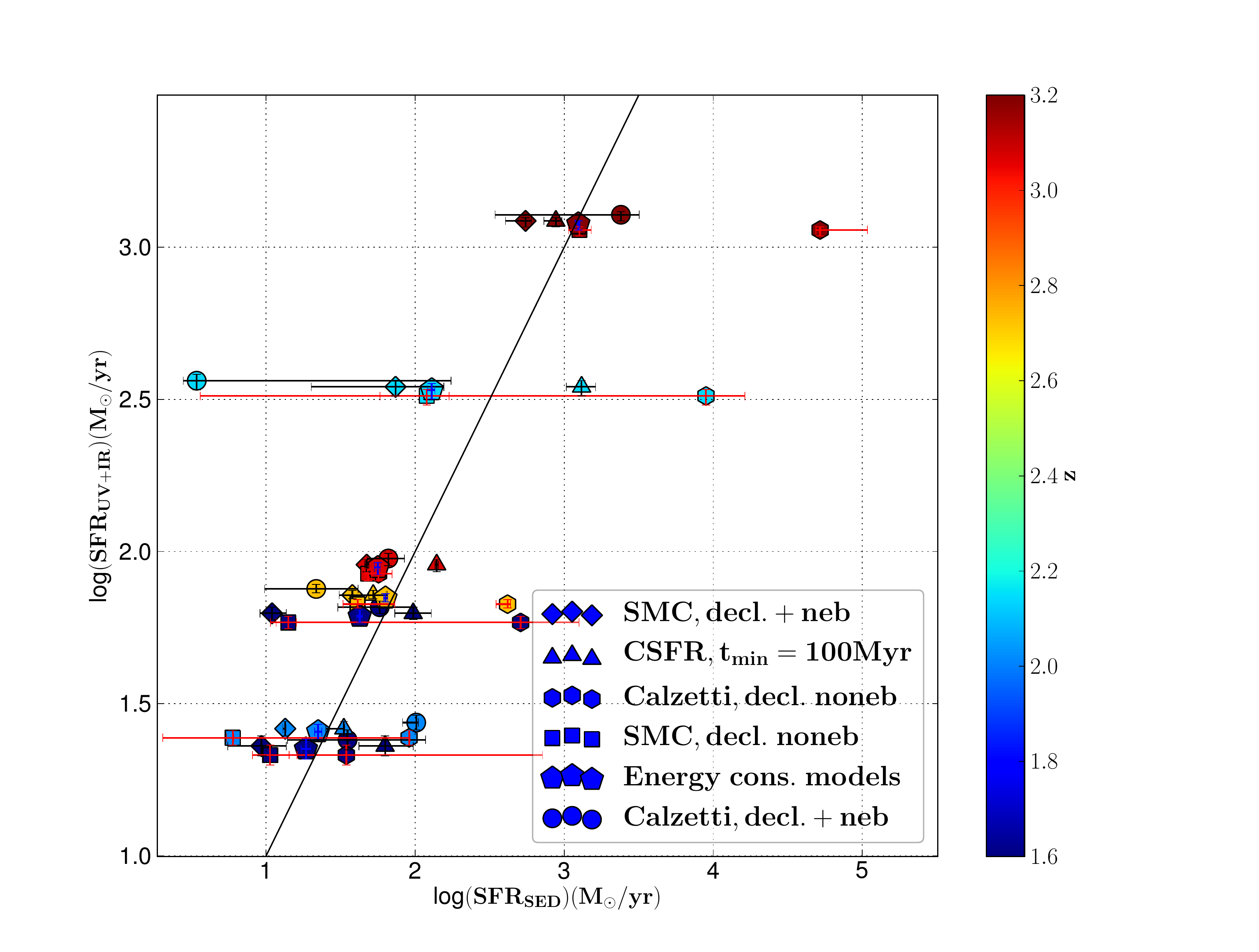} 
	\caption{
         \sfruvir\ vs SFR$ _{\rm SED} $ (\sfrbc\ in the rest of the text) diagram.
         The straight line indicates unity. The various symbols correspond to the models shown in the key, and colors 
         correspond to redshift according to the colorbar. For more clarity the errorbars of the models without nebular emission are in red
         and slight vertical shifts are imposed for all objects to better separate them.
         }
                                                                \label{sfr_compar}
\end{figure} 
%-----------FIG SFR_comparisons--------------

In Fig. \ref{sfr_compar} we compare our instantaneous SFRs that our stellar models produce with observation based total SFR, which we 
take as \sfruvir, with the two terms estimated separately via the \cite{1998ARA&A..36..189K} conversions. 
Overall we have a good agreement in our sample, with only the strongly degenerated case of h7 which presents the largest 
spread. Our preferred model (Calzetti, declining SFH, nebular emission) reproduces \sfruvir\ within 0.3 dex, and produces much 
more reasonable values in that regard than the same models without nebular emission that over-predict it of up to $\sim 1 $ dex.

Despite our use of exponentially declining SFHs with variable timescales from $ \tau =  $ 0.05 to 3 Gyr
we do not find any particular under-estimation of \sfruvir\ by the SED-derived SFRs, in contrast to the findings of 
\cite{2011ApJ...738..106W}.

For the energy conserving models (shown as pentagons in Fig. \ref{sfr_compar}),
we obtain the best correspondence between \sfrbc\ and \sfruvir.
The good agreement is due to the fact that for most sources
our fits yield no very large values of $t/\tau  $ and no extremely young
ages, cases in which the assumptions made for the standard SFR
calibrations may break down. The largest difference is found for
A68/h7, which appears in a declining phase, with the largest
value of $t/\tau\sim3.6$. 

In any case the relevant quantities which should be compared are the observed luminosities (IR+UV),
not the SFR values derived from those using calibrations assuming a fixed SF timescale (constant SFR
for this matter). Differences between \sfrbc\  and \sfruvir\  may naturally be found for 
models with variable SFHs, as discussed in detail in \cite{2013A&A...549A...4S}.

\subsection{Mass--SFR relation and specific SFR of our sample}
\label{s_stellarpops}

\begin{figure}[tb]
\centering
	\includegraphics[width=8.8cm]{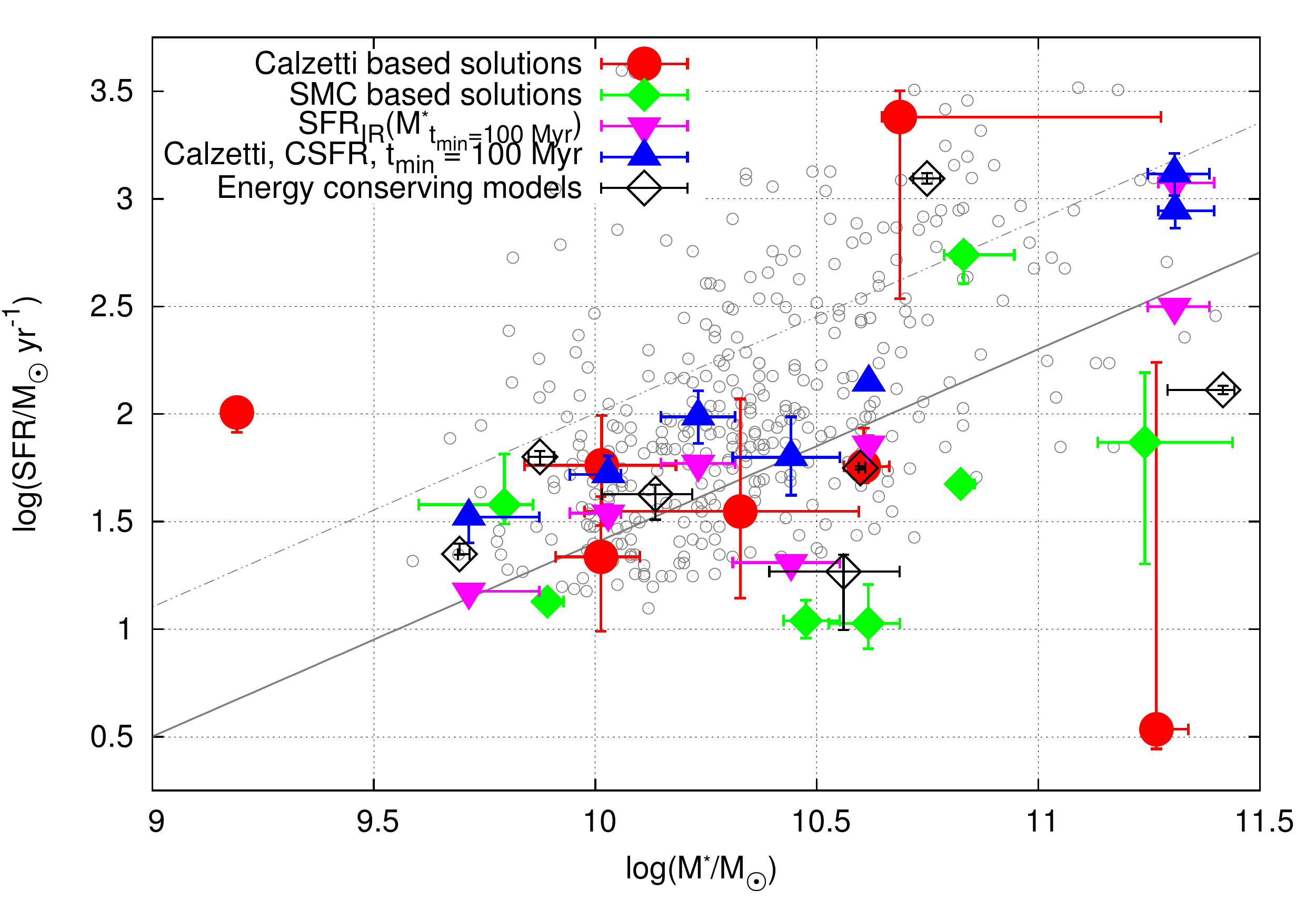} 
	\caption{
        Mass-SFR diagram, with the values obtained by the various stellar population models plus the IR-inferred SFRs (red circles: 
        Calzetti based exp. declining models with nebular emission, green diamonds: same but with SMC; upwards blue triangles:
        the "classic" models assuming Calzetti, constant SFR and $ t_{min}=100 $Myr; downwards purple triangles: IR-inferred SFRs versus
        mass from SFR=const model; black hollow diamonds: energy conserving models). Also 
        plotted are \protect\citet{2007ApJ...670..156D}'s main sequence for z $ \sim2 $ from the GOODS field (grey continuous line)
        , and the relation 4 times above this main sequence as shown in \protect\citet{2011ApJ...739L..40R}
        (grey dot-dot-dashed line). 
        } 
                                                                \label{mass_sfr}
\end{figure}

In Fig. \ref{mass_sfr} we show the SFR as a function of the stellar mass of the seven lensed galaxies
studied here, including values obtained from different models.
Also plotted is the sample from \citet{2007ApJ...670..156D} for comparison, and the mean SFR--mass relation
--- the so-called SF main-sequence --- obtained for star-forming galaxies at $z \sim 2$.
Most of the objects lie close to (i.e.\ within a factor $\sim 2$) the ``main sequence", and for most model assumptions.
We can see that the ``classic" scenario, i.e.\ constant SFR, gives the smallest dispersion as seen also
in \citet{2013A&A...549A...4S}, but most other solutions lie still within the ``tight" main sequence \cite[cf.][defined as 4 times above 
the main sequence]{2011ApJ...739L..40R}. 
Possible outliers are MACS0451N and nn4 (h7) whose specific SFR,
sSFR=SFR/\mstar, exceeds (falls below) the median sSFR $\approx 2$ Gyr$^{-1}$.
A68/nn4 being an extreme starburst, it is located unsurprisingly in the starburst regime \cite[cf.][10 times above 
the main sequence]{2011ApJ...739L..40R}. Globally the position of our objects for varying SFHs on the \mstar -- SFR
diagram can be understood in terms of the median age over e-folding timescale ratio, $t/\tau$. Galaxies with a median
$t/\tau \sim 1 - 2$ in our sample lie very close to the MS, whereas A68/h7 with $t/\tau \sim 7$ lies far below, and  the ones in the 
starburst regime with  $t/\tau < 1$. A68/h7's median solution obviously must not be representative of the current star-formation in that 
galaxy as we've seen also (Fig. \ref{Lir_plot}) that it also under-predicts the observed \lir\ which imposes for an 
upward correction to its SFR and would bring it close to the MS.

However, the extreme sSFR values of these 2 objects are only obtained for models assuming variable, declining SF histories.
Whether the true sSFR values are as high/low, could be tested with accurate, reddening corrected
SFR(\ha) measurements. In any case we note that for none of our objects do we find specific star formation
rates as high as obtained for $z \ga 4$ LBGs in our work examining the role of nebular emission
and variable SFHs for these galaxies \citep{dBSS12,2013A&A...549A...4S}, although the same model assumptions were made.

As expected from the discussion of the individual objects, models with the SMC extinction law
yield somewhat higher masses and a lower SFR, due to the preference of fits with older populations.
They are, however, less favored, given the inconsistency with the observed IR luminosity for the majority
of them.

%== 
The energy conserving models displace the solutions a little on the \mstar-SFR space, relatively to the unconstrained ones,
more so for the more extreme cases like MACS0451 and A68/h7 that become less extreme. 
The galaxies are still found in coherent positions in respect with their
starbursting state (the $t/\tau$ ratio), with cB58 and A68/nn4 as starbursts, while  A68/h7 and the Cosmic Eye (A68/C0) as post-starbursts 
(quiescently star-forming).
%==

Compared to {\it Herschel}-detected galaxies from blank field studies including the deep GOODS-South data, which are 
restricted to SFR $\ga 100$ \msunyr\ and to stellar masses $\mstar \ga 10^{10}$ \msun\ \citep[cf.][]{2011ApJ...739L..40R}, 
our sample includes IR-detected objects with lower SFR (up to $ \sim 1 $ dex) and also somewhat lower masses.

%-----------FIG MASS_SFR--------------
%-----------FIG MASS_SFR--------------

For most of the galaxies the sSFR is in the range of $ \sim1 $ to 10 Gyr$^{-1}$ with
a median value very similar to the weighted mean specific SFR of 2.4 Gyr$^{-1}$ derived 
for masses $\mstar \sim 10^{8.5}-10^{11}$ \msun\ by \cite{2012ApJ...754...25R} for $z \sim$ 1.5--2.6
galaxies combining individual IR detections and stacking results. Our sSFR values are also comparable
to other determinations, e.g.\ for $z \sim$2--3 LBGs by \cite{2006ApJ...647..128E} using SFR(\ha).
As already mentioned above, the two cases of MACS0451 and A68/h7 with extreme sSFR values
for some model assumptions should be taken with caution.

%-----------FIG IRX-UV--------------
\begin{figure}[Htb] 
	\centering
	\includegraphics[width=8.8cm]{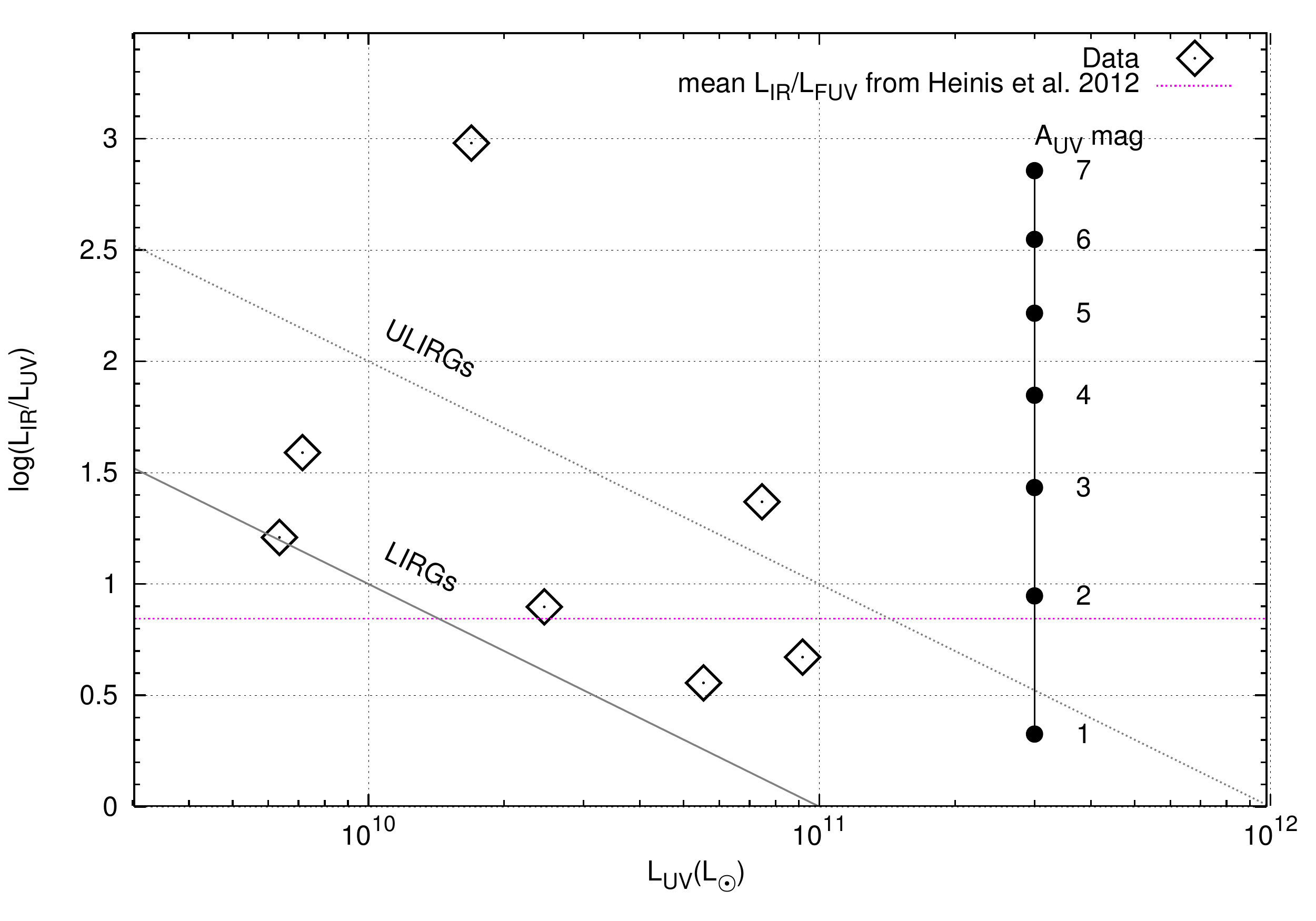} 
	\caption{
	Observed $\lir/\luv$ luminosity ratio versus \luv\ for our sample.
         Solid and dashed gray lines show the limits for LIRGs and ULIRGs 
         respectively, and the pink dotted line the mean relation derived from stacking for z $\sim$1.5 UV-selected galaxies by
          \protect\citet{2013MNRAS.429.1113H}. On the embedded vertical axis we have the corresponding UV attenuation from 
          \protect\citet{2013A&A...549A...4S}. The individual objects are ordered from left to right as following: C0, HLS115, nn4,
          MACS0451, cB58, h7 and the eye.
         } 
                                                                       \label{lirluv}
\end{figure} 
%-----------FIG IRX-UV-------------- 

%-----------FIG A_FUV-MASS--------------
\begin{figure}[htb]
	\centering
	\includegraphics[width=8.8cm]{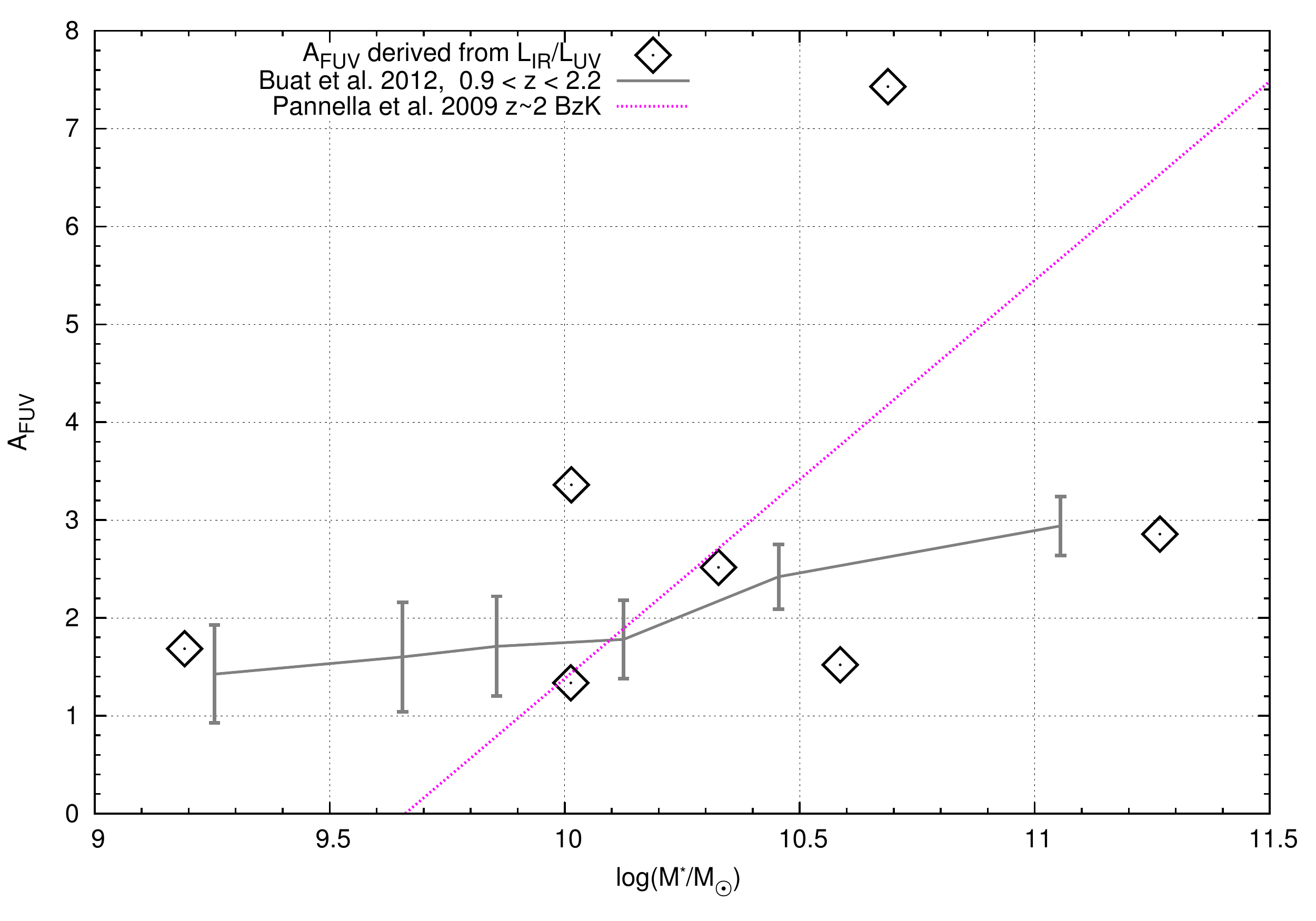} 
	\caption{
           UV extinction versus stellar mass diagram. Plotted with the average relation and standard deviation obtained with the 
           GOODS-\textit{Herschel} sample by \protect\citet{2012A&A...545A.141B} in the redshift range $ 0.9 < z < 2.2$, after
           correcting it with a factor of 1.8 to account for the difference  in the choice of IMFs. Also plotted is the relation from 
           \citet{2009ApJ...698L.116P} for a sample of radio observed BzKs.
           The values shown in black diamonds are derived from the observed IR-to-UV ratio via the correspondence from
           \protect\citet{2013A&A...549A...4S}.}
                                                                \label{mass_afuv}
\end{figure} 
%-----------FIG A_FUV-MASS--------------

\subsection{Dust extinction as a function of stellar mass and UV luminosity}	
\label{s_av} 

In Fig. \ref{lirluv} we show the observed ratio of \lir/\luv\ of our sample, a good measure of the 
UV attenuation, as a function of the UV luminosity. On the embedded vertical axis we show
the corresponding UV extinction, $A_{\rm FUV}$, following the parametrisation
given by \cite{2013A&A...549A...4S}. Note that the UV attenuation derived in this way
is independent of the extinction law, as it depends on the energy redistribution
between the UV and IR. 
Our objects span a range of $\sim 1$ dex in \lir/\luv, except for A68/nn4, which stands out
by its high IR/UV ratio of \lir/\luv $\approx 10^3$. The corresponding UV attenuation
is between $A_{\rm FUV} \sim$ 1--4 (or $A_V \sim$ 0.5--1.6 for the Calzetti law).
As shown in this figure, our individual IR/UV measurements are in good agreement with
those derived from stacking results as a function of UV magnitude by for $z \sim 1.5$ UV selected
galaxies by \citet{2013MNRAS.429.1113H}.
Our small sample and the complicated selection function does not allow us to draw
any conclusions on a possible trend with \luv.

In recent years it has become clear that galaxies show also a trend of increasing extinction
with stellar mass. We therefore show our results at $z \sim$ 1.6--3 in Fig.\ \ref{mass_afuv},
which are compared to recent results from radio and UV-stacking $z \sim 2$ BzK galaxies
by \cite{Pannella2009}, and to the median value of $A_{\rm FUV}$ as a function of stellar mass
derived with {\it Herschel}/PACS data from UV selected galaxies \citep{2012A&A...545A.141B}.
As before, A68/nn4 stands out by its very high extinction.
Besides this, our individual measurements are in good agreement with these independent 
results obtained also for galaxies selected with different criteria. This suggests that in general
star-forming galaxies show a comparable relation between extinction and stellar mass.
Indeed the $z \sim$ 1.5--3 galaxies plotted in this figure show a similar extinction to low-z
galaxies \citep{2007ApJS..173..342M,2009A&A...507..693B,Whitaker2012,2013ApJ...763..145D,2013ApJ...763...92Z}.
However, given different selection criteria and the small sample size, it is difficult to address
if there is a possible redshift evolution of the dust extinction -- stellar mass relation,
as also discussed in \citet{2012A&A...545A.141B}.

% % % % % % % % % % % % % % % % % % % % % % % % % % % % % % % 

\subsection{Dust properties}
\label{s_dust}

We now discuss the physical properties we derive for the dust content of our sample, temperature and mass,
from the exploitation of our IR/submm observations.
%-----------FIG  --------------
\begin{figure*}[!ht] 
%\begin{center} 
  \vspace*{-0.8in} 
   \includegraphics[width=18 cm]{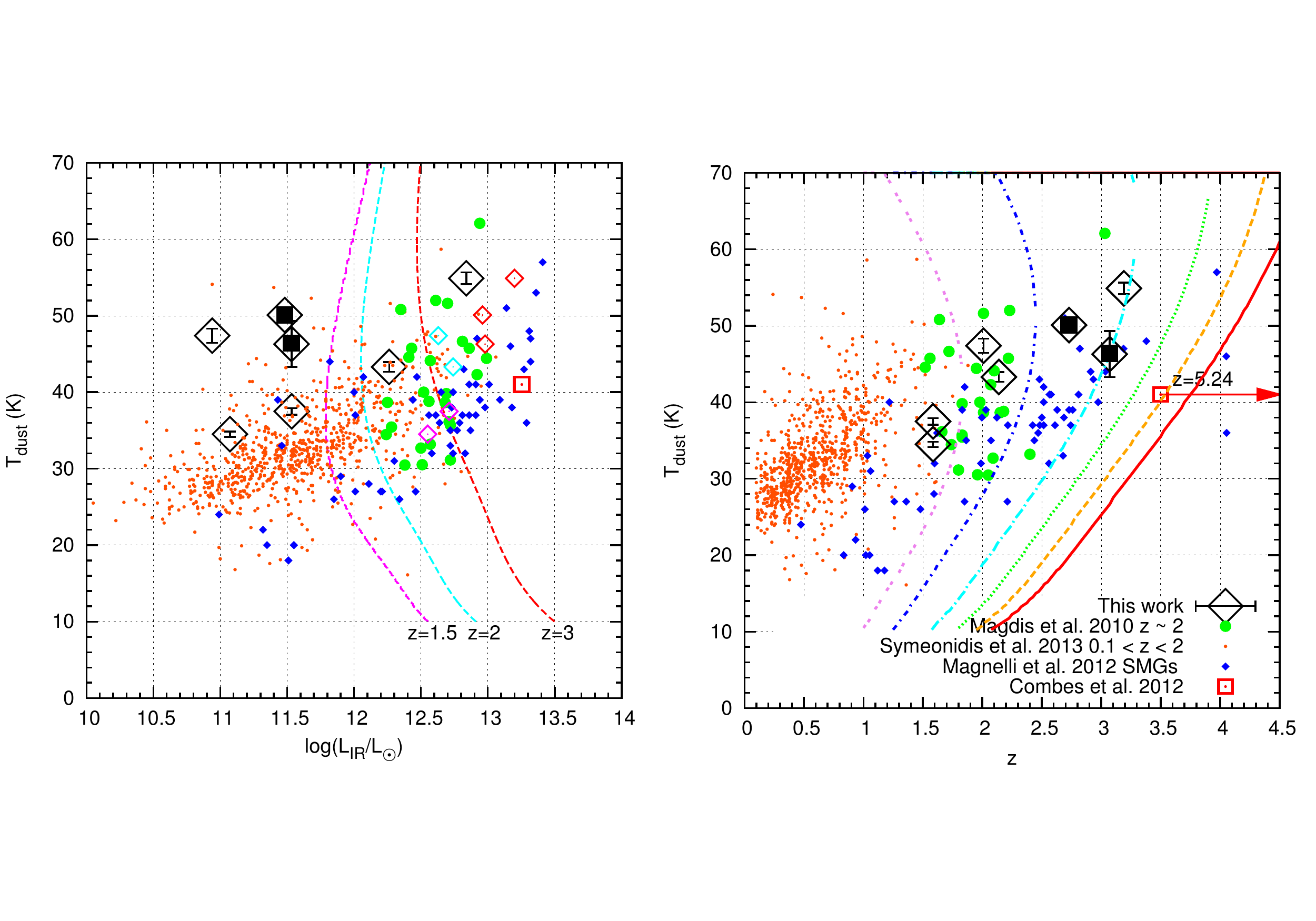}
   \vspace*{-0.9in}  
     \caption{ \textit{Left:}  \lir\ vs \tdust\ diagram, plotted against the larger sample of \protect\citet{2013arXiv1302.4895S} (orange dots), 
     			 that of \protect\citet{2010MNRAS.409...22M} (green circles) and the SMG sample of \cite{2012A&A...539A.155M} (blue diamonds). 
			 cB58 and the poorly constrained Cosmic Eye are shown by black squares surrounded by a diamond.
     			 Our sample (large hollow diamonds) has warmer
     			 \tdust\ compared to same luminosity lower redshift galaxies. Actually our sample has similar temperatures with the 
     			 ULIRG sample of \protect\citet{2010MNRAS.409...22M}, indicating overall warmer temperatures at $ z\sim2 $ or more than in
     			 the more local Universe. The SMGs that lie closest to our sample in this parameter space and appear warmer than the main
     			 SMG trend are among the most strongly lensed ones in \cite{2012A&A...539A.155M}'s sample, suggesting we have similar objects
     			 in our two surveys. The three dashed curves represent the PACS160 $ 5\sigma $ detection limits at the redshifts mentioned below them.
     			 The smaller hollow color-coded diamonds represent the observed (lensed) luminosities of our sources, which we can see lie well above 
     			 their corresponding limiting curves.
     			 \textit{Right:} Redshift vs \tdust\ diagram. Our sample lies together the the $ z\sim2 $ ULIRGs and with many of the SMGs at
     			 same redshift. Compared to the latter in particular, our sample has similar temperatures with the SMGs that are in the ULIRG -
     			 HyLIRG regime, again indicating no strong evolution of \tdust\ with \lir\ at the considered redshifts. The rainbow colored curves
     			 represent from left to right the $ 5\sigma $ limits in \tdust\ per redshift 
     			 for an object of respective $ \lir = [1,2,4,6,8,10\times 10 ^{12} \lsun] $.
     			 This illustrates the fact that for their corresponding luminosities and redshifts, 
     			 our objects lie way above the minimal detectable \tdust.
     			}
     						\label{fig_tdust}
%  \end{center}
\end{figure*} 
%-----------FIG --------------   

\subsubsection{Dust temperature}
\label{s_tdust}

We have performed modified black body fits on our FIR/submm
photometry (starting from restframe 40 \micron\ and longwards), 
for two values of the cold end slope $ \beta =$ 1.5 and 2.0. As we do not have strong constraints from the 
submm and longwards, both values produce fits of similar quality.To allow meaningful comparison with the other studies
discussed in Sect. \ref{s_tdust}, we present the temperatures for the $ \beta= 1.5 $ fits, as it is the value used by the mentioned publications
\footnote{\tdust\ depends on the chosen value of $ \beta $ as the Wien's displacement law is modified as $ \tdust(K) \sim \frac{hc}
{(3+\beta)k\lambda_{\rm peak}} $ (in the case of a black body in the $ F_{\nu} $ formalism). For instance, the temperatures we derive for 
$ \beta = 2$ are $ \approx 10-13\%$ lower than the ones shown in Table \ref{OBSERVED-GLOBAL} and the figures.}.

The dust temperatures obtained for our sample (\tdust\ in Table \ref{OBSERVED-GLOBAL}, as prescribed in the beginning of the section) 
cover a range between $ \sim $ 35 and $ \sim $ 55 K, typical for star-forming galaxies and starbursts. 
There seems to be no particular correlation between \tdust\ and \lir, in contrast with the trends often found from various samples
(Fig. \ref{fig_tdust}).
Our galaxies occupy mostly the same temperature range as the star-forming ULIRGs of 
\citet{2010MNRAS.409...22M} and as the brightest (and warmest) SMGs of \cite{2012A&A...539A.155M} at z $ \sim 2-3 $. 
Our median temperature for the sample is $\approx $ 46 K, which is very similar to the median value of 42.3 K 
inferred by \citet{2010MNRAS.409...22M}. In comparison with the large sample of \citet{2013arXiv1302.4895S} (this sample was carefully selected
among the COSMOS and GOODS-Herschel fields in order for the properties of these galaxies to be representative of the whole IR-luminous population 
up to z $ \sim $2)
our objects lie systematically on the warm side of its distribution, or above.
Concerning the recent publication of \cite{2013arXiv1309.3281S} and our two galaxies in common -- cB58 and the Cosmic Eye -- 
we find our estimates of \tdust\ to be $ \sim $5 K cooler but within reasonable range of our respective uncertainties.

This tends to show that although lensing allows us to probe much fainter galaxies at z $ \geq 1.5 $, these galaxies are not colder than the bright ones 
at these redshifts. This is either indicative of a selection bias towards higher temperatures, or of a redshift -- \tdust\ relation.

Our objects all peak around the SPIRE 250\micron\ band. In the redshift 
range of z $ \sim $ 2--3 objects of similar observed luminosity 
($ \lir\times\mu\geq 10^{12}\lsun $) and temperatures $ \sim 15 $ K lower 
than ours would still peak within the SPIRE bands, would have
higher intrinsic fluxes and thus would be detectable. This means 
that only colder galaxies with lower intrinsic \lir\ and/or magnification would fall
undetected. As an example, we can consider a galaxy at z $ \sim 3$ with $ 
\lir\times\mu\approx 10^{13}\lsun $, like the Cosmic Eye, which is well detectable
whether it peaks at 47 K (like the Eye) or at say 30 K. If it was 10 times fainter, 
it would still  be detectable in the PACS bands if at $ \sim 47 $
K, but would go undetected by all bands if it were at $ \sim30$ K. It seems so 
that for the detection levels of our objects we have not reached the
lowest detectable temperatures, indicating that we are not biased in that sense, 
but more that such IR-bright but colder objects were not to be 
found in the HLS fields we'explored so far.
This is shown quantitatively in Fig. \ref{fig_tdust}b. in the rainbow 
colored curves where we have computed the minimal dust temperatures detectable at a given
redshift for \lir 's from $10^{12}$ to $10^{13}\lsun $, corresponding to the global luminosity range of our sample.
Specifically the curves represent  $ 5\sigma $ detection limits of $ \sim 6 $ mJy in PACS160. 
In Fig. \ref{fig_tdust}a. we also show the same limits for luminosity and temperature at three
given redshifts, representative of of our sample.
Clearly, our various observed luminosities  
lie well above their respective limits. This shows that our sample is not limited neither by temperature nor
luminosity.
Also, given the hand-picked 
selection of our objects and the consideration of all {\it Herschel} bands, sources
detected in SPIRE but dropping out in PACS would not have gone unnoticed.
The SMGs that are biased to lower temperatures and low luminosity \citep{2012A&A...539A.155M}
occupy a strip that is only a little bit lower than our objects in the z - \tdust\ space, and provide a hint of the eventual coldest galaxies
we might detect in the global HLS sample. 

\cite{Magdis2012} find a similar trend with redshift in a sample of main sequence galaxies.
The same trend may not apply, though, to other galaxies such as SMGs.

%%% %%% %%% %%% %%% 
\subsubsection{Dust masses}
\label{s_mdust}

%-----------FIG --------------   
\begin{figure}[htb]
	\centering
	%\vspace*{-1in}
	\includegraphics[width=8.8cm]{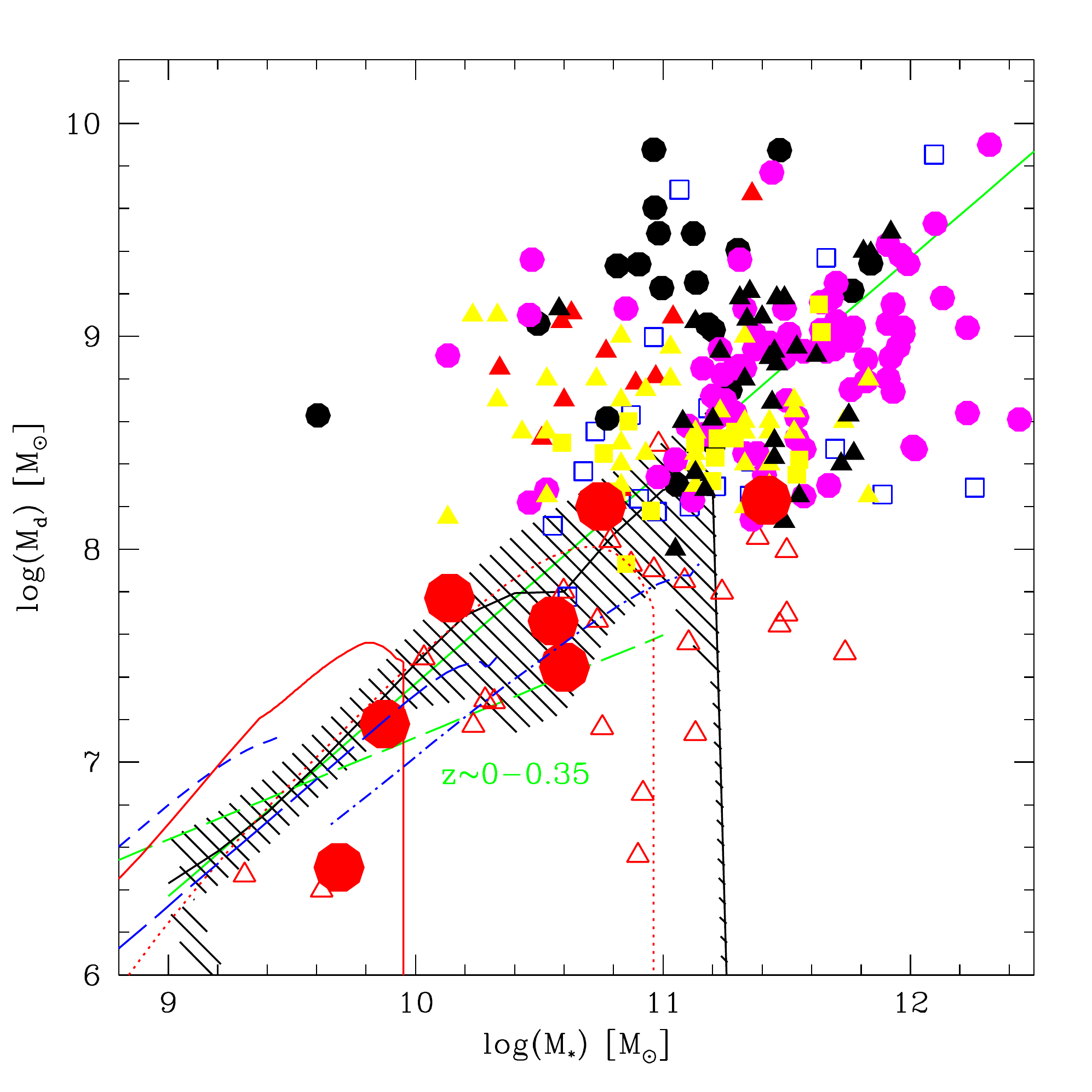}
	%\includegraphics[width=8.8cm]{plot_mstar_mdust_michalowski.pdf}
	 %\includegraphics[width=8.8cm]{plot_mstar_mdust_lofaro.pdf}
    %  \vspace*{-0.9in} 	
	\caption{Dust mass as a function of the stellar mass of our lensed galaxy sample (red circles) compared
	to other high-z (filled symbols) and local galaxy samples (open symbols).
	The stellar masses used here come from the energy conserving models, presented in Table \ref{energy_consist_table}.
	 Red triangles and blue squares
	(both open) show local spirals and ULIRGs; filled black circles show high-z SMGs; all taken from Fig.\ 1
	of \citet{Santini2010}.
	Filled magenta circles show the $z \sim 0.5-3.5$ SMG sample of \cite{2010A&A...514A..67M},
	yellow squares 17 $z>1$ galaxies observed with {\it Herschel} in GOODS-N 	\citep{Magnelli2012}; 
	filled magenta triangles the $z \sim$ 0.5--3 galaxies from \citet{Magdis2012};
	filled black triangle the $z \sim$ 1--2 (U)LIRG from \citet{Lofaro2013}.
	The green dashed line shows the location of the sequence observed by the H-ATLAS/GAMA survey at
	$z \sim$ 0--0.35 \citep{Bourne2012}; the green solid line the median value of $M_d/\mstar=-2.63$ obtained
	by \cite{Smith2012} from the H-ATLAS survey after adjustment to the Salpeter IMF used here.
	The red solid (dotted) line shows the predictions from the chemical- and dust-evolution models of
     \citet{2011A&A...525A..61P} for galaxies leading to the formation of ellipticals with masses
     of $10^{10}$ ($10^{11}$) \msun\ at $z=0.$; blue lines the dwarf, M101, and Milky Way models of
     \citet{Calura2008}.
     	The blacked shaded area (black line) shows the 68\% confidence interval (median) predicted by the semi-analytical models
	of \citet{Lagos2012} for $z \sim 1.5$.	The inferred dust and stellar masses of our galaxies seem to follow a 
	simple \mstar-$M_d$ relation extending also	to the SMGs . 
	They do not show significant offsets from low 
	redshift galaxies, and are in good agreement with the models.
	}
 
\label{fig_mdust}
\end{figure}
%-------------------------

In this section we present the dust masses $M_{\rm d} $ for our objects we derived from the {\it Herschel} data.
One straight-forward way to do this is with the help of the flux -- $M_{\rm d} $ calibration \citep{2003pid..book.....K}:

\begin{equation}
			M_{\rm d} = \frac{S_{\nu}(\lambda_{\rm obs}) D{_L}^2}{(1+z) \kappa(\lambda_{\rm rest}) B_{\nu}(\lambda_{\rm rest}, \tdust) }		
		\label{mdust_relation}
\end{equation}
where  $S_{\nu}(\lambda_{obs})$ is the flux at a given observed wavelength, $D{_L}  $ the luminosity distance, $ \kappa(\lambda_{rest}) $ the dust grain opacity
per unit of dust mass and $  B_{\nu} $ the Planck function. For the opacities we follow the calibration of \citet{2001ApJ...554..778L}
\footnote{Compared to Draine et al. (2003), this calibration yields dust mass a factor 1.2 smaller.},
given by:
\begin{equation}
		\kappa(\lambda) \approx 2.92 \times 10^5 \left(\frac{\lambda}{\mu m}\right)^{-2}
		\label{LD1_a}
\end{equation}
for $ 20 < \lambda < 700 $ \micron, and
\begin{equation}
		\kappa(\lambda) \approx 3.58 \times 10^4 \left(\frac{\lambda}{\mu m}\right)^{-1.68}
		\label{LD1_b}
\end{equation}
for $700 < \lambda < 10^4 $ \micron. 
We estimate $S_{\nu}$ from a modified black-body fit to the \textit{Herschel}, and, when available, longer wavelength data with a $\beta$-slope fixed to~2, which is the form compatible with the above calibration.
We present the dust masses thus obtained in Table~\ref{dust_mass}.
We note that this is a different fit from the one used to estimate dust temperatures, which assumes a $\beta$ of~1.5.  The values of \tdust\ and $M_{d}$ given here should, therefore, not be used in conjonction.  This duality was needed in order to have temperatures determined in a manner consistent with those of the comparison samples taken from the litterature, although these might not necessarily represent the actual temperature of any physical component.  We note also that this approach is valid within the quality of our data which does not allow us to constrain well the Reyleigh-Jeans slope.  We find that fits are equally good with either value of $\beta$.

\begin{table}
	\centering
	\caption{Dust masses derived from MBB fits with $ \beta = 2 $ for our sample. We recall here too
			that for MACS0451 we list the mass derived for the northern segment.}
												\label{dust_mass}
    	\begin{tabular}{l c }% c c }
    	\hline\hline	
    	ID	  &  $M_{\rm d} $ (MBB)		\\ 		 
    	&	[$  10^7 $\msun]\\ 
    	\hline
    A68/C0	&		$ 4.6 \pm 0.2 $ \\ 
	A68/h7	&	$ 17   \pm   1.27		 $	 \\
	A68/HLS115 &		$ 5.89  \pm  0.34	 $	   \\  
	A68/nn4  &		$ 	15.9   \pm 1.17 $	\\ 
	MACS0451 N  &$ 0.32  \pm 0.05	 $	 \\ 
	cB58	&	1.51 $ \pm 0.11	 $	 \\ 
	Cosmic Eye&		$ 	2.79   \pm 1.27 $	\\
    \hline
    \end{tabular} 
\end{table}

In Fig.\ \ref{fig_mdust} we show the derived dust masses of our lensed targets as a function of stellar mass
and compare them with other recent samples, both at low and high redshift.
The figure clearly shows that our measurements extend the currently available dust masses at $z>1.5$
to lower values (down to $M_{\rm d}  \approx 10^7$ \msun), again, because of the strong lensing. Our galaxies appear to show
a similar relation between dust and stellar mass as the one found for lower redshift galaxies, like the ones observed with the recent
H-ATLAS/GAMA survey, or like nearby spirals or ULIRGs \citep[cf.][]{Santini2010,Bourne2012}.
They are consistent with a constant dust-to-stellar mass ratio of $M_d/\mstar \approx -2.6$, the 
median value obtained by \citet{Smith2012} from the H-ATLAS survey.
Our data indicate a continuity in \mstar--$M_d$ with the $z \sim$ 0.5--3.5 ($z_{\rm median} \approx 2.1$)
sub-mm galaxy (SMG) sample of \cite{2010A&A...514A..67M} (magenta points in Figure~\ref{fig_mdust}), although \citet{Santini2010} suggest 
that the high-z SMGs show a higher dust/stellar mass ratio, as shown
by the filled black circles.

We compare our measurements with predictions of the dust content of galaxies from the chemical evolution models of \citet{Calura2008} and \citet{2011A&A...525A..61P} which include dust production and destruction.  These are also shown in Fig.\ \ref{fig_mdust}.
The model ``tracks" plotted here correspond to the evolution of galaxies that become ellipticals of mass $10^{10}$ and $10^{11}$
\msun\ at $z=0$ as well as to the modelled evolution of the Milky Way and M101.  
During their star-forming phase, the models cover well the observed range of our
observations and their approximate $M_d-\mstar$ slope, and are in good agreement with the
observed SFR--\mstar\ values.
We also show the predictions from the semi-analytical galaxy models of \cite{Lagos2012} for galaxies
at $z \sim 1.5$. Models at other redshifts trace a very similar locus in  \mstar-$M_{d}$.
Again, these models describe quite well  the observed \mstar-$M_{d}$ relation of our galaxies.
In short, the bulk of the data at high-z seems to follow a simple relation between stellar and dust mass,
and this relation does not seem to differ significantly from the one observed at lower redshift.

We  note that our galaxies show larger ratios of IR luminosity to dust mass, \lir/$M_{\rm d} $,
typically by a factor $\sim 5$, than derived by \citet{Magdis2012} from stacking for $z \sim 2$ galaxies with masses $\mstar \sim
10^{10}-10^{11}$ \msun. Although again systematics for the dust masses may be a factor $\sim$ 2 approximately,
this probably does not explain this difference. A more detailed analysis is deferred to a subsequent publication,
where we also include measurements of the (molecular) gas mass.

\subsection{Other implications}
 
As discussed in Sects.\ \ref{s_cb58} and \ref{s_eye}, our work has led
to revised stellar masses for the well-studied lensed galaxies cB58 and the Cosmic Eye found at redshift 2.7 and 3.07.
Our masses are broadly in agreement with other recent studies, but are found to be a factor $\sim$ 5--6 lower
than earlier studies often used in the literature. The origin of the discrepancies is mostly understood 
(Sects.\ \ref{s_cb58} and \ref{s_eye}).
For example, this implies that while \cite{2010ApJ...724L.153R} find that cB58 and the Cosmic Eye show high specific star formation
rates compared to the ``normal'' population at this redshift, this is not the case anymore with our results.
Also, the underestimate of the stellar masses of these galaxies imply for example that quantities as estimated gas fraction,
gas depletion timescale, and other related quantities may need to be revised. This will be discussed in a companion
paper presenting new molecular gas measurements of the lensed galaxy sample considered in this paper.
 
 \subsection{Other multi-wavelength SED models of high-z galaxies and future improvements}
 
We now briefly discuss other  recent multi-wavelength SED models and their methodology, and 
discuss future improvements of our method.
 
Although various codes allowing for multi-wavelength SED fits exist,
relatively few studies of distant galaxies analyzing the combined optical, near-IR and IR SED  observed 
with {\it Herschel} have presently been published. 
For example, \cite{2012ApJ...758L...9M} have analyzed a small sample of $z \sim 2$ LBGs, but do not discuss in detail
the stellar populations and properties such as the SFH and attenuation law discussed here.
\cite{2012A&A...545A.141B} have fitted 750 UV-selected galaxies at $z=$ 1--2 with CIGALE, a code doing 
energetically-consistent SED fits. They adopt a model including two separate stellar populations 
with an old/young component (each with a separate attenuation), and also fit a parametrised attenuation law.
For $\sim$ 20\% of their galaxies they find indications for an attenuation law steeper than the Calzetti law.
Some of their results have already been compared with ours (cf.\ Sect.\ \ref{s_av}); others are difficult to compare. 
\cite{Lofaro2013} have presented detailed fits of 31 (U)LIRG, 20 of them at $z \sim 2$, with the GRASIL code.
This energetically-consistent code takes various stellar and dust emission components into account, accounts for a variety of SFHs, 
and is described by a large number of parameters; see \cite{Silva2011} for more information.
From their analysis they find that all of the galaxies requires the presence of an old ($>1$ Gyr) population,
and at the same time host a moderate ongoing SF activity, i.e.\ a higher SFR in the past.
Their model also predicts a lower SFR than expected from the IR luminosity with standard \sfrir\ calibrations, 
since a non-negligible fraction of the IR emission originates from cirrus heated by evolved stellar populations.
Finally, they find that the stellar masses derived from SED fits with simple models, similar to ours, 
may be underestimated --- by 0.36 dex for their ULIRG sample --- especially for the most dusty galaxies.
From their extinction, stellar mass, and \lir, only two of our objects, A68/h7 and A68/HLS115, are in a similar
domain than the ULRIGs of \cite{Lofaro2013}, where their results/caveats may apply. 
However, given the different methodologies and different object selection, it is difficult to compare
their results with ours. In particular, none of these studies compares systematically different 
SFHs and attenuation laws, as done here, and includes nebular emission.

In any case, we note that the CIGALE and GRASIL models applied by \cite{2012A&A...545A.141B} and 
\cite{Lofaro2013} are more complex than ours, involving in particular several stellar populations and dust components,
and more free parameters. In a first, conservative step we have here chosen much more simple
SED models with a minimal number of parameters (three: age, $\tau$, and \av), and we explore the consistency between
the stellar part of the SED and the IR by verifying the energy balance between absorbed stellar and
re-emitted IR radiation. Such a simple model is also motivated by the
fact that similar models have often been applied for the analysis of galaxies at higher redshift,
including our study of a large sample of LBGs at $z \sim$ 3--8 \citep{2012arXiv1207.3663D,schaerer&debarros2010}.
Although our approach has lead to some interesting insight e.g.\ on the star formation histories and
attenuation laws of high-z galaxies, it is, however, not possible to demonstrate that our conclusions
may not be altered if a different, more complex model was adopted.
Also, in a next step our model could include the energy balance as a constraint in the fitting
procedure, in a similar fashion to CIGALE, GRASIL or the MAGPHYS code \cite{dacunha2008},
or it could also include available emission line measurements as constraints.
Obviously such an approach would reduce the uncertainties on the derived physical parameters.
However, some methodological issues remain, e.g.\ whether to include observed fluxes in all the IR
bands or more fundamentally the derived IR luminosity and others. Such improved models
will be applied in the future.

%%%%%%%%%%%%%%%%%%%%%%%%%%%%%%%%%%%%%%%%%%%%%%%%%%%%%%%%%%%%%%%%%%%%%%%%%%%%%%%%%
\section{Summary and conclusions}
\label{s_conclude}

We have studied in detail a small new sample of {\it Herschel} detected lensed galaxies in the redshift interval $z \sim$ 1.6--3.2.
We extended our initial 5-object sample from the \textit{HLS} to include 2 other strongly lensed star-forming galaxies that have robust
Herschel detections (the ``Cosmic Eye", and cB58). We have extracted and compiled the photometry for our sample, covering a large
wavelength range (from rest-frame UV to the FIR/submm), including observations from {\it HST}, CFHT, {\it Spitzer}, {\it Herschel} and SCUBA2.
Lensing has enabled us to extend {\it Herschel} blank field studies to lower luminosity. As seen in Fig. \ref{fig_lir_z}, our cases of strong
lensing allow us to measure IR luminosities up to more that one order of magnitude lower than for non lensed objects.

We have performed SED-fitting of these galaxies using our modified version of the \textit{Hyperz} code, modeling their stellar populations
and dust emission. The large wavelength coverage enabled us to perform interesting 
tests in trying to distinguish various SFHs, and extinction laws,
with the \lir\ serving as an \textit{a posteriori} consistency check for the validity 
of the scenarios explored. The main conclusions we derive from this approach are:
\begin{itemize}
\item SED models with nebular emission and variable SF histories provide good fits to $z \sim$ 1.6--3 galaxies.
They do not predict too high IR luminosity, as might have been suspected if models yielded too young ages. 
Nebular emission does not lower the age to ``unrealistic'' values, at least not
if they were not preferred even before its inclusion. It also contributes to reduce the age-extinction degeneracy
in some cases (see Sect. \ref{sec_h7}). 
\item Although for some cases the use of the SMC extinction law produces slightly better fits in terms of \ki2, it is not appropriate 
to describe our sample as it favors old populations which underpredict the observed IR luminosity. 
Two cases seem to achieve a correct prediction (nn4 and cB58), and their only common feature is that their corresponding SMC based models invoke
young ages in opposition with the rest of the sample ($ > 90 $ Myr).

\item IR luminosity in combination with emission line measurements can constrain the SFH and extinction
law of SF galaxies (at least in some cases), as proposed by \cite{2013A&A...549A...4S}. In particular for the Cosmic eye,
rising SFH or constant SFR can be excluded as they strongly overpredict the observed line fluxes and \lir. 
A declining SFH in addition 
with stronger line extinction \citep{Calzetti2001} is the model reproducing the most accurately 
the observed spectrum. The case of cB58 that also has
a wide spectroscopic coverage in the literature is more mitigated in the sense that most 
models and both SMC and Calzetti extinction laws reproduce accurately 
the line measurements and observed \lir. This is probably linked to the fact that 
this galaxy's population appears to be very young in age (thus not allowing 
for much difference to build up between the various SFHs), and has a very 
blue slope in the rest-frame UV (thus not allowing much distinction between
the 2 extinction laws we explored).

\item Our ``normal'' star-forming galaxies lie close to the Main Sequence 
\citep{elbaz2011} even for the variable SFHs we have explored. Notable exceptions 
are nn4 that is a very intense starburst, MACS0451 which seems to be starbursting 
too but is also located in the very low stellar mass regime, and finally A68/h7 
that lies far beneath the MS, but still close to it with the 1$ \sigma $ 
confidence level due to its strong degeneracy. As it could be expected, the
``classical'' recipe SFH (CSFR, age prior, no nebular emission) shows less spread, 
and gives a flat relation in the \mstar-- sSFR plane only slightly higher
than the mean derive value of \cite{2012ApJ...754...25R}. Lensing has enabled us 
to extend the \mstar-SFR diagram of IR-detected galaxies at  z $ \sim 2 $ towards lower 
masses and SFR.

\item The comparison of the SFR indicators between the SED-inferred ones, SFR(SED),
and the ones derived from the observed UV+IR luminosities via 
straightforward application of the Kennicutt calibrations overall agrees within 0.3 dex
when considering the models with nebular emission. Calzetti-based free SFH models and no 
emission give systematically SFR(SED) that are  too large
when compared to the observationally inferred ones ($ \sim 1 $ dex).
In any case, direct observables (\lir, \luv\ etc.) which can be consistently
derived from SED fits should be compared instead of comparing with 
SFR calibrations making specific assumptions.

\item The UV extinctions inferred from the $\lir /\luv$ measurements is in broad agreement with the main trend derived in \cite{2012A&A...545A.141B}, but 
the small statistics of our sample do not allow us to state that there is a universal $ {\rm A_{UV}} $ -- \mstar\ relation from low to high z.

\item Furthermore, the use of the observed $\lir /\luv$ ratio to constrain  \av\ proves very useful in breaking the 
age-extinction degeneracy that many of our red-sloped galaxies suffer from, and produces population models that are coherent with the 
observationally derived SFR estimates. Among the declining SFHs explored, our sample shows a bimodal tendency to either prefer fast decaying 
bursts ($ \tau \leq 100 $ Myr) or constant star formation.

\end{itemize}

Next we have sought to characterize the dust properties, namely temperature and mass of our sample. For that we have performed 
modified black body fits of the FIR/submm photometry.
In our temperature analysis, we observe that our objects appear to be warmer than other star forming galaxies at low (z $< 1  $)
redshift with similar luminosity (above the median trends of \cite{2013arXiv1302.4895S}'s sample but mostly within the scatter)
. They actually seem to occupy the temperature ranges of more luminous objects at their corresponding
redshift \citep{2010MNRAS.409...22M}, indicating a possible trend with z. Our sample is not temperature limited but rather luminosity limited,
and although we probe luminosities up to 1 dex lower than for blank field surveys 
at z $ \sim 2-3 $ we find no evident correlation between \tdust\ and \lir. However, in order to robustly claim the observation of a general
shift towards higher dust temperatures between the local Universe and higher redshifts, we need to conduct our analysis on the largest HLS sample
possible.

The dust mass study shows our galaxies to occupy the same space as local 
spirals in the \mstar--$M_d$ plane and are extending samples of other studies
at higher z towards lower regime ($M_d \leq 10 ^{7} \msun $ ). The stellar-dust mass relation is found
in good agreement with the chemical- and dust-evolution models of
\cite{Calura2008,2011A&A...525A..61P},the semi-analytical galaxy models of \cite{Lagos2012} and with
observations of nearby spirals and ULIRGs \citep[][]{Santini2010}.

This work is the first from the \emph{HLS} survey to exploit the {\it Herschel} observations together with a large
amount of ancillary data, covering the SED from the rest-frame UV to the IR/mm.  It has shown the strength of this survey
to probe the faint Universe thanks to lensing. Although the sample is very limited, our present 5 sources come from 
only 2 out of the  54 targeted clusters. The next step for our study in the immediate future consists 
of course to increase our sample's size by including sources of the other clusters.

%%%%%%%%%%%%%%%%%%%%%%%%%%%%%%%%%%%%%%%%%%%%%%%%%%%%%%%%%%%%%%%%%%%%%%%%%%%%%%%%%
\begin{acknowledgements}

We thank Claudio Lagos and Antonio Pipino for predictions
from their models and for advice on their use. 
We acknowledge support for the International Team 181
from the International Space Science Institute in Berne.
This work was supported by the Swiss National Science Foundation.
SCUBA2 observations have been done on the James Clerk Maxwell Telescope.
The James Clerk Maxwell Telescope is operated by the Joint Astronomy
Centre on behalf of the Science and Technology Facilities Council of the
United Kingdom, the National Research Council of Canada, and (until 31
March 2013) the Netherlands Organisation for Scientific Research.

\end{acknowledgements}

\bibliographystyle{aa}
\bibliography{refs,references_ds}

\begin{appendix}
\section{Photometry tables}

\begin{table*}
\caption{Restframe UV to FIR photometry sets for our sample (AB magnitudes up to the IRAC bands included,
then in mJy) with 1$ \sigma $ uncertainties. B and R bands correspond to CFHT/12k for A68, whereas B, V 
and Ic in the case of MACS0451 correspond to Subaru's SuprimeCam. Ks photometry comes from UKIRT, z is from VLT/FORS2, and g,r,i from SDSS. 
}
\begin{center}

\begin{tabular}{c c c c c c }
\hline\hline
Observing Band & C0 & h7 & HLS115 & nn4  & MACS0451\tablefootmark{a} \\ \hline
B		& 22.300 $\pm$ 0.104	& 23.298 $\pm$ 0.131	& 22.968 $\pm$ 0.048	& $ < $26.60	& 20.838 $ \pm $ 0.020	\\
g		&-	&-	&	&-	& 20.606 $ \pm $ 0.045	\\
V		&-	&-	&	&-	& 20.425 $ \pm $ 0.020	\\
F602W 	&-	&	-&	&-	& 20.347 $ \pm $ 0.020	\\
R		&-	&21.994 $\pm$ 0.050	&	& $ < $26.28	& 	-\\
r		&-	&-	&	& -		& 20.175 $ \pm $ 0.050	\\
F702W	&21.154 $\pm$  0.083	&-	&21.896 $\pm$ 0.050	&-	& -	\\
i		&-	&-	&&-	& 20.299 $ \pm $ 0.076	\\
Ic		&-	&-	&	&-	&  20.148 $ \pm $ 0.020\\
F814W	&20.802  $\pm$ 0.078	&21.602 $\pm$ 0.027	&21.450 $\pm$ 0.060&-	& 20.062 $ \pm $ 	0.020\\
 z		&-	&21.360 $\pm$ 0.053	& 21.144 $\pm$ 0.072	& $24.70\pm 0.21$	&-	\\
J/ISAAC	&	-&20.104 $\pm$ 0.029	& 20.278 $\pm$ 0.049	& $22.98\pm 0.06$	&-	\\	
F110W	&19.643 $\pm$ 0.031	&	&	&-	&19.675 $ \pm $ 0.020	\\	
F140W	&-	&-	&	&-	&19.181 $ \pm $ 0.020	\\	
F160W 	&19.118 $\pm$ 0.030	&-	&	&    &	\\	
H/ISAAC	&-	&19.775 $\pm$ 0.040	&19.928 $\pm$ 0.052 & $22.33 \pm 0.06$ 	&-  	\\
Ks		&18.643 $\pm$ 0.091	&19.470 $\pm$ 0.036	&19.285 $\pm$ 0.030	& $ 21.45\pm 0.04$	&-  	\\	
IRAC 3.6\micron	& 17.925 $\pm$ 0.061	& 18.635 $\pm$ 0.027	&$18.885 \pm  0.029$&$ 20.53\pm0.04 $	& 18.913 $ \pm $ 0.020	\\	
IRAC 4.5\micron	& 17.693 $\pm$ 0.067	& 18.441 $\pm$ 0.038 	&	$ 18.677\pm 0.031 $  &$ 20.15\pm0.04 $	& 18.898 $ \pm $ 0.020	\\	
\hline
MIPS 24\micron\ (mJy)	&~~0.945  $\pm$ 0.060	&~~$0.341   \pm 0.060  $	&	~~$0.69   \pm 0.06 $		&$ < $0.20	&-	\\
PACS 100\micron\ (mJy)	&~~7.73  $\pm$ 0.38	&~~$8.37	 \pm 0.37  $	&	$	19.96   \pm 0.37  $	&~~$8.58	\pm  0.37$	&   ~~$ 5.92\pm  0.35  $	\\
PACS 160\micron\ (mJy)	&31.26  $\pm$ 0.95	&$24.61   \pm 0.95  $	&	$ 41.48   \pm 0.96 $	&$25.26	\pm 0.95 $	&	$18.60 \pm 1.57 $\\
SPIRE 250\micron\ (mJy)	&46.68  $\pm$ 1.68	&$40.16   \pm 1.33  $	&	$53.59   \pm 1.75  $	&$45.30	\pm 1.87 $	&	$29.24\pm  2.81 $\\ 
SPIRE 350\micron\ (mJy)	&39.88  $\pm$ 1.68	&$27.36   \pm 1.48  $	&	$ 37.84   \pm 1.64 $	&$38.04	\pm  2.19 $	&  $15.58 \pm 3.98 $	\\
SPIRE 500\micron\ (mJy)	&21.47  $\pm$ 1.13	&$12.01   \pm 1.29  $	&	$18.64   \pm 1.22  $	&$17.74	\pm 1.50 $	&  $  ~~6.54\pm  4.24 $	\\
SCUBA2 850\micron\ (mJy)&~~5.39  $\pm$ 0.23	&~~$3.26   \pm 0.70  $	& 	-	&~~$ 3.02	\pm 0.65 $	&-	\\
\hline
\end{tabular}
\end{center}
\tablefoottext{a}{For MACS0451 we present the FIR photometry of the northern part, and the stellar emission of the  
full arc. We've established that the northern part accounts for $ \sim40\% $ of the total emission in the visible.} \label{tab_photometry}
\end{table*}

\begin{table*}
\caption{ \emph{Herschel} and submm/mm fluxes measured for cB58 and the Cosmic Eye after deblending (in mJy, 1$ \sigma $ uncertainties).
1.2mm data come from the IRAM 30m telescope, and the 3.5mm from the IRAM  PdBI.
}
\begin{center}
\begin{tabular}{c c c c c c c c c c }
\hline\hline
ID & 70\micron\ & 100\micron\ & 160\micron\ & 250\micron\ & 350\micron\ & 500\micron\ & 850\micron\ & 1.2mm & 3.5mm \\\hline
cB58 &  3.107$\pm$0.496& 7.988$\pm$0.482 & 19.74$\pm$0.71  &  29.70$\pm$1.42  &  24.54$\pm$1.48  & 11.78$\pm$0.93 & 4.2$\pm$0.9 & 1.06$\pm$0.35 & - \\
Cosmic Eye & 3.51$\pm$0.51 & 6.08$\pm$0.56 & 12.86$\pm$0.78 & 22.32$\pm$1.39 & $ <33.64  $ & $ <28.81 $ & - & 1.6$ \pm $0.3\tablefootmark{a} & $ <0.14 $  \tablefootmark{b}\\
\hline
\end{tabular}
\end{center}
\begin{tabular}{l}
\tablefoottext{a}{from \citet{2013arXiv1309.3281S}}, 
\tablefoottext{b}{from \citet{2007ApJ...665..936C}}.
\end{tabular}
\end{table*}

\end{appendix}

\end{document}